\documentclass[preprint,3p]{elsarticle}

\usepackage{lineno}
\modulolinenumbers[5]

\usepackage[utf8]{inputenc}
\usepackage[T1]{fontenc}
\usepackage{amsmath,amssymb,array}
\usepackage{booktabs}

\usepackage{graphicx}
\usepackage[pagebackref,backref, linktocpage=true, pdfpagelabels=true, hyperindex,pageanchor,plainpages]{hyperref}
\usepackage{verbatim}
\usepackage{listings}
\usepackage{caption}
\usepackage{multirow}
\usepackage{xcolor}

\usepackage{color, colortbl}
\usepackage{longtable}

\usepackage{footnote}
\usepackage{perpage} 
\MakePerPage{footnote} 

\usepackage{listings}

\usepackage[T1]{fontenc}  
\usepackage{textcomp}     

\usepackage{lipsum}

\usepackage{fancyvrb}

\newcommand{\TODO}[1]{{\color{blue}~\textsf{[TODO: #1]}}}
\newcommand{\CMT}[1]{{\color{red}~\textsf{[{\sc comment} #1]}}}

\newcommand{\Rlisting}{%
\lstset{frame=single, language=R, basicstyle=\small\ttfamily, numberstyle=\tiny, breaklines, backgroundcolor=\color{gray!20},numbers=none}
}

\newcommand{\configlisting}{%
\lstset{frame=single, language=HTML, basicstyle=\small\ttfamily, numberstyle=\tiny, breaklines, backgroundcolor=\color{gray!10}, numbers=none}
}

\newcommand{\outputlisting}{%
\lstset{frame=single, language=R, basicstyle=\tiny\ttfamily, numberstyle=\tiny, breaklines=FALSE, backgroundcolor=\color{gray!5}, numbers=none}
}

\newcommand{\CRANpkg}[1]{\texttt{#1}}
\newcommand{\code}[1]{\texttt{#1}}

\newcommand{\covidPckg}{\CRANpkg{covid19.analytics}~}

\usepackage[utf8]{inputenc}

\usepackage{etoolbox}

\makeatletter
\newcounter{backrefinst}

\protected\def\rugk@ref@fbackref#1#2{%
  \ifcsundef{rugk@ref@backreflist@\detokenize{#1}}
    {\global\cslet{rugk@ref@backreflist@\detokenize{#1}}\@empty}
    {}%
  \ifinlistcs{#2}{rugk@ref@backreflist@\detokenize{#1}}
    {}
    {\listcsgadd{rugk@ref@backreflist@\detokenize{#1}}{#2}}}

\def\rugk@write@ref@fbackref#1{%
  \if@filesw
    \protected@write\@mainaux{}{\string\rugk@ref@fbackref
      {#1}{\the\value{backrefinst}}}%
  \fi}

\newcommand*{\rugk@create@fbackref@label}[1]{%
  \begingroup
    \refstepcounter{backrefinst}%
    \label{fbackref.\the\value{backrefinst}}%
    \rugk@write@ref@fbackref{#1}%
  \endgroup
}

\newcommand{\fbackref}[1]{%
  \ref{#1}%
  \rugk@create@fbackref@label{#1}%
}
\newcommand{\autobackref}[1]{%
  \ref{#1}%
  \rugk@create@fbackref@label{#1}%
}

\newcounter{backrefpages}
\newcounter{totalbackrefpages}

\newcommand*{\printbackrefpage}[1]{%
  \stepcounter{backrefpages}%
  \pageref{fbackref.#1}%
  \ifnumless{\value{backrefpages}}{\value{totalbackrefpages}}
    {, }
    {}}

\newrobustcmd*{\printlabelbackrefs}[1]{%
  \setcounter{backrefpages}{0}%
  \setcounter{totalbackrefpages}{0}%
  \ifcsundef{rugk@ref@backreflist@\detokenize{#1}}
    {}	
    {\def\do##1{\stepcounter{backrefpages}}%
     \dolistcsloop{rugk@ref@backreflist@\detokenize{#1}}%
     \setcounter{totalbackrefpages}{\value{backrefpages}}%
     \setcounter{backrefpages}{0}%
     \ifnumgreater{\value{totalbackrefpages}}{1}
       {(pp.}
       {(p.}~%
     \forlistcsloop{\printbackrefpage}{rugk@ref@backreflist@\detokenize{#1}})}}
\makeatother

\newrobustcmd{\backcaption}[3][]{%
  \if\relax\detokenize{#1}\relax
    \def\rugk@tmpcapt{\caption[#2]}%
  \else
    \def\rugk@tmpcapt{\caption[#1]}%
  \fi
  \rugk@tmpcapt{#2 ~\printlabelbackrefs{#3} }%
  \label{#3}}

\makeatletter

\newcommand*\figreflistout[1]{}

\newcommand*\makefigreflist{%
    \begingroup
        \makeatletter
        \@input{\jobname.frl}%
        \newwrite\frl@outfile
        \immediate\openout\frl@outfile=\jobname.frl\relax
    \endgroup
    \let\makefigreflist\@empty
    \renewcommand*\figreflistout{%
        \@bsphack
        \begingroup
            \@sanitize
            \@figreflistout}%
    }

\newcommand*\@figreflistout[1]{%
        \protected@write\frl@outfile{}%
            {\string\addtofigreflist{#1}{\thepage}}%
    \endgroup
    \@esphack}

\newcommand*\addtofigreflist[2]{%
    \@ifundefined{frl@#1}%
        {\expandafter\protected@xdef\csname frl@#1\endcsname{#2}}%
        {\expandafter\protected@xdef\csname frl@#1\endcsname{\csname frl@#1\endcsname, #2}}}

\newcommand*\printfigreflist[1]{%
    \@ifundefined{frl@#1}%
	{\ }
	{\ \footnotesize({pp.~\@nameuse{frl@#1}}) }}

\newcommand\defaultcapt{}

\newcommand\figcaption[3][\defaultcapt]{%
    \renewcommand\defaultcapt{#2}%
    \caption[#1\protect\printfigreflist{#3}]{#2}%
    \label{#3}}

\newcommand\figref[1]{%
    \ref{#1}\figreflistout{#1}}

\makeatother

\makefigreflist

\begin{document}

\begin{frontmatter}

\title{covid19.analytics: An R Package to Obtain, Analyze and Visualize Data from the Coronavirus Disease Pandemic}

\author{Marcelo Ponce\corref{mycorrespondingauthor}}
\address{SciNet HPC Consortium, University of Toronto \\
661 University Ave., Suite 1140. Toronto, ON M5G 1M1 - Canada}
\ead{mponce@scinet.utoronto.ca}
\cortext[mycorrespondingauthor]{Corresponding author}

\author{Amit Sandhel}
\address{Brampton, Ontario. Canada}
\ead{amit.sandhel@gmail.com}

\begin{abstract}
With the emergence of a new pandemic worldwide, a novel strategy to approach
it has emerged.
Several initiatives under the umbrella of ``open science'' are contributing to
tackle this unprecedented situation.
In particular, the ``R Language and Environment for Statistical Computing''
\cite{citeR,ihaka:1996} offers an excellent tool and ecosystem for
approaches focusing on open science and reproducible results.
Hence it is not surprising that with the onset of the pandemic, a large number
of R packages and resources were made available for researches working in the
pandemic.
In this paper, we present an R package that allows users to access and analyze
worldwide data from resources publicly available.
We will introduce the \covidPckg package
\cite{covid19analytics}, focusing in its capabilities and presenting a particular
study case where we describe how to deploy the \textit{COVID19.ANALYTICS Dashboard Explorer}.
\end{abstract}

\begin{keyword}
CoViD19 \sep R package \sep ...
\end{keyword}

\end{frontmatter}

\tableofcontents


\section{Introduction}
\label{sec:intro}


In 2019 a novel type of Corona Virus was first reported, originally in the
province of Hubei, China.
In a time frame of months this new virus was capable of producing a global
pandemic of the \textit{Corona Virus Disease} (CoViD19), which can end up in a
\textit{Severe Acute Respiratory Syndrome} (SARS-COV-2).
The origin of the virus is still unclear
\cite{Cyranoski_2020,Mallapaty_2020,Zhou_2020}, although some studies
based on genetic evidence, suggest that it is quite unlikely that this virus
was human made in a laboratory, but instead points towards cross-species
transmission \cite{origin-SARS-CoV-2,Letko:2020aa}.
Although this is not the first time in the human history when humanity faces a
pandemic, this pandemic has unique characteristics.
For starting the virus is ``peculiar'' as not all the infected individuals
experience the same symptoms. Some individuals display symptoms that are similar to the ones of a
common cold or flu while other individuals experience serious symptoms that can cause death or
hospitalization with different levels of severity, including staying in
intensive-care units (ICU) for several weeks or even months.
A recent medical survey shows that the disease can transcend pulmonary manifestations
affecting several other organs \cite{Gupta_2020}.
Studies also suggest that the level of severity of the disease can be linked to
previous conditions \cite{Williamson_2020}, gender \cite{Scully:2020aa}, or
even blood type \cite{Willyard_2020} but the fundamental and underlying reasons
still remain unclear.
Some infected individuals are completely asymptomatic, which makes them ideal
vectors for disseminating the virus.
This also makes very difficult to precisely determine the transmission rate of
the disease, and it is argued that in part due to the peculiar characteristics
of the virus, that some initial estimates were underdetermining the actual
value \cite{Silvermaneabc1126}.
Elderly are the most vulnerable to the disease and reported mortality rates
vary from 5 to 15\% depending on the geographical location.
In addition to this, the high connectivity of our modern societies, make
possible for a virus like this to widely spread around the world in a
relatively short period of time.
Moreover the actual way of transmission is still uncertain, being the most likely
explanations to be droplets or airborne \cite{airborneLancet}.

What is also unprecedented is the pace at which the scientific community has
engaged in fighting this pandemic in different fronts \cite{Fry_2020}.
Technology and scientific knowledge are and will continue playing a fundamental
role in how humanity is facing this pandemic and helping to reduce the risk of
individuals to be exposed or suffer serious illness.
Techniques such as DNA/RNA sequencing, computer simulations, models generations
and predictions, are nowadays widely accessible and can help in a
great manner to evaluate and design the best course of action in a situation
like this \cite{Singh_Chawla_2020}.
Public health organizations are relying on mathematical and data-driven models (e.g. \cite{Duque:2020aa}),
to draw policies and protocols in order to try to mitigate the impact on
societies by not suffocating their health institutions and resources \cite{Adam_2020}.
Specifically, mathematical models of the evolution of the virus spread,
have been used to establish strategies, like \textit{social distancing},
\textit{quarantines}, \textit{self-isolation} and \textit{staying at home},
to reduce the chances of transmission among individuals.
Usually, \textit{vaccination} is also another approach that emerges as a possible
contention strategy, however this is still not a viable possibility in the case
of CoViD19, as there is not vaccine developed yet \cite{Hotez:2020ab,Ahmed:2020aa}.

Simulations of the spread of virus have also shown that among the most
efficient ways to reduce the spread of the virus are \cite{Block:2020aa}:
	increasing \textit{social distancing}, which refers to staying apart
from individuals so that the virus can not so easily disperse among
individuals;
	improving \textit{hygiene routines}, such as proper hand washing, use of
hand sanitizer, etc. which would eventually reduce the chances of the virus
to remain effective;
	\textit{quarantine} or \textit{self-isolation}, again to reduce
unnecessary exposure to other potentially infected individuals.
Of course these recommendations based on simulations and models can be as
accurate and useful as the simulations are, which ultimately depend on the
value of the parameters used to set up the initial conditions of the models.
Moreover these parameters strongly depend on the actual data which can be also
sensitive to many other factors, such as data collection or reporting
protocols among others \cite{doi:10.1063/5.0008834}.
Hence counting with accurate, reliable and up-to-date data is critical when
trying to understand the conditions for spreading the virus but also for
predicting possible outcomes of the epidemic, as well as, designing
proper containment measurements.
\\
Similarly, being able to access and process the huge amount of genetic
information associated with the virus has proben to shred light into
the disease's path \cite{KORBER2020,Boni_2020}.

Encompassing these unprecedented times,
another interesting phenomenon has also occurred, in part related to a
contemporaneous trend in how science can be done by emphasizing transparency,
reproducibility and robustness: an \textit{open} approach to the methods and the
data; usually refer as \textit{open science}.
In particular, this approach has been part for quite sometime of the software
developer community in the so-called \textit{open source} projects or codes.
This way of developing software, offers a lot of advantages in comparison to
the more traditional and closed, proprietary approaches.
For starting, it allows that any interested party can look at the actual
implementation of the code, criticize, complement or even contribute to the project.
It improves transparency, and at the same time, guarantees higher standards due to
the public scrutiny; which at the end results in benefiting every one: the
developers by increasing their reputation, reach and consolidating a widely
validated product and the users by allowing direct access to the sources and
details of the implementation.
It also helps with reproducibility of results and bugs reports and fixes.
Several approaches and initiatives are taking the openness concepts and
implementing in their platforms.
Specific examples of this have drown the Internet, e.g. the surge of open
source powered dashboards \cite{Dong:2020aa}, open data repositories, etc.

Another example of this is for instance the number of scientific papers related
to CoViD19 published since the beginning of the pandemic \cite{Palayew:2020aa},
the amount of data and tools developed to track the evolution of pandemic, etc.
\cite{Callaway_2020}.
As a matter of fact, scientists are now drowning in publications related to the 
CoViD19 \cite{Kwon_2020,Vabret2020}, and some collaborative and community
initiatives are trying to use machine learning techniques to facilitate
identify and digest the most relevant sources for a given topic
\cite{epidyHealth,coronaWhy,coronaAbs}.

The ``R Language and Environment for Statistical Computing''
\cite{citeR,ihaka:1996} is not exception here.
Moreover, promoting and based on the open source and open community principles,
R has empowered scientists and researchers since its inception. 
Not surprisingly then, the R community has contributed to the official CRAN
\cite{cran} repository already with more than a dozen of packages related to
the CoViD19 pandemic since the beginning of the crisis.
In particular, in this paper we will introduce and discuss the \covidPckg
R package \cite{covid19analytics}, which is mainly designed and focus in an open
and modular approach to provide researchers quick access to the latest reported
worldwide data of the CoViD19 cases, as well as, analytical and visualization
tools to process this data.

This paper is organized as follow:
	in Sec.~\ref{sec:covid19Pckg} we describe the \covidPckg,
	in Sec.~\ref{sec:examples} we present some examples of data analysis and visualization,
	in Sec.~\ref{sec:studyCase} we describe in detail how to deploy a web dashboard employing the capabilities of the \covidPckg package providing full details on the implementation so that this procedure can be repeated and followed by interested users in developing their own dashboards.
Finally we summarize some conclusions in Sec.~\ref{sec:conc}.


\section{The \covidPckg R Package}
\label{sec:covid19Pckg}


The \covidPckg R package \cite{covid19analytics} allows users to obtain
live\footnote{The data usually is accessible from the repositories with a 24
hours delay.} worldwide data from the novel CoViD19.
It does this by accessing and retrieving the data publicly available and
published by several sources:
\begin{itemize}
	\item the ``COVID-19 Data Repository by the Center for Systems Science and
Engineering (CSSE) at Johns Hopkins University'' \cite{JHUCSSErepo} for the
worldwide and US data

	\item Health Canada \cite{HealthCanada}, for Canada specific data

	\item the city of Toronto for the Toronto data \cite{TorontoData}

	\item Open Data Toronto for Toronto data \cite{OpenDataToronto}
\end{itemize}

The package also provides basic analysis and visualization tools and functions
to investigate these datasets and other ones structured in a similar fashion.

The \covidPckg package is an open source tool, which its main implementation and API
is the R package \cite{covid19analytics}.
In addition to this, the package has a few more adds-on:
\begin{itemize}
	\item a central GitHUB repository, \url{https://github.com/mponce0/covid19.analytics}
where the latest development version and source code of the package are available.
	Users can also submit tickets for bugs, suggestions or comments using the "issues" tab.
	\item a rendered version with live examples and documentation also hosted at GitHUB pages,
\url{https://mponce0.github.io/covid19.analytics/};
	\item a dashboard for interactive usage of the package with extended capabilities
for users without any coding expertise, \url{https://covid19analytics.scinet.utoronto.ca}.
We will discuss the details of the implementation in Sec.~\ref{sec:studyCase}.
	\item a ``backup'' data repository hosted at GitHUB,
\url{https://github.com/mponce0/covid19analytics.datasets} --
	where replicas of the live datasets are stored for redundancy and
robust accesibility sake (see Fig.~\figref{fig:dataSources}).
\end{itemize}


\subsection{Data Accessibility}
\label{sec:dataAccess}

One of the main objectives of the \covidPckg package is to make the latest data
from the reported cases of the current CoViD19 pandemic promptly available to
researchers and the scientific community
In what follows we describe the main functionalities from the package regarding
data accessibility.

The \texttt{covid19.data} function allows users to obtain realtime data about
the CoViD19 reported cases from the JHU's CCSE repository, in the following
modalities:
\begin{itemize}
	\item \textit{aggregated} data for the latest day, with a great 'granularity' of geographical regions (ie. cities, provinces, states, countries)

	\item \textit{time series} data for larger accumulated geographical regions (provinces/countries)

	\item \textit{deprecated}: we also include the original data style in which these datasets were reported initially.
\end{itemize}

The datasets also include information about the different categories (status)
"confirmed"/"deaths"/"recovered" of the cases reported daily per
country/region/city.

This data-acquisition function, will first attempt to retrieve the data
directly from the JHU repository with the latest updates.
If for what ever reason this fails (eg. problems with the connection) the
package will load a preserved ``image'' of the data which is \textbf{not} the
latest one but it will still allow the user to explore this older dataset.
In this way, the package offers a more robust and resilient approach to the
quite dynamical situation with respect to data availability and integrity.

The package also provides access to historical pandemics records, as well,
as historical time lines for vacination developments and current vaccination
records for CoViD-19.
Historical pandemic and timeline vaccine development records are obtained from
``Visualizing the History of Pandemics'' \&
``The Race to Save Lives: Comparing Vaccine Development Timelines'' infographics \cite{VCpandemics},
while up-to-date current vaccination records are obtained from
``Our World In Data'' CoViD19 data repository \cite{OWIDvaccination}.

In addition to the data of the reported cases of CoViD19, the \covidPckg
package also provides access to \textit{genomics data} of the virus.
The data is obtained from the National Center for Biotechnology Information (NCBI)
databases \cite{NCBI,NCBIdatabases}.

\subsubsection{Data retrieval options}

Table~\ref{table:dataRetrieval} shows the functions available in the \covidPckg
package for accessing the reported cases of the CoViD19 pandemic.
The functions can be divided in different categories, depending on what data
they provide access to.
For instance, they are distinguished between \textit{agreggated} and
\textit{time series} data sets.
They are also grouped by specific geographical locations, i.e. worldwide,
United States of America (US) and the City of Toronto (Ontario, Canada) data.

\begin{table}
\begin{tabular}{l | l}
	argument	&	description	\\
	\hline\hline
	aggregated	&	latest number of cases aggregated by country	\\
	\hline
	\rowcolor{lightgray}
	\multicolumn{2}{c}{Time Series data}	\\
	\hline
	ts-confirmed	&	time series data of confirmed cases	\\
	ts-deaths	&	time series data of fatal cases	\\
	ts-recovered	&	time series data of recovered cases	\\
	ts-ALL		&	all time series data combined	\\
	\hline
	\rowcolor{lightgray}
	\multicolumn{2}{c}{Deprecated data formats}	\\
	\hline
	ts-dep-confirmed	&	time series data of confirmed cases as originally reported (deprecated)	\\
	ts-dep-deaths		&	time series data of deaths as originally reported (deprecated)	\\
	ts-dep-recovered	&	time series data of recovered cases as originally reported (deprecated)	\\
	\hline
	\rowcolor{lightgray}
	\multicolumn{2}{c}{Combined}	\\
	\hline
	ALL		&	all of the above	\\
	\hline
	\rowcolor{lightgray}
	\multicolumn{2}{c}{Time Series data for specific locations}	\\
	\hline
	ts-Toronto	&	time series data of confirmed cases for the city of Toronto, ON - Canada	\\
	ts-confirmed-US	&	time series data of confirmed cases for the US detailed per state	\\
	ts-deaths-US	&	time series data of fatal cases for the US detailed per state	\\
	\hline\hline
\end{tabular}
\caption{List of \textit{data retrieval} functions available in the \covidPckg package.}
\label{table:dataRetrieval}
\end{table}

\subsubsection{Data Structure}
\label{sec:dataStructure}

The \textit{Time Series} data is structured in an specific manner with a given
set of fields or columns, which resembles the following format:

\fbox{\texttt{\fbox{"Province.State"}~|~\fbox{"Country.Region"}~|~\fbox{"Lat"}~|~\fbox{"Long"}~|~\fbox{...~sequence of dates~...}}}

One of the modular features this package offers is that if an user has data
structured in a data.frame organized as described above, then most of the
functions provided by the \covidPckg package for analyzing \textit{Time Series}
data will just work with the user's defined data. In this way it is possible to
add new data sets to the ones that can be loaded using the repositories
predefined in this package and extend the analysis capabilities to these new
datasets.

Sec.~\ref{sec:syntheticData} presents an example of how \textit{external} or
\textit{synthetic} data has to be structured so that can use the function from
the \covidPckg package.
It is also recommended to check the compatibility of these datasets using the
Data Integrity and Consistency Checks functions described in the following
section.

\subsubsection{Data Integrity and Consistency}

Due to the ongoing and rapid changing situation with the CoViD-19 pandemic,
sometimes the reported data has been detected to change its internal format or
even show some \textit{anomalies} or \textit{inconsistencies}\footnote{See
\url{https://github.com/CSSEGISandData/COVID-19/issues/}.}.

For instance, in some cumulative quantities reported in time series datasets,
it has been observed that these quantities instead of continuously increase
sometimes they decrease their values which is something that should not happen\footnote{See for instance, \url{https://github.com/CSSEGISandData/COVID-19/issues/2165}.}.
We refer to this as an inconsistency of \textbf{``type II''}.

Some negative values have been reported as well in the data, which also is not
possible or valid; we call this inconsistency of \textbf{``type I''}.

When this occurs, it happens at the level of the origin of the dataset, in our
case, the one obtained from the JHU/CCESGIS repository \cite{JHUCSSErepo}.
In order to make the user aware of this, we implemented two consistency and
integrity checking functions:
\begin{itemize}
	\item \texttt{consistency.check}: this function attempts to determine
whether there are consistency issues within the data, such as, negative
reported value (inconsistency of ``type I'') or anomalies in the cumulative
quantities of the data (inconsistency of ``type II'')

	\item \texttt{integrity.check}: this determines whether there are integrity
issues within the datasets or changes to the structure of the data
\end{itemize}

Alternatively we provide a \texttt{data.checks} function that will execute the
previous described functions on an specified dataset.

\paragraph{Data Integrity}

It is highly unlikely that the user would face a situation where the internal
structure of the data or its actual integrity may be compromised.
However if there are any suspicious about this, it is possible to use the
\code{integrity.check} function in order to verify this.
If anything like this is detected we urge users to contact us about it, e.g.
\url{https://github.com/mponce0/covid19.analytics/issues}.

\paragraph{Data Consistency}

Data consistency issues and/or anomalies in the data have been reported several
times\footnote{See \url{https://github.com/CSSEGISandData/COVID-19/issues/}.}
These are claimed, in most of the cases, to be missreported data and usually
are just an insignificant number of the total cases.
Having said that, we believe that the user should be aware of these situations
and we recommend using the \texttt{consistency.check} function to verify the
dataset you will be working with.

\paragraph{Nullifying Spurious Data}

In order to deal with the different scenarios arising from incomplete, inconsistent
or missreported data, we provide the \texttt{nullify.data} function, which will
remove any potential entry in the data that can be suspected of these incongruencies.
In addition ot that, the function accepts an optional argument \texttt{stringent=TRUE},
which will also prune any incomplete cases (e.g. with NAs present).

\subsubsection{CoViD19 Genomic Data}
\label{sec:genomicData}
Similarly to the rapid developments and updates in the reported cases of the disease,
the sequencing of the virus is moving almost at equal pace.
That's why the \covidPckg package provides access to good number of the genomics
data currently available.

The \texttt{covid19.genomic.data} function allows users to obtain the CoViD19's
genomics data from NCBI's databases \cite{NCBIdatabases}.
The type of genomics data accessible from the package is described in
Table~\ref{table:genomicTypes}.

\begin{table}
\begin{tabular}{p{.2\textwidth}|p{.5\textwidth}|p{.25\textwidth}}
	type	&	description	&	source	\\
	\hline\hline
	\texttt{genomic}	
			&	a composite list containing different indicators and elements of the SARS-CoV-2 genomic information
			&	\url{https://www.ncbi.nlm.nih.gov/sars-cov-2/}
	\\
	\hline
	\texttt{genome}		
			&	genetic composition of the reference sequence of the SARS-CoV-2 from GenBank
			&	\url{https://www.ncbi.nlm.nih.gov/nuccore/NC_045512}
	\\
	\hline
	\texttt{fasta}		
			&	genetic composition of the reference sequence of the SARS-CoV-2 from a fasta file
			&	\url{https://www.ncbi.nlm.nih.gov/nuccore/NC_045512.2?report=fasta}
	\\
	\hline
        \texttt{ptree}      
                        &       phylogenetic tree as produced by NCBI data servers
                        &       \url{{https://www.ncbi.nlm.nih.gov/labs/virus/vssi/#/precomptree}}
        \\
	\hline
	\texttt{nucleotide}
	\texttt{protein}
			&	list and composition of nucleotides/proteins from the SARS-CoV-2 virus
			&	\multirow{2}{.25\textwidth}{
					\href{https://www.ncbi.nlm.nih.gov/labs/virus/vssi/\#/virus?SeqType\_s=Nucleotide&VirusLineage\_ss=SARS-CoV-2,\%20taxid:2697049}{NCBI Virus DataHub}
					}
	\\
	\cline{1-2}
	\texttt{nucleotide-fasta}
	\texttt{protein-fasta}
			&	FASTA sequences files for nucleotides, proteins and coding regions
	\\
	\hline
\end{tabular}
\caption{Different types of genomic data retrieved by the \covidPckg package.}
\label{table:genomicTypes}
\end{table}

Although the package attempts to provide the latest available genomic data, there are
a few important details and differences with respect to the reported cases data.
For starting, the amount of genomic information available is way larger than
the data reporting the  number of cases which adds some additional constraints
when retrieving this data.
In addition to that, the hosting servers for the genomic databases impose
certain limits on the rate and amounts of downloads.

In order to mitigate these factors, the \covidPckg package employs a couple of
different strategies as summarized below:
\begin{itemize}
	\item most of the data will be attempted to be retrieved live from NCBI
databases
		-- same as using \code{src='livedata'}.
	\item if that is not possible, the package keeps a local version of
some of the largest datasets (i.e. genomes, nucleotides and proteins) which
might not be up-to-date 
		-- same as using \code{src='repo'}.
	\item the package will attempt to obtain the data from a mirror server
with the datasets updated on a regular basis but not necessarily with the
latest updates
		-- same as using \code{src='local'}.
\end{itemize}

These sequence of steps are implemented in the package using \code{tryCath()}
exceptions in combination with \textit{recursivity}, i.e. the retrieving data
function calling itself with different variations indicating which data source
to use.

\subsubsection{Data Repositories}
\label{sec:dataRepos}

\begin{figure}
	\includegraphics[width=\textwidth]{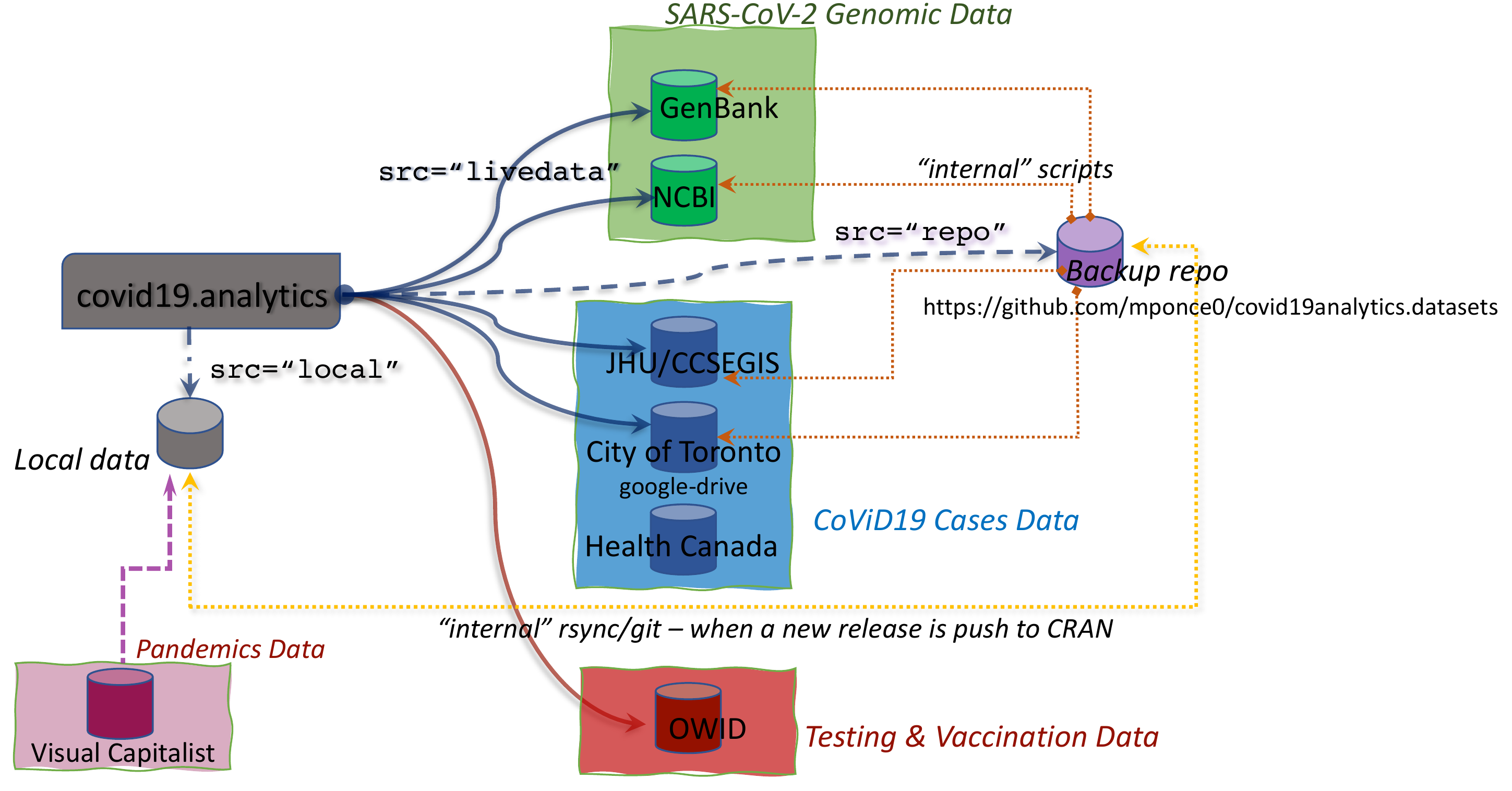}
	\figcaption{Schematic of the data acquision flows between the \covidPckg package and the different sources of data.
		Dark and solid/dashed lines represent API functions provided by the package accesible to the users.
		Dotted lines are ``internal'' mechanisms employed by the package to synchronize and update replicas of the data.
		Data acquisition from NCBI servers is mostly done utilizing the \CRANpkg{ape}~\cite{apePckg} and \CRANpkg{rentrez}~\cite{rentrezPckg} packages.}{fig:dataSources}
	\label{fig:dataSources}
\end{figure}

As the \covidPckg package will try present the user with the latest data sets
possible, different strategies (as described above) may be in place to achieve
this.
One way to improve the realiability of the access to and avialability of the data
is to use a series of replicas of the datasets which are hosted in different
locations.
Fig.~\figref{fig:dataSources} summarizes the different data sources and point of
access that the package employs in order to retrieve the data and keeps the latest
datasets available.

Genomic data as mentioned before is accessed from NCBI databases.
This is implemented in the \code{covid19.genomic.data} function employing the
\CRANpkg{ape}~\cite{apePckg} and \CRANpkg{rentrez}~\cite{rentrezPckg} packages.
In particular the proteins datasets, with more than 100K entries, is quite
challenging to obtain ``live''.
As a matter of fact, the \code{covid19.genomic.data} function accepts an
argument to specify whether this should be the case or not.
If the \texttt{src} argument is set to \texttt{'livedata'} then the function
will attempt to download the proteins list directly from NCBI databases.
If this fail, we recommend using the argument \texttt{src='local'} which
will provide an stagered copy of this dataset at the moment in which the
package was submitted to the CRAN repository, meaning that is quite likely this
dataset won't be complete and most likely outdated.
Additionaly, we offer a second replica of the datasets, located at
\url{https://github.com/mponce0/covid19analytics.datasets}
where all datasets are updated periodically, this can be accessed using the
argument \texttt{src='repo'}.


\subsection{Analytical \& Graphical Indicators Functions}
\label{sec:Fns}

In addition to the access and retrieval of the data, the \covidPckg package
includes several functions to perform basic analysis and visualizations.
Table~\ref{table:covid19Fns} shows the list of the main functions in the package.

\begin{longtable}{p{.285\textwidth} |  p{.35\textwidth} | p{.35\textwidth} }
	Function	&	Description	&	Main Type of Output	\\
	\hline\hline
	\rowcolor{lightgray}
	\multicolumn{3}{c}{Data Acquisition}	\\
	\hline
	\multicolumn{3}{c}{\it CoViD-19 Cases}    \\
	\hline
	\texttt{covid19.data}	&	obtain live* worldwide data for covid19 cases, from the JHU's CCSE repository \cite{JHUCSSErepo}	&	return dataframes/list with the collected data	\\
	\hline
	\texttt{covid19.Canada.data}        &       obtain live* Canada specific data for covid19 cases, from the Health Canada data \cite{HealthCanada}      &       return dataframe with the collected data        \\
	\hline
	\texttt{covid19.Toronto.data}	&	obtain live* data for covid19 cases in the city of Toronto, ON Canada, from the City of Toronto reports	\cite{TorontoData} or Open Data Toronto \cite{OpenDataToronto}	&	return dataframe/list with the collected data	\\
	\hline
	\texttt{covid19.US.data}	&	obtain live* US specific data for covid19 cases, from the JHU's CCSE repository \cite{JHUCSSErepo}	&	return dataframe with the collected data	\\
	\hline
        \multicolumn{3}{c}{\it Vaccination \& Testing}    \\
        \hline
        \texttt{covid19.vaccination}
				&	obtain up-to-date CoViD-19 vaccination records from ``Our World In Data'' CoViD19 data repository \cite{OWIDvaccination}
				&	depending on the type of query, it can be a dataframe or list with the collected data
                \\
        \hline
	\texttt{covid19.testing.data}
				&	obtain up-to-date CoViD-19 testing records from ``Our World In Data'' CoViD19 data repository \cite{OWIDvaccination}
				&	return dataframe with the testing data or testing data details
		\\
	\hline
	\multicolumn{3}{c}{\it Genomics}    \\
	\hline
	\texttt{covid19.genomic.data}
        \texttt{c19.refGenome.data  c19.fasta.data  c19.ptree.data  c19.NPs.data  c19.NP\_fasta.data}
					&	obtain genomic data from NCBI databases -- see Table~\ref{table:genomicTypes} for details	&	depending on the type of query, it can be a list or another type of composed object 
		\\
	\hline
        \multicolumn{3}{c}{\it Pandemics}    \\
        \hline
        \texttt{pandemics.data}
			&	obtain pandemics and pandemics vaccination \textit{historical} records from \cite{VCpandemics}
			&	return dataframe with the collected data
                \\
        \hline
	\rowcolor{lightgray}
	\multicolumn{3}{c}{Data Quality Assessment}	\\
	\hline
	\texttt{data.checks}		&	run integrity and consistency checks on a given dataset		&	diagnostics about the dataset integrity and consistency	\\
	\hline
	\texttt{consistency.check}	&	run consistency checks on a given dataset	&	diagnostics about the dataset consistency	\\
	\hline
	\texttt{integrity.check}	&	run integrity checks on a given dataset		&	diagnostics about the dataset integrity	\\
	\hline
	\texttt{nullify.data}		&	remove inconsistent/incomplete entries in the original datasets	&	original dataset (dataframe) without ``suspicious'' entries	\\
	\hline
	\rowcolor{lightgray}
	\multicolumn{3}{c}{Analysis}	\\
	\hline
	\texttt{report.summary}		&	summarize the current situation, will download the latest data and summarize different quantities	&	on screen table and static plots (pie and bar plots) with reported information, can also output the tables into a text file	\\
	\hline
	\texttt{tots.per.location}	&	compute totals per region and plot time series for that specific region/country		&	static plots: data + models (exp/linear, Poisson, Gamma), mosaic and histograms when more than one location are selected	\\
	\hline
	\texttt{growth.rate}		&	compute changes and growth rates per region and plot time series for that specific region/country	&	static plots: data + models (linear,Poisson,Exp), mosaic and histograms when more than one location are selected \\
		&	&	interactive figures: heatmap and 3d-surface representation	\\
	\hline
	\texttt{\begin{minipage}{\columnwidth} single.trend \\ mtrends\end{minipage}}	&	visualize different indicators of the ``trends'' in daily changes for a single or multiple locations	&	compose of static plots: total number of cases vs time, daily changes vs total changes in different representations	\\
	\hline
	\texttt{estimateRRs}	&	compute estimates for fatality and recovery rates on a rolling-window interval	& list with values for the estimates (mean and sd) of reported cases and recovery and fatality rates	\\
	\hline
	\rowcolor{lightgray}
	\multicolumn{3}{c}{Graphics and Visualization}	\\
	\hline
	\texttt{total.plts}	&	plots in a static and interactive plot total number of cases per day, the user can specify multiple locations or global totals	&	static and interactive plot	\\
	\hline
	\texttt{itrends}	&	generates an interactive plot of daily changes vs total changes in a log-log plot, for the indicated regions	&	interactive plot	\\
	\hline
	\texttt{live.map}	&	generates an interactive map displaying cases around the world	&	static and interactive plot	\\
	\hline
	\rowcolor{lightgray}
	\multicolumn{3}{c}{Modelling}	\\
	\hline
	\texttt{generate.SIR.model}	&	generates a \textit{Susceptible-Infected-Recovered} (SIR) model	&	list containing the fits for the SIR model	\\
	\hline
	\texttt{plt.SIR.model}		&	plot the results from the SIR model	&	static and interactive plots	\\
	\hline
	\texttt{sweep.SIR.models}	&	generate multiple SIR models by varying parameters used to select the actual data 	&	list containing the values  parameters, $\beta, \gamma$ and $R_0$	\\
	\hline
	\rowcolor{lightgray}
	\multicolumn{3}{c}{Data Exploration}   \\
	\hline
	\texttt{covid19Explorer}     &       launches a dashboard interface to explore the datasets provided by covid19.analytics package	&	interactive web-based dashboard	\\
	\hline
	\rowcolor{lightgray}
	\multicolumn{3}{c}{Auxiliary Functions}   \\
	\hline
	\texttt{geographicalRegions}	&	determines which countries compose a given continent	&	list of countries	\\
	\hline\hline
\caption{Overview of the main functions of the \covidPckg package}
\label{table:covid19Fns}
\end{longtable}

\subsubsection{Details and Specifications of the Analytical \& Visualization Functions}

\paragraph{Geographical Locations}
An important element in the recorded data as shown in Sec.~\ref{sec:dataStructure}
is the indication of the corresponding geographical location.
In the reported data, this is mostly given by the Province/City and/or Country/Region.
In order to facilitate the processing of locations that are located geo-politically
close, the \covidPckg package provides a way to identify regions by indicating the
corresponding continent's name where they are located.
I.e. "South America", "North America", "Central America", "America", "Europe",
"Asia" and "Oceania" can be used to process all the countries within each of
these regions.

The \code{geographicalRegions} function is the one in charge of determining
which countries are part of what continent and will display them when
executing \code{geographicalRegions()}.

In this way, it is possible to specify a particular continent and all the
countries in this continent will be processed without needing to explicitly
specifying all of them.

\paragraph{Reports}
As the amount of data available for the recorded cases of CoViD19 can be
overwhelming, and in order to get a quick insight on the main statistical indicators,
the \covidPckg package includes
the \texttt{report.summary} function, which will generate an overall report
summarizing the main statistical estimators for the different datasets.
It can summarize the "Time Series" data (when indicating
\texttt{cases.to.process="TS"}), the "aggregated" data
(\texttt{cases.to.process="AGG"}) or both (\texttt{cases.to.process="ALL"}).
The default will display the top 10 entries in each category, or the number
indicated in the \texttt{Nentries} argument, for displaying all the records
just set \texttt{Nentries=0}.

The function can also target specific geographical location(s) using the
\texttt{geo.loc argument}.
When a geographical location is indicated, the report will
include an additional "Rel.Perc" column for the confirmed cases indicating the
\textit{relative percentage} among the locations indicated. Similarly the totals
displayed at the end of the report will be for the selected locations.

In each case ("TS" or/and "AGG") will present tables ordered by the different
cases included, i.e. confirmed infected, deaths, recovered and active cases.

The dates when the report is generated and the date of the recorded data will
be included at the beginning of each table.

It will also compute the totals, averages or mean values, standard deviations
and percentages of various quantities, i.e.
\begin{itemize}
	\item it will determine the number of unique locations processed within the dataset

	\item it will compute the total number of cases per case type

	\item Percentages -- which are computed as follow:
	\begin{itemize}
		\item for the "Confirmed" cases, as the ratio between the
corresponding number of cases and the total number of cases, i.e. a sort of
"global percentage" indicating the percentage of infected cases with respect
to the rest of the world

		\item for "Confirmed" cases, when geographical locations are
specified, a \textit{"Relative percentage"} is given as the ratio of the
confirmed cases over the total of the selected locations

		\item for the other categories, "Deaths"/"Recovered"/"Active",
the percentage of a given category is computed as the ratio between the number
of cases in the corresponding category divided by the "Confirmed" number of
cases, i.e. a \textit{relative percentage} with respect to the number of confirmed
infected cases in the given region
	\end{itemize}

	\item For "Time Series" data:
	\begin{itemize}
		\item it will show the \textit{delta} (change or variation) in
the last day, daily changes day before that ($t-2$), three days ago ($t-3$), a
week ago ($t-7$), two weeks ago ($t-14$) and a month ago ($t-30$)
		\item when possible, it will also display the percentage of
"Recovered" and "Deaths" with respect to the "Confirmed" number of cases
		\item the column "GlobalPerc" is computed as the ratio between
the number of cases for a given country over the total of cases reported
		\item The "Global Perc. Average (SD: standard deviation)" is
computed as the average (standard deviation) of the number of cases among all
the records in the data
		\item The "Global Perc. Average (SD: standard deviation) in top
$X$" is computed as the average (standard deviation) of the number of cases among
the top $X$ records
	\end{itemize}
\end{itemize}

\begin{figure}
	\includegraphics[width=0.325\textwidth]{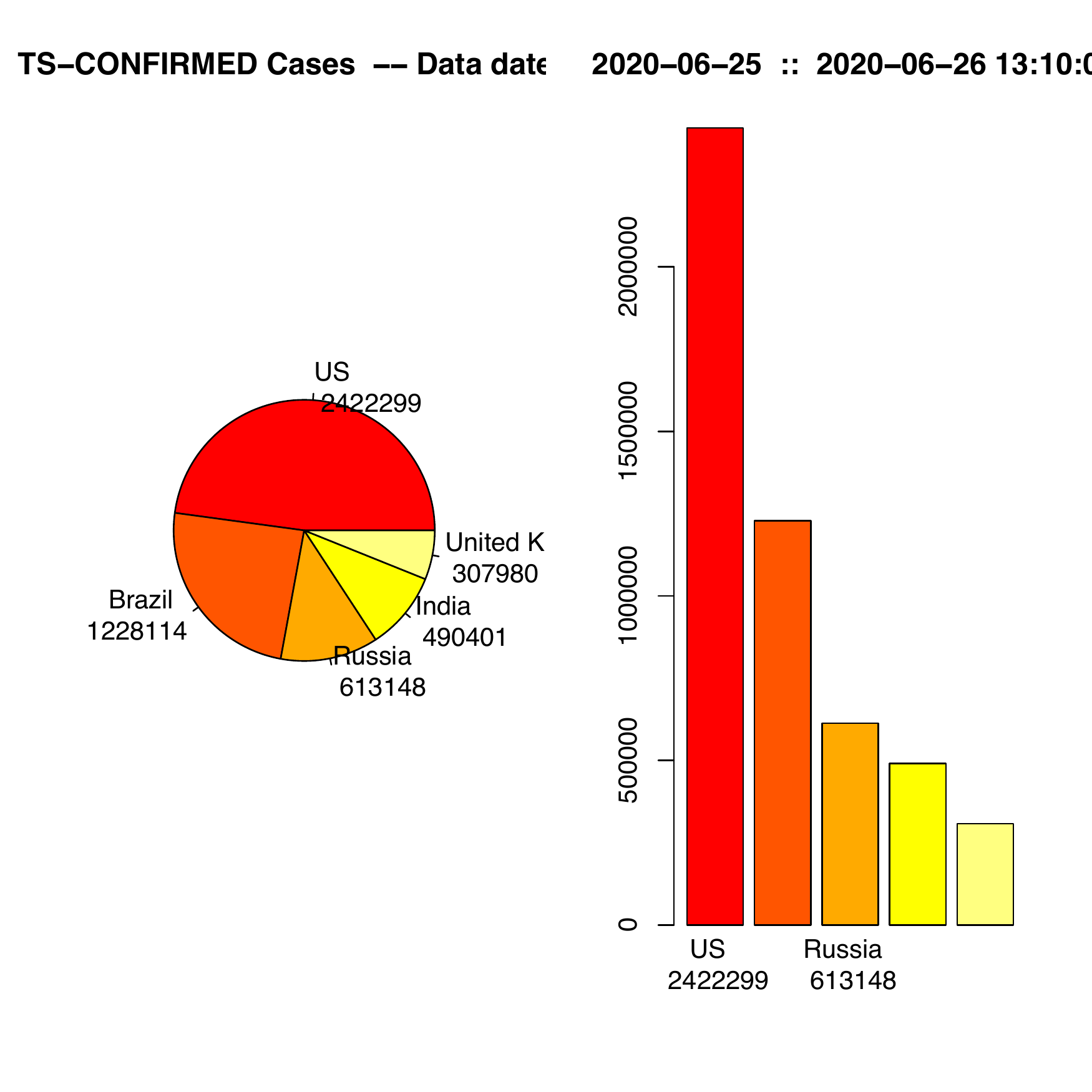}
        \includegraphics[width=0.325\textwidth]{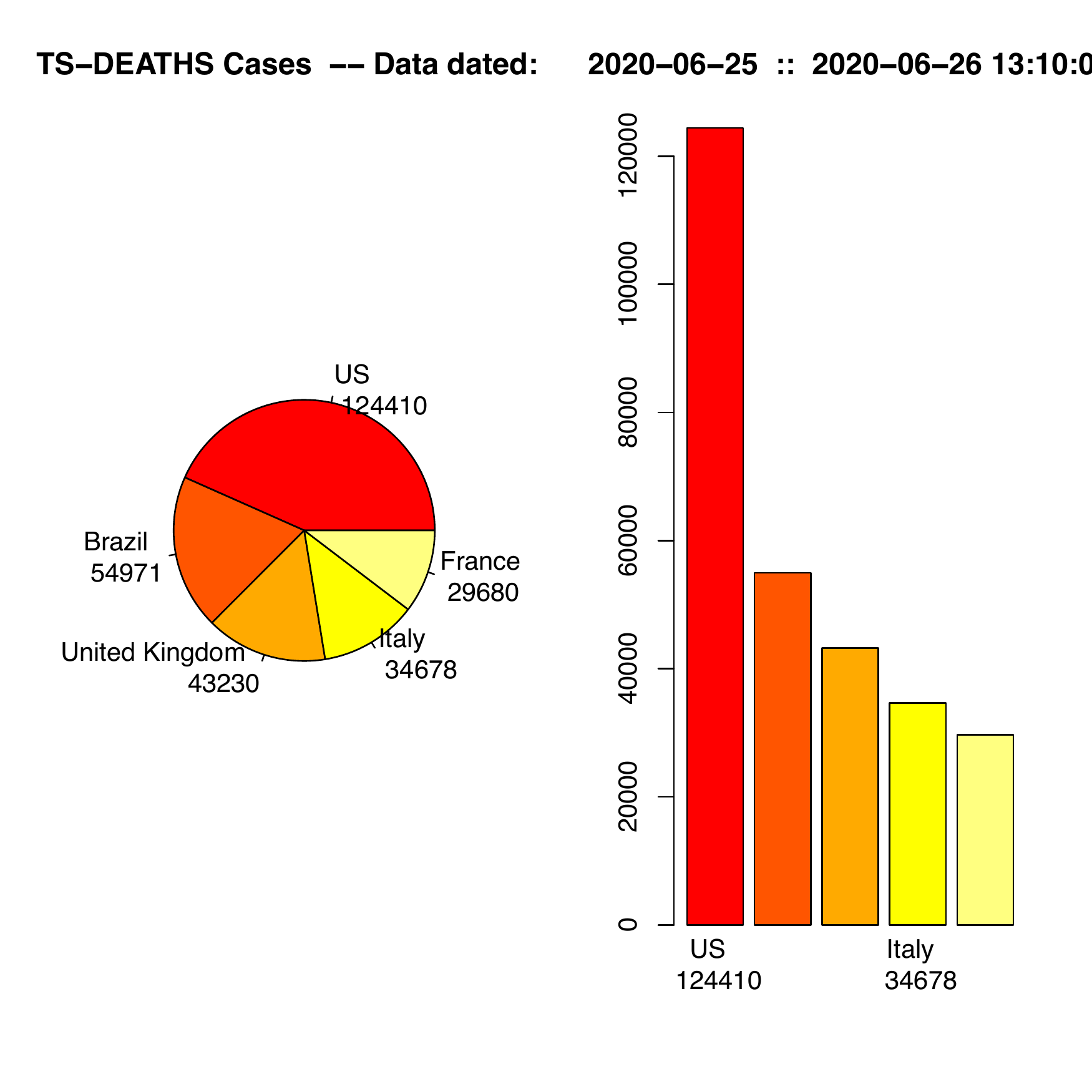}
        \includegraphics[width=0.325\textwidth]{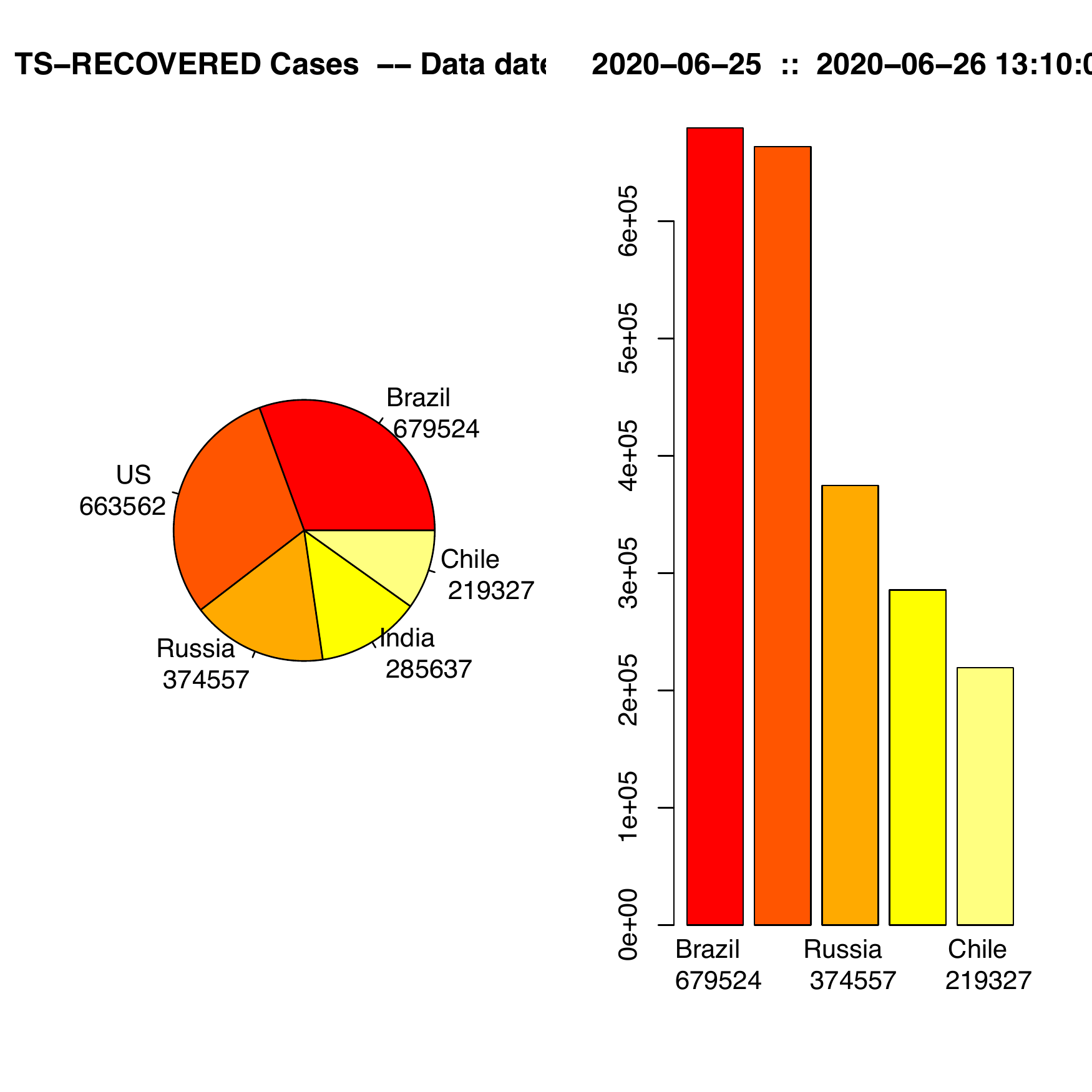}
	\centering
        \includegraphics[width=0.85\textwidth]{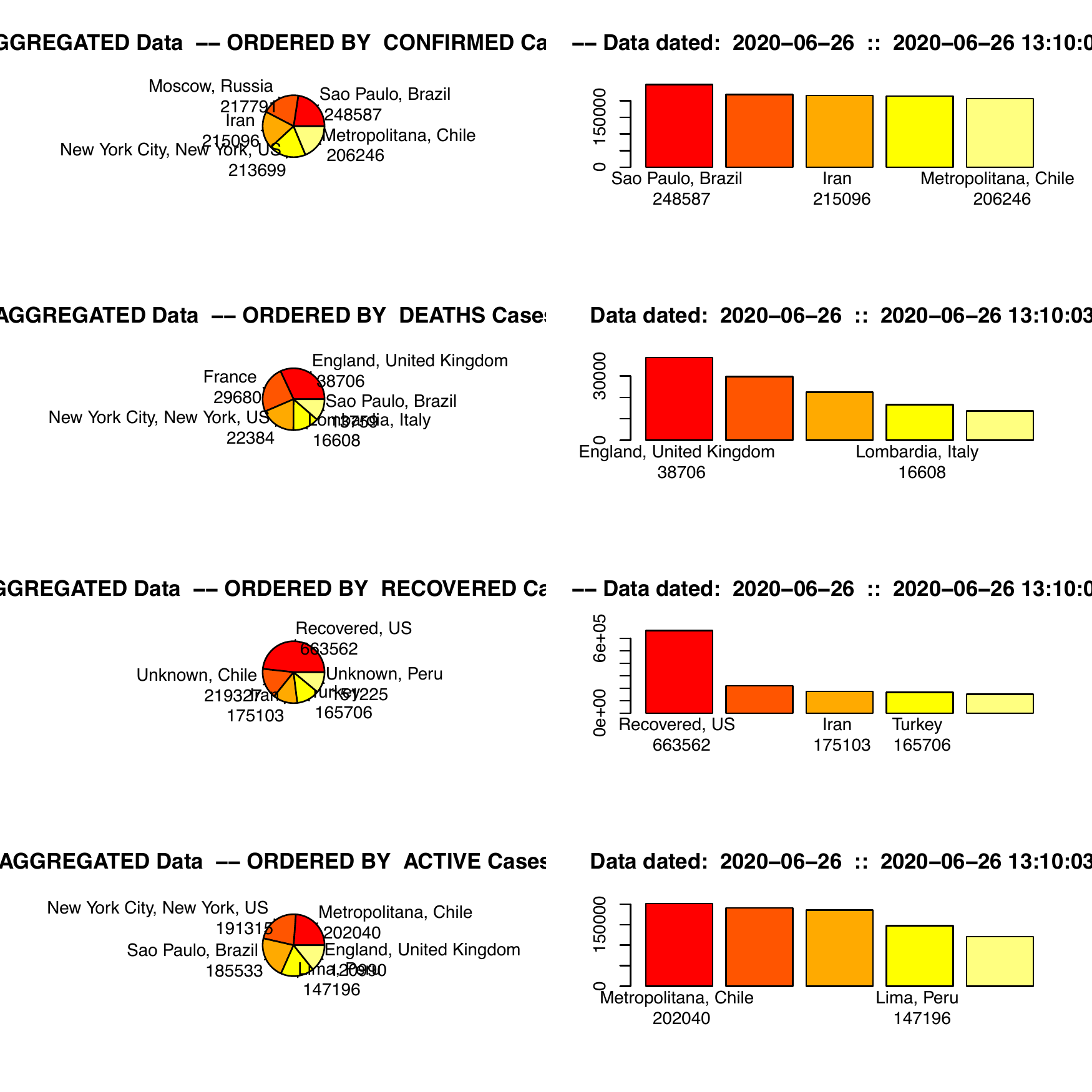}

	\caption{Graphical output produced by the \code{report.summary} function.
		The top row shows bar plots and pie charts for each respective category of reported cases,
		"confirmed", "deaths" and "recovered" for the top 5 entries for time series data.
		The bottom row shows a combined plot for the aggregated data.
		This graphical output aim to complement the text report generated, as shown in Lst.~\ref{report_output}.
		The plots show the distribution of cases in the corresponding category
		for the locations list in the top entries, in this case the top 5.}
		{fig:report}
	\label{fig:report}
\end{figure}

A typical output of the \texttt{summary.report} for the "Time Series" data,
is shown in the example \ref{report_output} in Sec.~\ref{sec:examples}.
In addition to this, the function also generates some graphical outputs,
including pie and bar charts representing the top regions in each category;
see Fig.~\figref{fig:report}.

\paragraph{Totals per Location \& Growth Rate}
It is possible to dive deeper into a particular location by using the
\texttt{tots.per.location} and \texttt{growth.rate} functions.
These functions are capable of processing different types of data, as far as
these are "Time Series" data.
It can either focus in one category (eg. "TS-confirmed", "TS-recovered", "TS-deaths",) or all ("TS-all"). When these
functions detect different types of categories, each category will be processed
separately. Similarly the functions can take multiple locations, ie. just one,
several ones or even \texttt{"all"} the locations within the data. The locations can
either be countries, regions, provinces or cities. If an specified location
includes multiple entries, eg. a country that has several cities reported, the
functions will group them and process all these regions as the location
requested.

\paragraph{Totals per Location}
The \texttt{tots.per.location} function will plot the number of cases as a
function of time for the given locations and type of categories, in two plots:
a log-scale scatter one a linear scale bar plot one.

When the function is run with multiple locations or all the locations, the
figures will be adjusted to display multiple plots in one figure in a mosaic
type layout.

Additionally, the function will attempt to generate different fits to match the
data:
\begin{itemize}
	\item an exponential model using a Linear Regression method
	\item a Poisson model using a General Linear Regression method
	\item a Gamma model using a General Linear Regression method
\end{itemize}

The function will plot and add the values of the coefficients for the models to
the plots and display a summary of the results in the console. 
It is also possible to instruct the function to draw a "confidence band" based
on a \textit{moving average}, so that the trend is also displayed including a
region of higher confidence based on the mean value and standard deviation
computed considering a time interval set to equally dividing the total range of
time over 10 equally spaced intervals.

The function will return a list combining the results for the totals for the
different locations as a function of time.

\paragraph{Growth Rate}
The \texttt{growth.rate} function allows to compute daily changes and the
\textit{growth rate} defined as the ratio of the daily changes between two consecutive
dates.

The \texttt{growth.rate} function shares all the features of the
\texttt{tots.per.location} function as described above, i.e. can process the
different types of cases and multiple locations.

The graphical output will display two plots per location:
\begin{itemize}
	\item a scatter plot with the number of changes between consecutive
dates as a function of time, both in linear scale (left vertical axis) and
log-scale (right vertical axis) combined
	\item a bar plot displaying the growth rate for the particular region
as a function of time.
\end{itemize}

When the function is run with multiple locations or all the locations, the
figures will be adjusted to display multiple plots in one figure
in a mosaic type layout.
In addition to that, when there is more than one location the function will
also generate two different styles of heatmaps comparing the changes per day
and growth rate among the different locations (vertical axis) and time
(horizontal axis).
Furthermore, if the \code{interactiveFig=TRUE} argument is used, then 
interactive heatmaps and 3d-surface representations will be generated too.

Some of the arguments in this function, as well as in many of the other
functions that generate both static and interactive visualizations, can be used
to indicate the type of output to be generated.
Table~\ref{table:fns_args} lists some of these arguments.
In particular, the arguments controlling the interactive figures
--\code{interactiveFig} and \code{interactive.display}-- can be used in
combination to compose an interactive figure to be captured and used in another
application.
For instance, when \code{interactive.display} is turned off but
\code{interactiveFig=TRUE}, the function will return the interactive figure, so
that it can be captured and used for later purposes.
This is the technique employed when capturing the resulting plots in the
\textit{covid19.analytics Dashboard Explorer} as presented in
Sec.~\ref{sec:dashboard-backend}.

\begin{table}
\begin{tabular}{p{.25\textwidth} | p{.5\textwidth} | p{.2\textwidth}}
        argument                &       effect  	&	default value\\
        \hline\hline
        \code{staticPlt}        &       when active, enables static plots to be displayed in screen	& \code{TRUE}	\\
        \hline
        \code{interactiveFig}   &       when active, enables interactive visualizations features        &	\code{FALSE}	\\
        \hline
        \code{interactive.display}      &       when active, pushes the interactive figures into a browser &	\code{TRUE}	\\
                &       When is turned off, but \code{interactiveFig=TRUE}, the function will return the interactive figure, so that it can be captured and used for later purposes.    \\
        \hline
\end{tabular}
\caption{List of some of the arguments used in several functions to control the type of graphical output.}
\label{table:fns_args}
\end{table}

Finally, the \code{growth.rate} function when not returning an interactive figure,
will return a list combining the results for the "changes per day" and the
"growth rate" as a function of time, i.e. when \code{interactiveFig} is not
specified or set to \code{FALSE} (which its default value)
or when \code{interactive.display=TRUE}.

\paragraph{Trends in Daily Changes}

The \covidPckg package provides three different functions to visualize the
trends in daily changes of reported cases from time series data.

\begin{itemize}
	\item \code{single.trend}, allows to inspect one single location, this
could be used with the worldwide data sliced by the corresponding location, the
Toronto data or the user's own data formatted as "Time Series" data.

	\item \code{mtrends}, is very similar to the \code{single.trend}
function, but accepts multiple or single locations generating one plot per
location requested; it can also process multiple cases for a given location.

	\item \code{itrends} function to generate an interactive plot of the
trend in daily changes representing changes in number of cases vs total number
of cases in log-scale using \textit{splines} techniques to smooth the abrupt
variations in the data
\end{itemize}

The first two functions will generate "static" plots in a compose with different insets:
\begin{itemize}
	\item the main plot represents daily changes as a function of time
	\item the inset figures in the top, from left to right:
	\begin{itemize}
		\item total number of cases (in linear and semi-log scales),
		\item changes in number of cases vs total number of cases
		\item changes in number of cases vs total number of cases in log-scale
	\end{itemize}
	\item the second row of insets, represent the "growth rate" (as defined
above) and the \textit{normalized growth rate} defined as the growth rate divided by
the maximum growth rate reported for this location
\end{itemize}

\paragraph{Plotting Totals}
The function \texttt{totals.plt} will generate plots of the total number of cases as a
function of time. It can be used for the total data or for a specific or
multiple locations. The function can generate static plots and/or interactive
ones, as well, as linear and/or semi-log plots.

\paragraph{Plotting Cases in the World}
The function \texttt{live.map} will display the different cases in each corresponding
location all around the world in an interactive map of the world. It can be
used with time series data or aggregated data, aggregated data offers a much
more detailed information about the geographical distribution.

\subsubsection{Modelling the Evolution of the Virus Spread}
The \covidPckg package allows users to model the dispersion of the disease by
implementing a simple \textit{Susceptible-Infected-Recovered} (SIR) model
\cite{kermack1927contribution, smith2004sir}.
The model is implemented by a system of ordinary differential equations (ODE),
as the one shown by Eq.(\ref{eqn:SIR_model}).

\begin{equation}
	\left\{
	\begin{array}{l l l}
	\dfrac{dS}{dt} = -\dfrac{\beta I S}{N}	\\
	\\
	\dfrac{dI}{dt} = \dfrac{\beta I S}{N} - \gamma I	\\
	\\
	\dfrac{dR}{dt} = \gamma I
	\end{array}
	\right.
\label{eqn:SIR_model}
\end{equation}

where $S$ represents the number of \textit{susceptible} individuals to be infected,
$I$ the number of \textit{infected} individuals and $R$ the number of
\textit{recovered} ones at a given moment in time.
The coefficients $\beta$ and $\gamma$ are the parameters controlling the
transition rate from $S$ to $I$ and from $I$ to $R$ respectively;
$N$ is the total number of individuals, i.e. $N = S(t) + I(t) + R(t)$;
which should remain constant, i.e.

\begin{equation}
	\frac{dN}{dt} = \frac{dS}{dt} + \frac{dI}{dt} + \frac{dR}{dt} \equiv 0
\end{equation}

Eq.(\ref{eqn:SIR_model}) can be written in terms of the \textit{normalized} quantities,
	$s(t) \equiv S(t)/N$,
	$i(t) \equiv I(t)/N$,
and
	$r(t) \equiv R(t)/N$;
as

\begin{equation}
	\left\{
        \begin{array}{l l l}
        \dfrac{ds}{dt} = -\beta i(t) s(t)  \\
        \\
        \dfrac{di}{dt} = \beta i(t) s(t) - \gamma i(t)        \\
        \\
        \dfrac{dr}{dt} = \gamma i(t)
        \end{array}
        \right.
	\label{eqn:SIR_norm}
\end{equation}

%

Although the ODE SIR model is non-linear, analytical solutions have been found \cite{HARKO2014184}.
However the approach we follow in the package implementation is to solve the
ODE system from Eq.(\ref{eqn:SIR_model}) numerically.

The function \texttt{generate.SIR.model} implements the SIR model from
Eq.(\ref{eqn:SIR_model}) using the actual data from the reported cases.
The function will try to identify data points where the onset of the
epidemic began and consider the following data points to generate proper
guesses for the two parameters describing the SIR ODE system, i.e. $\beta$
and $\gamma$.

It does this by minimizing the \textit{residual sum of squares} (RSS) assuming
one single explanatory variable, i.e. the sum of the squared differences
between the number of infected cases $I(t)$ and the quantity predicted
by the model $\tilde{I}(t)$,

\begin{equation}
	RSS(\beta,\gamma) = \sum_t \left( I(t)-\tilde{I}(t) \right)^2
	\label{eqn:RSS_I}
\end{equation}

The ODE given by Eq.(\ref{eqn:SIR_model}) is solved numerically using the
\code{ode} function from the \CRANpkg{deSolve} and the minimization is tackled
using the \code{optim} function from base R.

After the solution for Eq.(\ref{eqn:SIR_model}) is found, the function
will provide details about the solution, as well as, plot the quantities
$S(t), I(t), R(t)$ in a static and interactive plot.

The \code{generate.SIR.model} function also estimates the value of the
\textit{basic reproduction number} or \textit{basic reproduction ratio},
$R_0$, defined as,

\begin{equation}
	R_0 = \frac{\beta}{\gamma}
	\label{eqn:R0}
\end{equation}

which can be considered as a measure of the average expected number of new
infections from a single infection in a population where all subjects can be
susceptible to get infected.

The function also computes and plots on demand, the \textit{force of infection},
defined as,
	$F_{infection} = \beta I(t)$,
which measures the transition rate from the compartment of susceptible
individuals to the compartment of infectious ones.

For exploring the parameter space of the SIR model, it is possible to produce a
series of models by varying the conditions, i.e. range of dates considered for
optimizing the parameters of the SIR equation, which will effectively ``sweep''
a range for the parameters $\beta, \gamma$ and $R_0$.
This is implemented in the function \code{sweep.SIR.models}, which takes a
range of dates to be used as starting points for the number of cases used to
feed into the \code{generate.SIR.model} producing as many models as different
ranges of dates are indicated.
One could even use this in combination to other resampling or Monte Carlo
techniques to estimate statistical variability of the parameters from the
model.


\section{Examples and Applications}
\label{sec:examples}

In this section we will present some basic examples of how to use the main
functions from the \covidPckg package.

\subsection{Installation}
\label{sec:installation_ex}
We will begin by installing the \covidPckg package.
This can be achieved in two alternative ways:
\begin{enumerate}
	\item installing the latest stable version of the package directly from
the CRAN repository. This can be done within an R session using the
\code{install.packages} function, i.e.

	\begin{verbatim}
	> install.packages("covid19.analytics")
	\end{verbatim}

	\item installing the development version from the package's GITHUB
repository, \url{https://github.com/mponce0/covid19.analytics} using the
\CRANpkg{devtools} package \cite{devtools} and its \texttt{install\_github}
function.
I.e.
	\begin{verbatim}
	# begin by installing devtools if not installed in your system
	> install.packages("devtools")
	# install the covid19.analytics packages from the GITHUB repo
	> devtools::install_github("mponce0/covid19.analytics")
	\end{verbatim}
\end{enumerate}

After having installed the \covidPckg package, for accessing its functions, the
package needs to be loaded using R's \texttt{library} function, i.e.
	\begin{verbatim}
	> library(covid19.analytics)
	\end{verbatim}

The \covidPckg uses a few additional packages which are installed automatically
if they are not present in the system.
In particular,
    \CRANpkg{readxl} is used to access the data from the City of Toronto \cite{TorontoData},
    \CRANpkg{ape} is used for pulling the genomics data from NCBI;
    \CRANpkg{plotly} and \CRANpkg{htmlwidgets} are used to render the interactive plots and save them in HTML documents,
    \CRANpkg{deSolve} is used to solve the differential equations modelling the spread of the virus,
    and \CRANpkg{gplots, pheatmap} are used to generate heatmaps.

\subsection{Retrieving and Accessing Data}
\label{sec:readingdata_ex}

Lst.~\ref{lst:ex_readingData} shows how to use the \code{covid19.data}
function to obtain data in different cases.

\Rlisting
\begin{lstlisting}[caption={Reading data from reported cases of CoViD19 using the \covidPckg package.},captionpos=b,label={lst:ex_readingData}]
# obtain all the records combined for "confirmed", "deaths" and "recovered" cases
# for the global (worldwide) *aggregated* data
 covid19.data.ALLcases <- covid19.data()

# obtain time series data for global "confirmed" cases
 covid19.confirmed.cases <- covid19.data("ts-confirmed")

# reads all possible datasets, returning a list
 covid19.all.datasets <- covid19.data("ALL")

# reads the latest aggregated data of the global cases
 covid19.ALL.agg.cases <- covid19.data("aggregated")

# reads time series data for global casualties
 covid19.TS.deaths <- covid19.data("ts-deaths")

# read "Time Series" data for the city of Toronto
 Toronto.TS.data <- covid19.data("ts-Toronto")

# this can be also done using the covid19.Toronto.data() fn
 Tor.TS.data <- covid19.Toronto.data() 

# or get the original data as reported by the City of Toronto
 Tor.DF.data <- covid19.Toronto.data(data.fmr="ORIG")

# retrieve US time series data of confirmed cases
 US.confirmed.cases <- covid19.data("ts-confirmed-US")

# retrieve US time series data of death cases
 US.deaths.cases <- covid19.data("ts-deaths-US")

# or both cases combined
 US.cases <- covid19.US.data()
\end{lstlisting}

In general, the reading functions will return data frames. 
Exceptions to this, are when the functions need to return a more complex output, e.g.
when combining "ALL" type of data or when requested to obtain the original data
from the City of Toronto (see details in Table~\ref{table:covid19Fns}).
In these cases, the returning object will be a list containing in each
element dataframes corresponding to the particular type of data.
In either case, the structure and overall content can be quickly assessed by
using R's \code{str} or \code{summary} functions.

\Rlisting
\begin{lstlisting}[caption={Reading testing and vaccination data of CoViD19 using the \covidPckg package.},captionpos=b,label={lst:ex_reading-vacc-test}]
# reading COVID19 testing data
c19.testing.data <- covid19.testing.data()

# reading COVID19 vaccination data
c19.vacc.data <- covid19.vaccination()
\end{lstlisting}

\Rlisting
\begin{lstlisting}[caption={Reading historical records from different pandemic using the \covidPckg package.},captionpos=b,label={lst:ex_readingPandemics}]
# Pandemic historical records
 pnds <- pandemics.data(tgt="pandemics")

# Pandemics vaccines development times
 pnds.vacs <- pandemics.data(tgt="pandemics_vaccines")
\end{lstlisting}

\Rlisting
\begin{lstlisting}[caption={Reading genomic data using the \covidPckg package.},captionpos=b,label={lst:ex_readingGenomics}]

# obtain covid19's genomic data
 covid19.gen.seq <- covid19.genomic.data()

# display the actual RNA seq
 covid19.gen.seq$NC_045512.2
\end{lstlisting}

%

\subsection{Basic Analysis}
\label{sec:ex_analysis}

\subsubsection{Identifying Geographical Locations}
\label{sec:ex_geoLocns}
One useful information to look at after loading the datasets, would be to
identify which locations/regions have reported cases.
There are at least two main fields that can be used for that, the columns
containing the keywords: 'country' or 'region' and 'province' or 'state'.
Lst.~\ref{lst:ex_geoLocns} show examples of how to achieve this using partial
matches for column names, e.g. "Country" and "Province".

\Rlisting
\begin{lstlisting}[caption={Identifying geographical locations in the data sets.},captionpos=b,label={lst:ex_geoLocns}]
# read a data set
data <- covid19.data("TS-confirmed")

# look at the structure and column names
str(data)
names(data)

# find 'Country' column
country.col <- pmatch("Country",names(data))
# slice the countries
countries <- data[,country.col]
# list of countries
print(unique(countries))
# sorted table of countries, may include multiple entries
print(sort(table(countries)))

# find 'Province' column
prov.col <- pmatch("Province",names(data))
# slice the Provinces
provinces <- data[,prov.col]

# list of provinces
print(unique(provinces))
# sorted table of provinces, may include multiple entries
print(sort(table(provinces)))

\end{lstlisting}

\subsubsection{Reports}
\label{sec:ex_reports}
An overall view of the current situation at a global or local level can be
obtained using the \code{report.summary} function.
Lst.~\ref{lst:ex_reports} shows a few examples of how this function can be used.

\Rlisting
\begin{lstlisting}[caption={Reports generation},captionpos=b,label={lst:ex_reports}]
# a quick function to overview top cases per region for time series and aggregated records
report.summary()
  

# save the tables into a text file named 'covid19-SummaryReport_CURRENTDATE.txt'
# where *CURRRENTDATE* is the actual date
report.summary(saveReport=TRUE)

# summary report for an specific location with default number of entries
report.summary(geo.loc="Canada")

# summary report for an specific location with top 5
report.summary(Nentries=5, geo.loc="Canada")

# it can combine several locations
report.summary(Nentries=30, geo.loc=c("Canada","US","Italy","Uruguay","Argentina"))
\end{lstlisting}

A typical output of the report generation tool is presented in Lst.~\ref{report_output}.

\outputlisting
\begin{lstlisting}[caption={Typical output of the \code{report.summary} function. This particular example was generated using
		\code{report.summary(Nentries=5,graphical.output=TRUE,saveReport=TRUE)}, which indicates to consider just
		the top 5 entries, generate a graphical output as shown in Fig.~\ref{fig:report} and to save a text file
		including the report which is the one shown here.},
		label={report_output}]
~~~~~~~~~~~~~~~~~~~~~~~~~~~~~~~~~~~~~~~~~~~~~~~~~~~~~~~~~~~~~~~~~~~~~~~~~~~~~~~~ 
-------------------------------------------------------------------------------- 
################################################################################ 
  ##### TS-CONFIRMED Cases  -- Data dated:  2020-06-25  ::  2020-06-26 13:10:01 
################################################################################ 
  Number of Countries/Regions reported:  188 
  Number of Cities/Provinces reported:  82 
  Unique number of distinct geographical locations combined: 266 
-------------------------------------------------------------------------------- 
  Worldwide ts-confirmed  Totals: 9609829 
-------------------------------------------------------------------------------- 
  Country.Region Province.State  Totals GlobalPerc LastDayChange   t-2   t-3   t-7  t-14  t-30
1             US                2422299      25.21         39972 34836 35189 31527 25396 18366
2         Brazil                1228114      12.78         39483 42725 39436 54771 25982 20599
3         Russia                 613148       6.38          7105  7165  7413  7971  8961  8338
4          India                 490401       5.10         17296 16922 15968 14516 11458  7293
5 United Kingdom                 307980       3.20          1118   652   921  1346  1541  2013
-------------------------------------------------------------------------------- 
  Global Perc. Average:  0.38 (sd: 1.85) 
  Global Perc. Average in top  5 :  10.53 (sd: 8.96) 
-------------------------------------------------------------------------------- 
================================================================================ 
~~~~~~~~~~~~~~~~~~~~~~~~~~~~~~~~~~~~~~~~~~~~~~~~~~~~~~~~~~~~~~~~~~~~~~~~~~~~~~~~ 
-------------------------------------------------------------------------------- 
################################################################################ 
  ##### TS-DEATHS Cases  -- Data dated:  2020-06-25  ::  2020-06-26 13:10:02 
################################################################################ 
  Number of Countries/Regions reported:  188 
  Number of Cities/Provinces reported:  82 
  Unique number of distinct geographical locations combined: 266 
-------------------------------------------------------------------------------- 
  Worldwide ts-deaths  Totals: 489312 
-------------------------------------------------------------------------------- 
  Country.Region Province.State Totals  Perc LastDayChange  t-2  t-3  t-7 t-14 t-30
1             US                124410  5.14          2425  754  829  692  846 1505
2         Brazil                 54971  4.48          1141 1185 1374 1206  909 1086
3 United Kingdom                 43230 14.04           149  154  280  173  202  412
4          Italy                 34678 14.47            34  -31   18   47   56  117
5         France                 29680 15.52            19    9   57   14   28   66
-------------------------------------------------------------------------------- 
-------------------------------------------------------------------------------- 
================================================================================ 
~~~~~~~~~~~~~~~~~~~~~~~~~~~~~~~~~~~~~~~~~~~~~~~~~~~~~~~~~~~~~~~~~~~~~~~~~~~~~~~~ 
-------------------------------------------------------------------------------- 
################################################################################ 
  ##### TS-RECOVERED Cases  -- Data dated:  2020-06-25  ::  2020-06-26 13:10:02 
################################################################################ 
  Number of Countries/Regions reported:  188 
  Number of Cities/Provinces reported:  68 
  Unique number of distinct geographical locations combined: 253 
-------------------------------------------------------------------------------- 
  Worldwide ts-recovered  Totals: 4838921 
-------------------------------------------------------------------------------- 
  Country.Region Province.State Totals LastDayChange   t-2   t-3   t-7  t-14  t-30
1         Brazil                679524         19055 32506 26227 17051 15158  8054
2             US                663562          7401  8613  7350  7600  7094  6606
3         Russia                374557          6335 12375 12000 10442  8213 11079
4          India                285637         13940 13012 10495  9120  7135  3472
5          Chile                219327          4234  4523  5173  5050  4914  2625
-------------------------------------------------------------------------------- 
-------------------------------------------------------------------------------- 
================================================================================ 
~~~~~~~~~~~~~~~~~~~~~~~~~~~~~~~~~~~~~~~~~~~~~~~~~~~~~~~~~~~~~~~~~~~~~~~~~~~~~~~~ 
################################################################################################################################## 
  ##### AGGREGATED Data  -- ORDERED BY  CONFIRMED Cases  -- Data dated:  2020-06-26  ::  2020-06-26 13:10:03 
################################################################################################################################## 
  Number of Countries/Regions reported: 188 
  Number of Cities/Provinces reported: 549 
  Unique number of distinct geographical locations combined: 3781 
---------------------------------------------------------------------------------------------------------------------------------- 
                     Location Confirmed Perc.Confirmed Deaths Perc.Deaths Recovered Perc.Recovered Active Perc.Active
1           Sao Paulo, Brazil    248587           2.59  13759        5.53     49295          19.83 185533       74.64
2              Moscow, Russia    217791           2.27   3669        1.68    142194          65.29  71928       33.03
3                        Iran    215096           2.24  10130        4.71    175103          81.41  29863       13.88
4 New York City, New York, US    213699           2.22  22384       10.47         0           0.00 191315       89.53
5        Metropolitana, Chile    206246           2.15   4206        2.04         0           0.00 202040       97.96
================================================================================================================================== 
################################################################################################################################## 
  ##### AGGREGATED Data  -- ORDERED BY  DEATHS Cases  -- Data dated:  2020-06-26  ::  2020-06-26 13:10:03 
################################################################################################################################## 
  Number of Countries/Regions reported: 188 
  Number of Cities/Provinces reported: 549 
  Unique number of distinct geographical locations combined: 3781 
---------------------------------------------------------------------------------------------------------------------------------- 
                     Location Confirmed Perc.Confirmed Deaths Perc.Deaths Recovered Perc.Recovered Active Perc.Active
1     England, United Kingdom    159696           1.66  38706       24.24         0           0.00 120990       75.76
2                      France    191288           1.99  29680       15.52     71322          37.29  90286       47.20
3 New York City, New York, US    213699           2.22  22384       10.47         0           0.00 191315       89.53
4            Lombardia, Italy     93431           0.97  16608       17.78     64831          69.39  11992       12.84
5           Sao Paulo, Brazil    248587           2.59  13759        5.53     49295          19.83 185533       74.64
================================================================================================================================== 
################################################################################################################################## 
  ##### AGGREGATED Data  -- ORDERED BY  RECOVERED Cases  -- Data dated:  2020-06-26  ::  2020-06-26 13:10:03 
################################################################################################################################## 
  Number of Countries/Regions reported: 188 
  Number of Cities/Provinces reported: 549 
  Unique number of distinct geographical locations combined: 3781 
---------------------------------------------------------------------------------------------------------------------------------- 
        Location Confirmed Perc.Confirmed Deaths Perc.Deaths Recovered Perc.Recovered  Active Perc.Active
1  Recovered, US         0           0.00      0         NaN    663562            Inf -739500        -Inf
2 Unknown, Chile         0           0.00      0         NaN    219327            Inf -219327        -Inf
3           Iran    215096           2.24  10130        4.71    175103          81.41   29863       13.88
4         Turkey    193115           2.01   5046        2.61    165706          85.81   22363       11.58
5  Unknown, Peru         0           0.00      0         NaN    151225            Inf -151225        -Inf
================================================================================================================================== 
################################################################################################################################## 
  ##### AGGREGATED Data  -- ORDERED BY  ACTIVE Cases  -- Data dated:  2020-06-26  ::  2020-06-26 13:10:03 
################################################################################################################################## 
  Number of Countries/Regions reported: 188 
  Number of Cities/Provinces reported: 549 
  Unique number of distinct geographical locations combined: 3781 
---------------------------------------------------------------------------------------------------------------------------------- 
                     Location Confirmed Perc.Confirmed Deaths Perc.Deaths Recovered Perc.Recovered Active Perc.Active
1        Metropolitana, Chile    206246           2.15   4206        2.04         0           0.00 202040       97.96
2 New York City, New York, US    213699           2.22  22384       10.47         0           0.00 191315       89.53
3           Sao Paulo, Brazil    248587           2.59  13759        5.53     49295          19.83 185533       74.64
4                  Lima, Peru    151225           1.57   4029        2.66         0           0.00 147196       97.34
5     England, United Kingdom    159696           1.66  38706       24.24         0           0.00 120990       75.76
================================================================================================================================== 
  	 Confirmed	Deaths	Recovered	Active 
  Totals 
  	 9609829	489312	4838921	4205658 
  Average 
  	 2541.61	129.41	1279.8	1112.31 
  Standard Deviation 
  	 13197.35	1071.37	13601.24	15132.39 
  

 * Statistical estimators computed considering 3781 independent reported entries 
 

******************************************************************************** 
********************************  OVERALL SUMMARY******************************** 
******************************************************************************** 
  ****  Time Series Worldwide TOTS **** 
  	 ts-confirmed	ts-deaths	ts-recovered 
  	 9609829	489312	4838921 
  			5.09% 		50.35% 
  ****  Time Series Worldwide AVGS **** 
  	 ts-confirmed	ts-deaths	ts-recovered 
  	 36127.18	1839.52	19126.17 
  			5.09% 		52.94% 
  ****  Time Series Worldwide SDS **** 
  	 ts-confirmed	ts-deaths	ts-recovered 
  	 177717.39	9498.41	72771.61 
  			5.34% 		40.95% 
  

 * Statistical estimators computed considering 266/266/253 independent reported entries per case-type 
******************************************************************************** 
\end{lstlisting}

A daily generated report is also available from the \covidPckg documentation site,
\url{https://mponce0.github.io/covid19.analytics/}.

\subsubsection{Totals per Geographical Location}
\label{sec:ex_totals}
The \covidPckg package allows users to investigate total cumulative quantities
per geographical location with the \code{totals.per.location} function.
Examples of this are shown in Lst.~\ref{lst:ex_totals}.

\Rlisting
\begin{lstlisting}[caption={Calculation of totals per Country/Region/Province.
		In addition to the graphical output as shown in Fig.~\ref{fig:ex_totals},
		the function will provide details of the models fitted to the data.},
		captionpos=b,
		label={lst:ex_totals}]
# totals for confirmed cases for "Ontario"
tots.per.location(covid19.confirmed.cases,geo.loc="Ontario")

# total for confirmed cases for "Canada"
tots.per.location(covid19.confirmed.cases,geo.loc="Canada")

# total nbr of confirmed cases in Hubei including a confidence band based on moving average
tots.per.location(covid19.confirmed.cases,geo.loc="Hubei", confBnd=TRUE)

# total nbr of deaths for "Mainland China"
tots.per.location(covid19.TS.deaths,geo.loc="China")

###

# read the time series data for all the cases
all.data <- covid19.data('ts-ALL')

# run on all the cases
tots.per.location(all.data,"Japan")

###

# total for death cases for "ALL" the regions
tots.per.location(covid19.TS.deaths)

# or just
tots.per.location(covid19.data("ts-confirmed"))
\end{lstlisting}

\begin{figure}
        \includegraphics[width=0.325\textwidth]{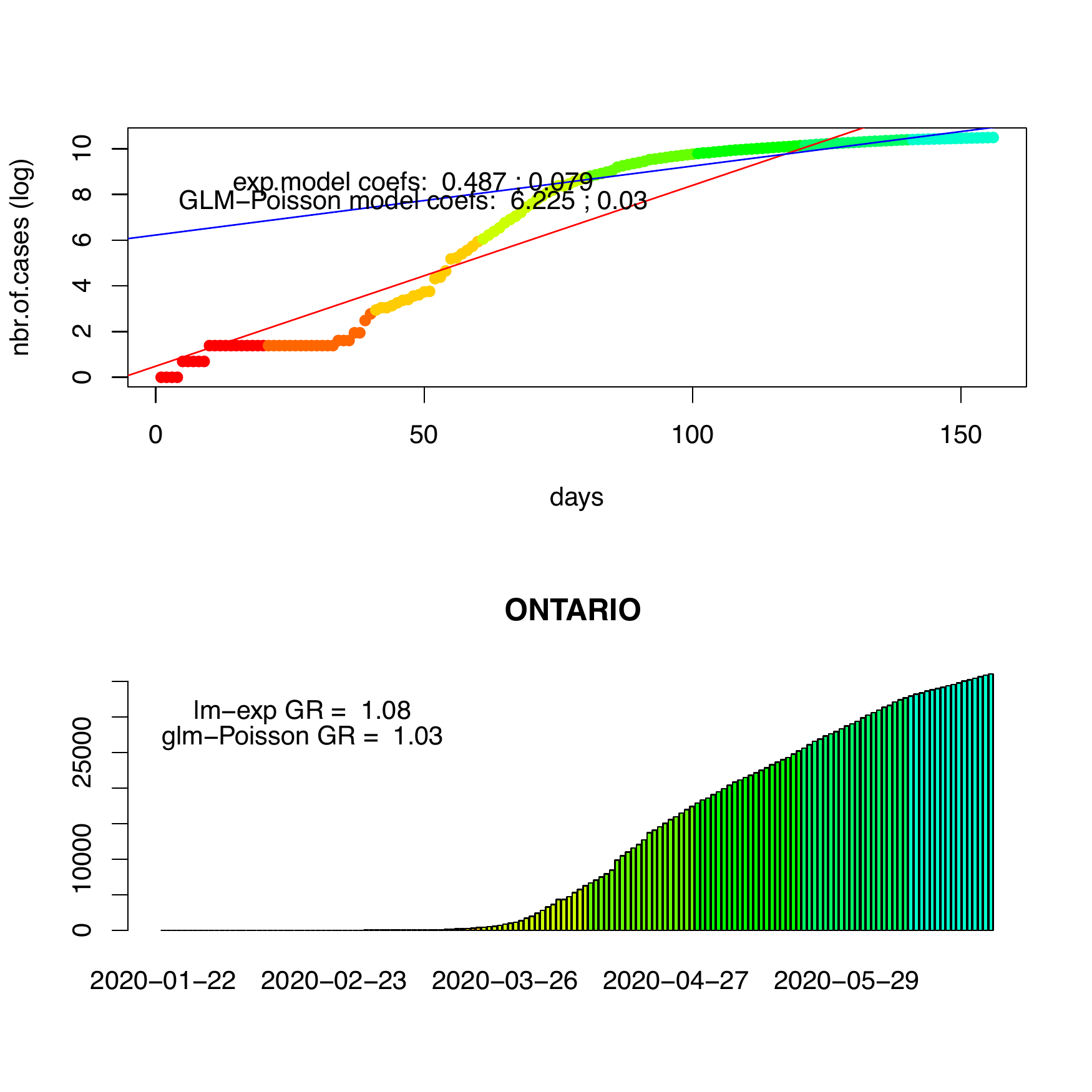}
        \includegraphics[width=0.325\textwidth]{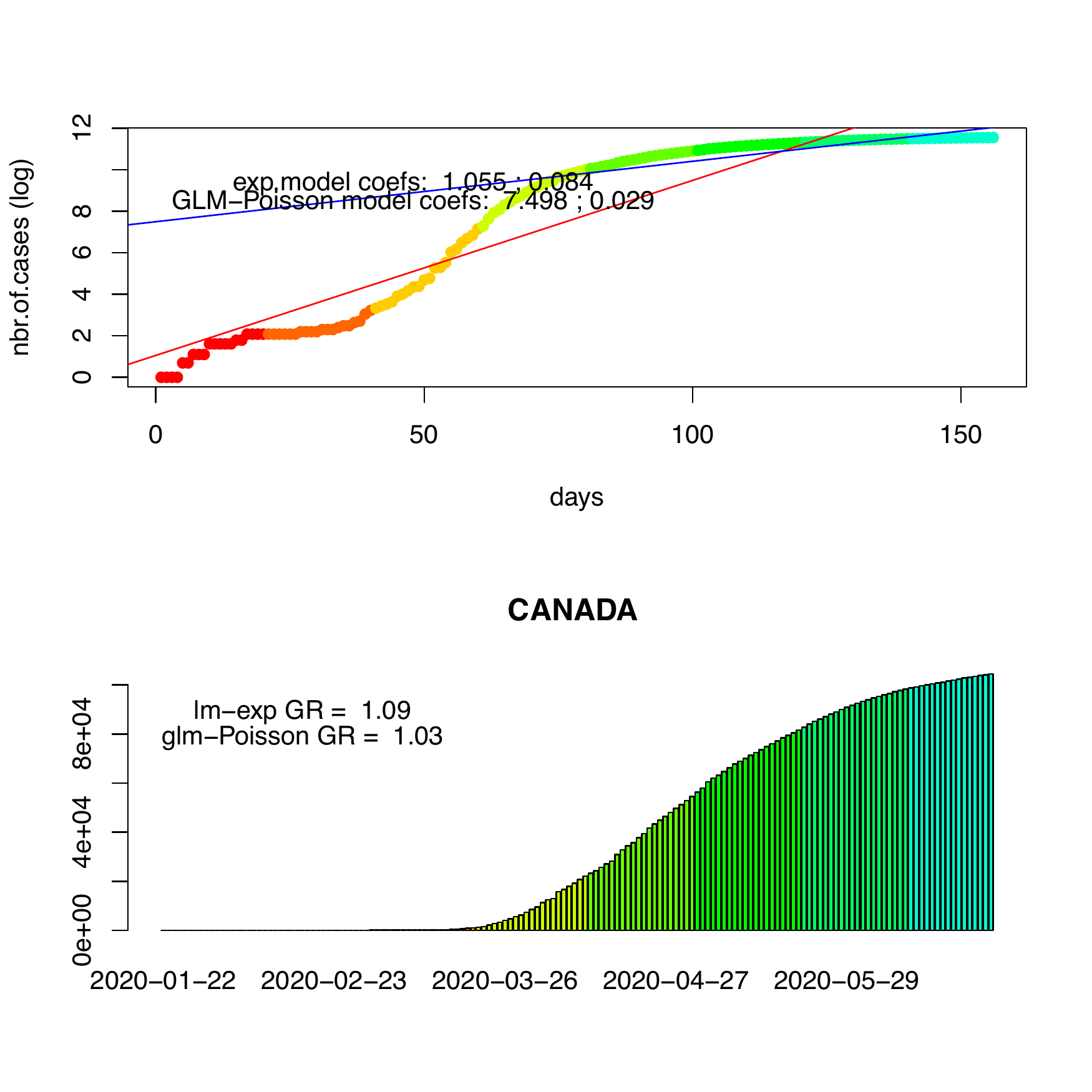}
        \includegraphics[width=0.325\textwidth]{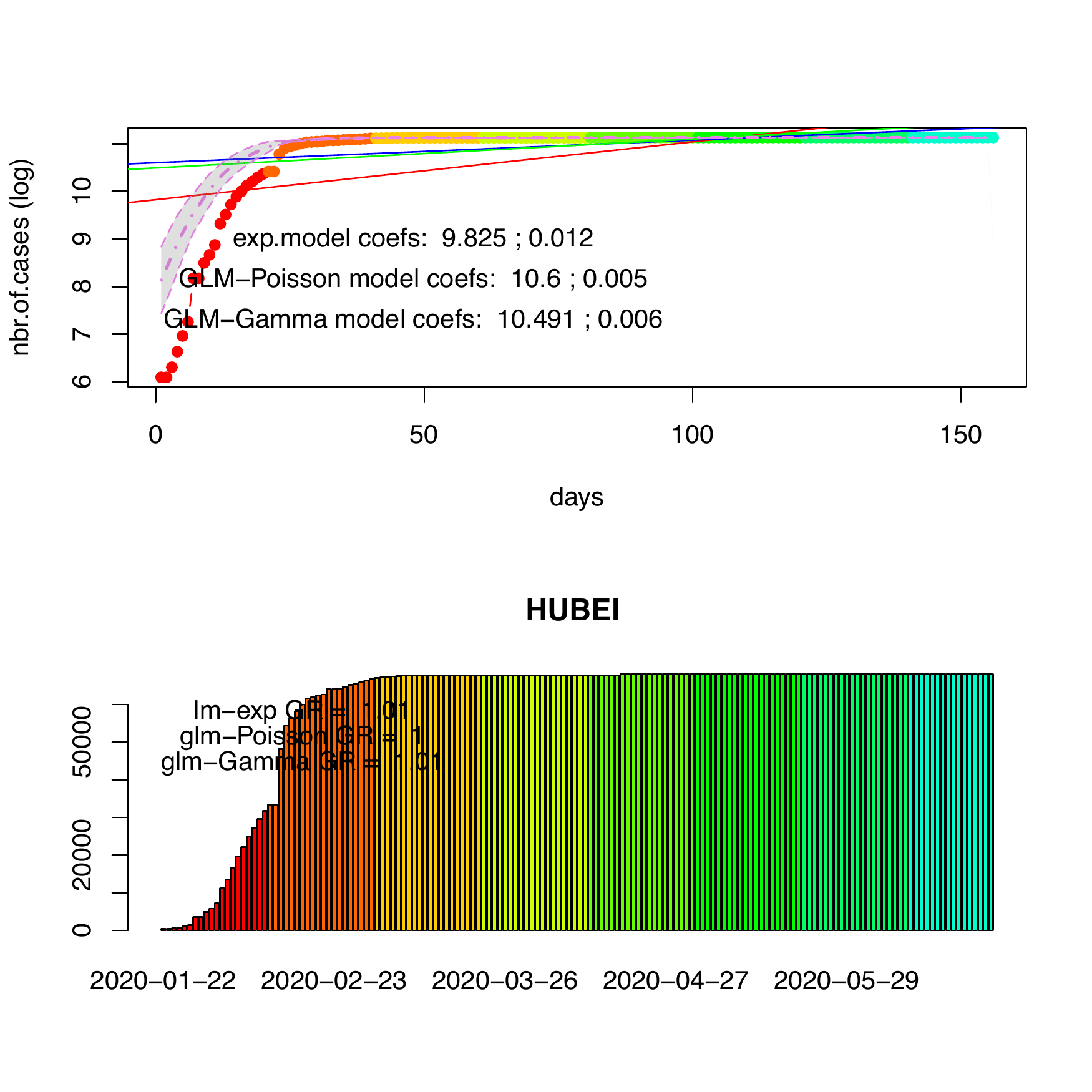}

        \caption{Graphical output produced by the \code{totals.per.location} function
		for the first three examples shown in Lst.~\ref{lst:ex_totals}.
		Each figure shows in the top row the number of cases in log-scale in
		the vertical axis and the number of days in the horizontal axis.
		The upper panel also includes the possible fits as described in
		Sec.~\ref{sec:Fns} that the function attempts to perform to the data.
		In the lower panel, the number of cases is presented in linear
		scale and the horizontal axis shows the actual dates.
		}
	\label{fig:ex_totals}
\end{figure}

\subsubsection{Growth Rate}
\label{sec:ex_growthrate}
Similarly, utilizing the \code{growth.rate} function is possible to compute
the actual \textit{growth rate} and \textit{daily changes} for specific locations,
as defined in Sec.~\ref{sec:Fns}.
Lst.~\ref{lst:ex_growthRate} includes examples of these.

\Rlisting
\begin{lstlisting}[caption={Calculation of growth rates and daily changes per Country/Region/Province.
		},
		captionpos=b,
		label={lst:ex_growthRate}]
# read time series data for confirmed cases
TS.data <- covid19.data("ts-confirmed")

# compute changes and growth rates per location for all the countries
growth.rate(TS.data)

# compute changes and growth rates per location for 'Italy'
growth.rate(TS.data,geo.loc="Italy")

# compute changes and growth rates per location for 'Italy' and 'Germany'
growth.rate(TS.data,geo.loc=c("Italy","Germany"))

#####

# Combining multiple geographical locations:

# obtain Time Series data
TSconfirmed <- covid19.data("ts-confirmed")

# explore different combinations of regions/cities/countries
# when combining different locations, heatmaps will also be generated comparing the trends among these locations
growth.rate(TSconfirmed,geo.loc=c("Italy","Canada","Ontario","Quebec","Uruguay"))

growth.rate(TSconfirmed,geo.loc=c("Hubei","Italy","Spain","United States","Canada","Ontario","Quebec","Uruguay"))

growth.rate(TSconfirmed,geo.loc=c("Hubei","Italy","Spain","US","Canada","Ontario","Quebec","Uruguay"))

# turn off static plots and activate interactive figures
growth.rate(TSconfirmed,geo.loc=c("Brazil","Canada","Ontario","US"), staticPlt=FALSE, interactiveFig=TRUE)

# static and interactive figures
growth.rate(TSconfirmed,geo.loc=c("Brazil","Italy","India","US"), staticPlt=TRUE, interactiveFig=TRUE)
\end{lstlisting}

\begin{figure}
	\centering
        \includegraphics[width=0.8\textwidth]{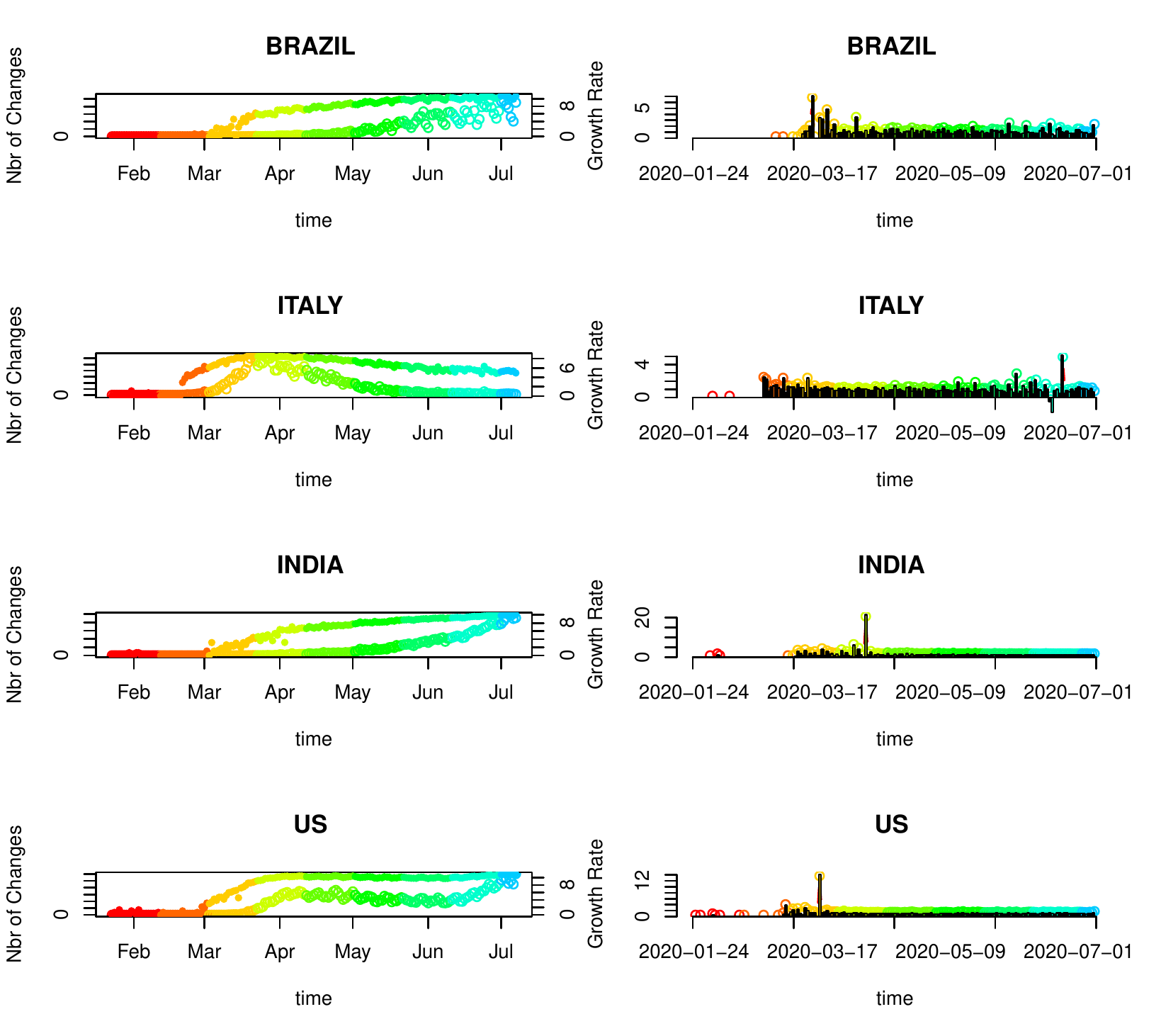}

	\includegraphics[width=0.45\textwidth]{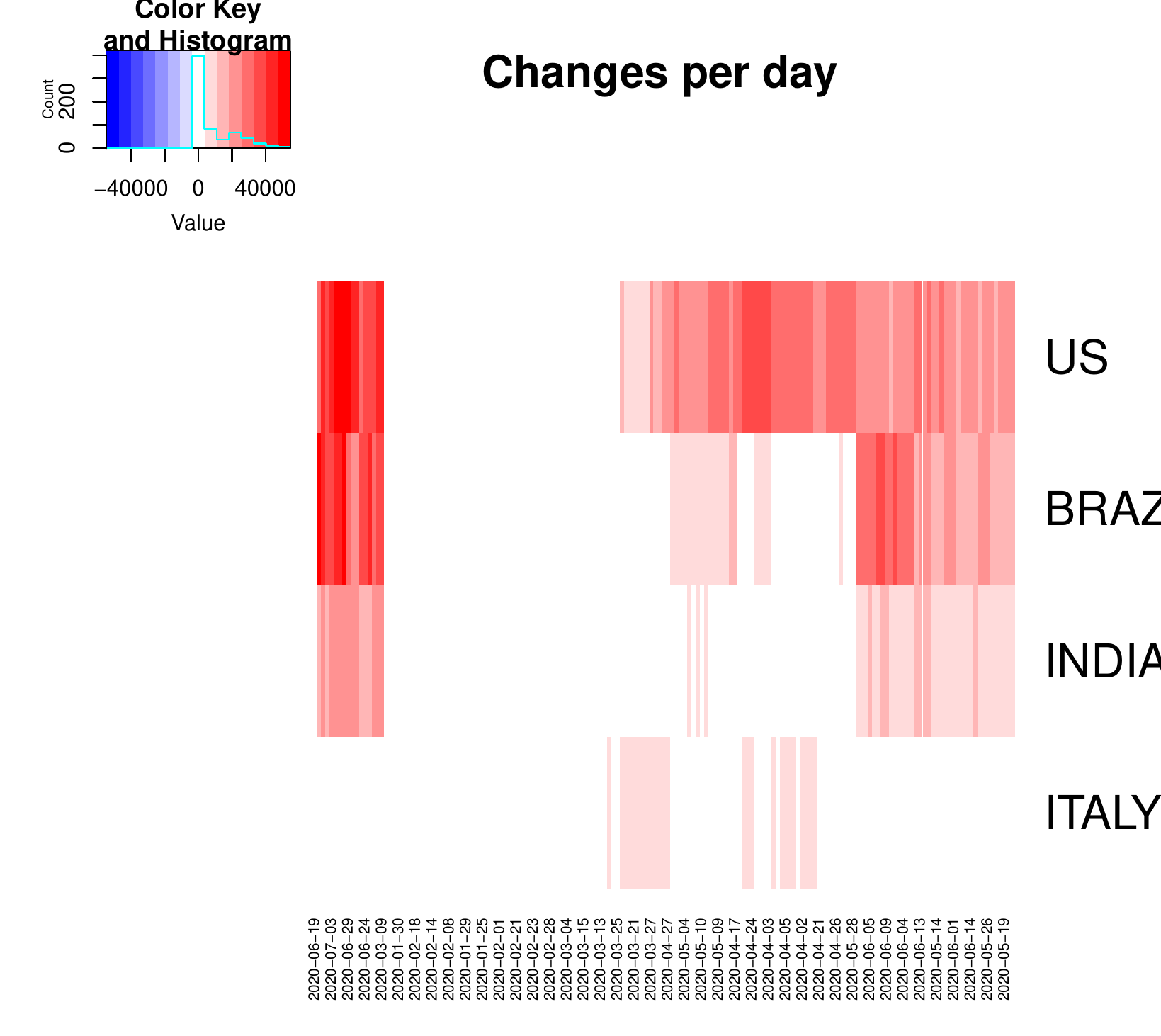}
	\includegraphics[width=0.45\textwidth]{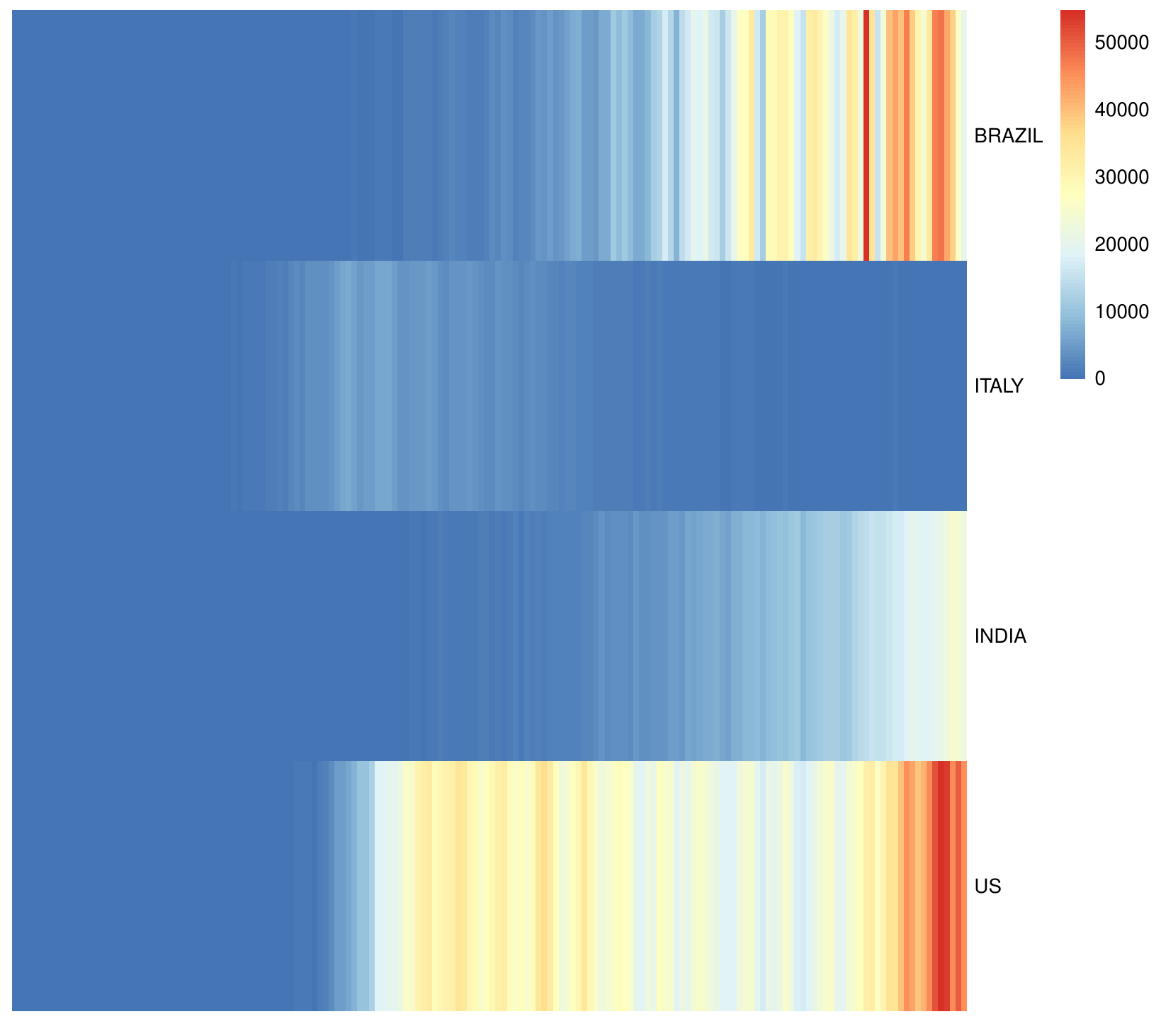}

	\includegraphics[width=0.45\textwidth]{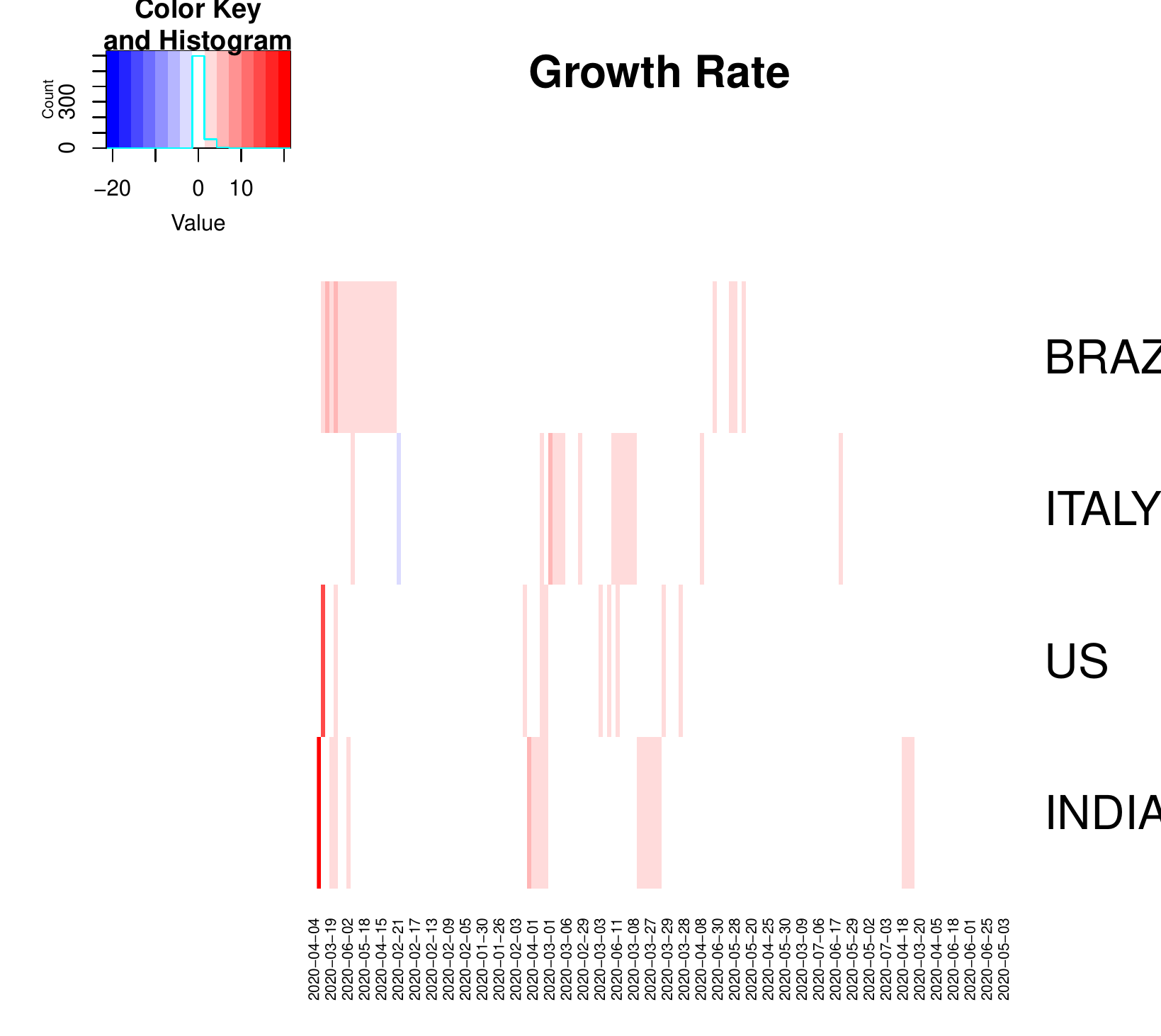}
	\includegraphics[width=0.45\textwidth]{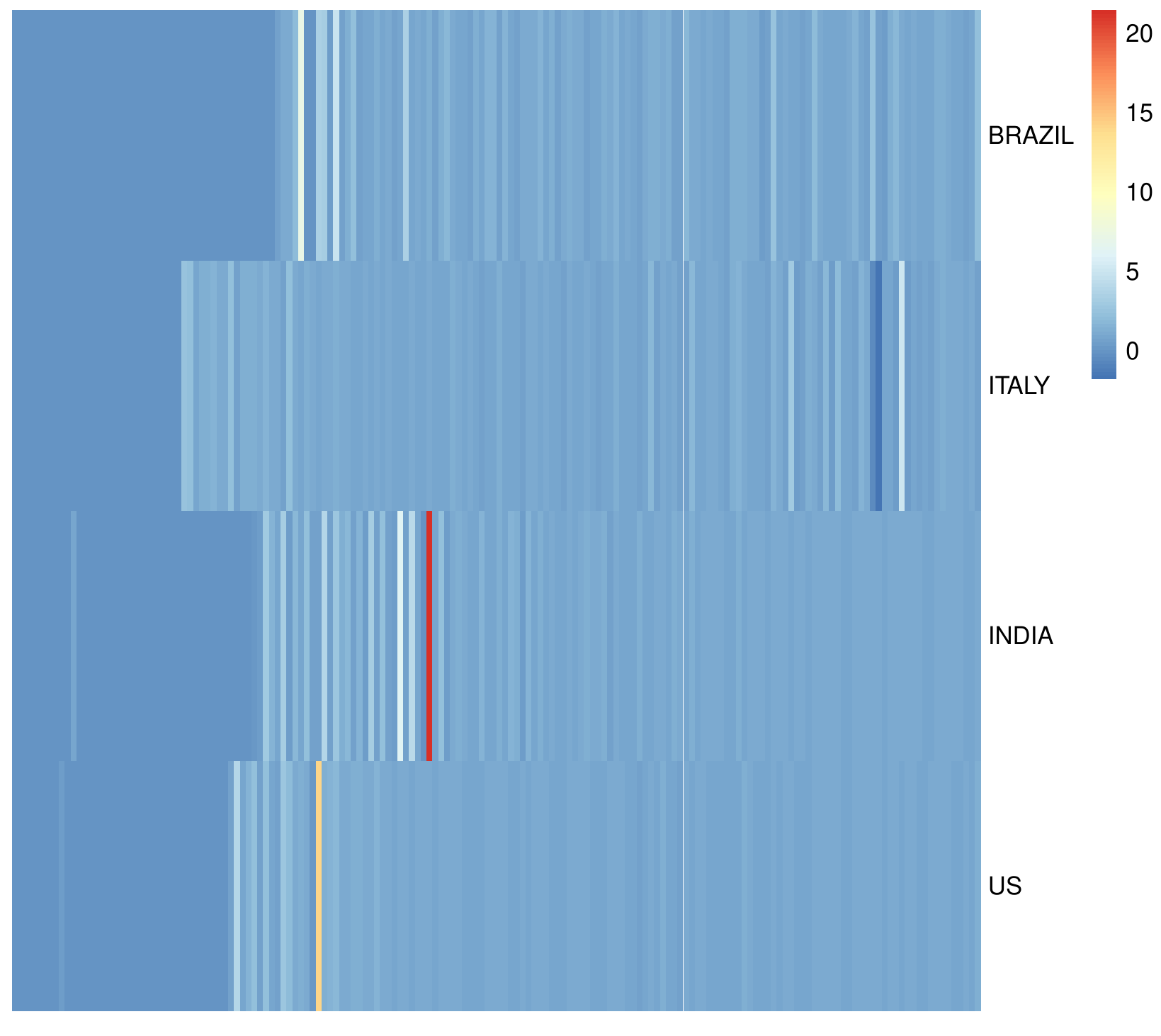}

        \backcaption{Graphical output produced by the \code{growth.rate} function when
		comparing the situation in "Brazil", "Italy", "India" and the "US".
		The first figure in the top row,
		displays the daily changes both in linear (left vertical axis) and
		log (right vertical axis) scales as a function of time (first column) --hence the two indicators in the plot--,
		and growth rate in linear scale (second column) as a function of time;
		each row within this specific figure represents each of the different locations.
		The remaining figures are heatmaps displayed in two different styles to emphasize
		different aspects of the daily changes and growth rates comparing the selected locations
		-- dates are represented in the horizontal direction, locations are placed along the vertical axis
		and color-coded are the corresponding quantities.
                }
	{fig:ex_growthRates}
	\label{fig:ex_growthRates}
\end{figure}

\begin{figure}
        \centering
        \includegraphics[width=0.45\textwidth]{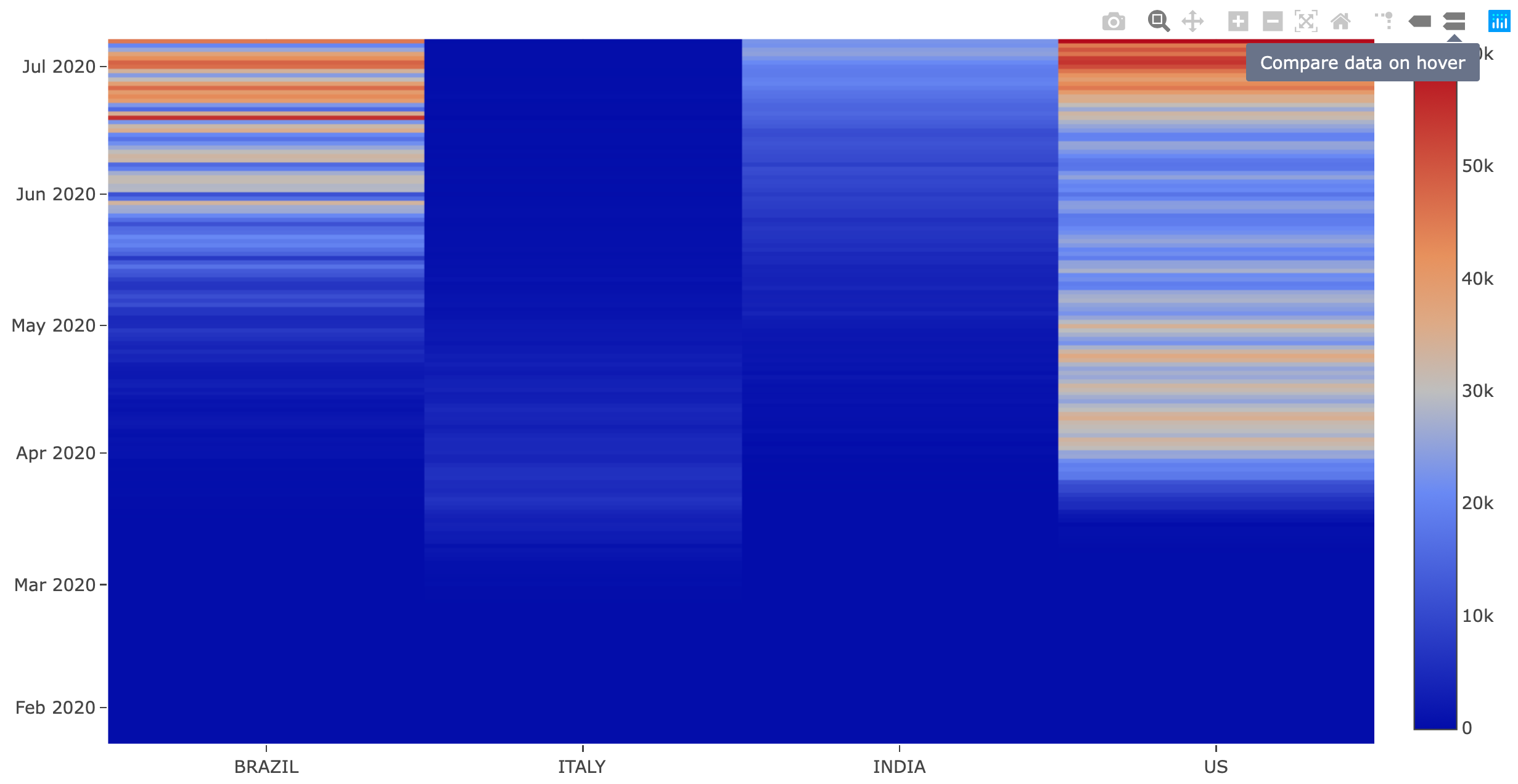}
        \includegraphics[width=0.45\textwidth]{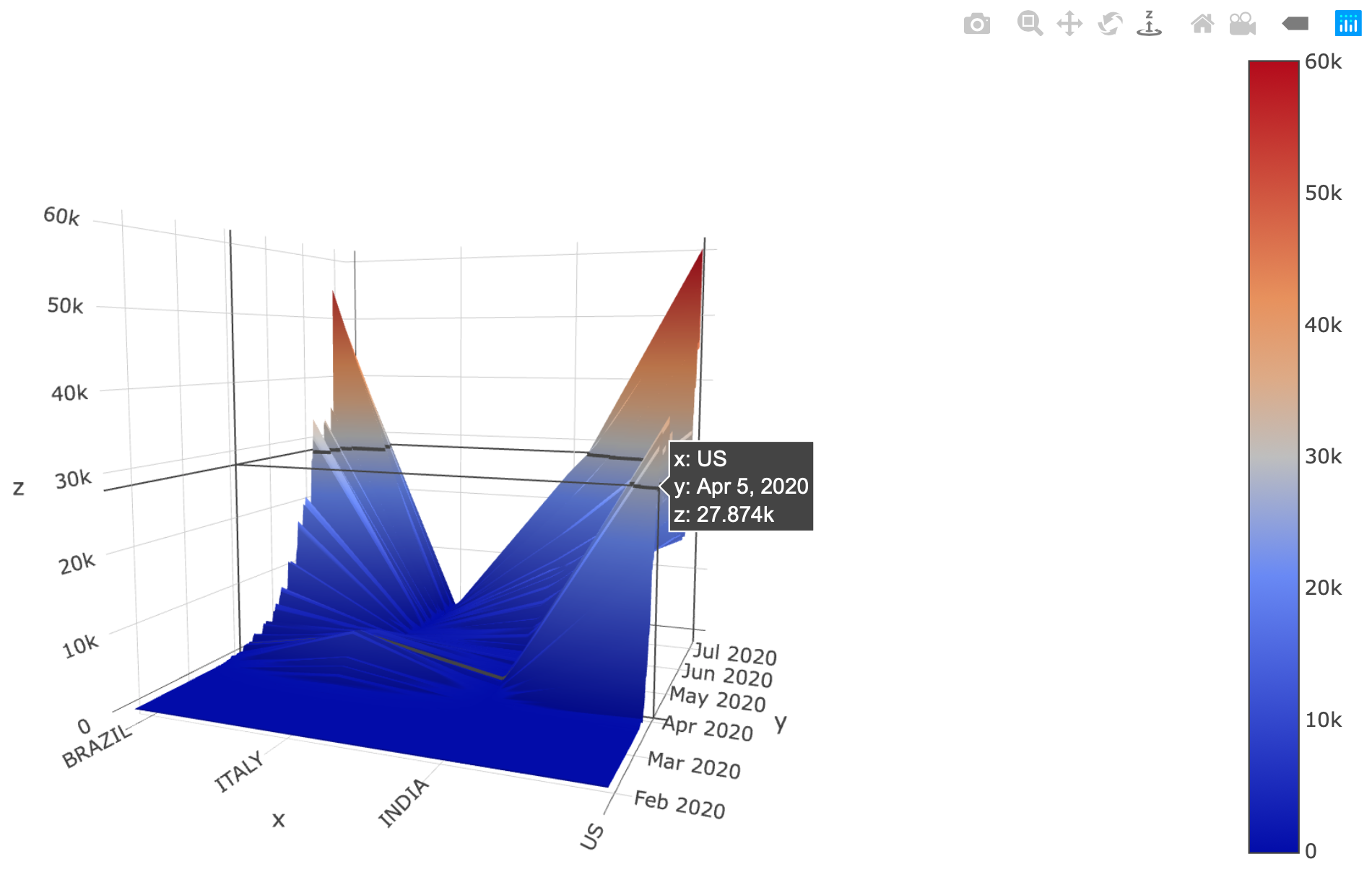}

	\backcaption{Interactive visualization of the daily changes in a heatmap (left) and
		3d-surface (right) representation, generated by the \code{growth.rate} function
		when activating the \code{interactiveFig=TRUE}.
	}
	{fig:changes_interactive}
	\label{fig:changes_interactive}
\end{figure}

\subsubsection{Trends}
\label{sec:ex_trends}
\begin{figure}
	\includegraphics[width=0.4\textwidth]{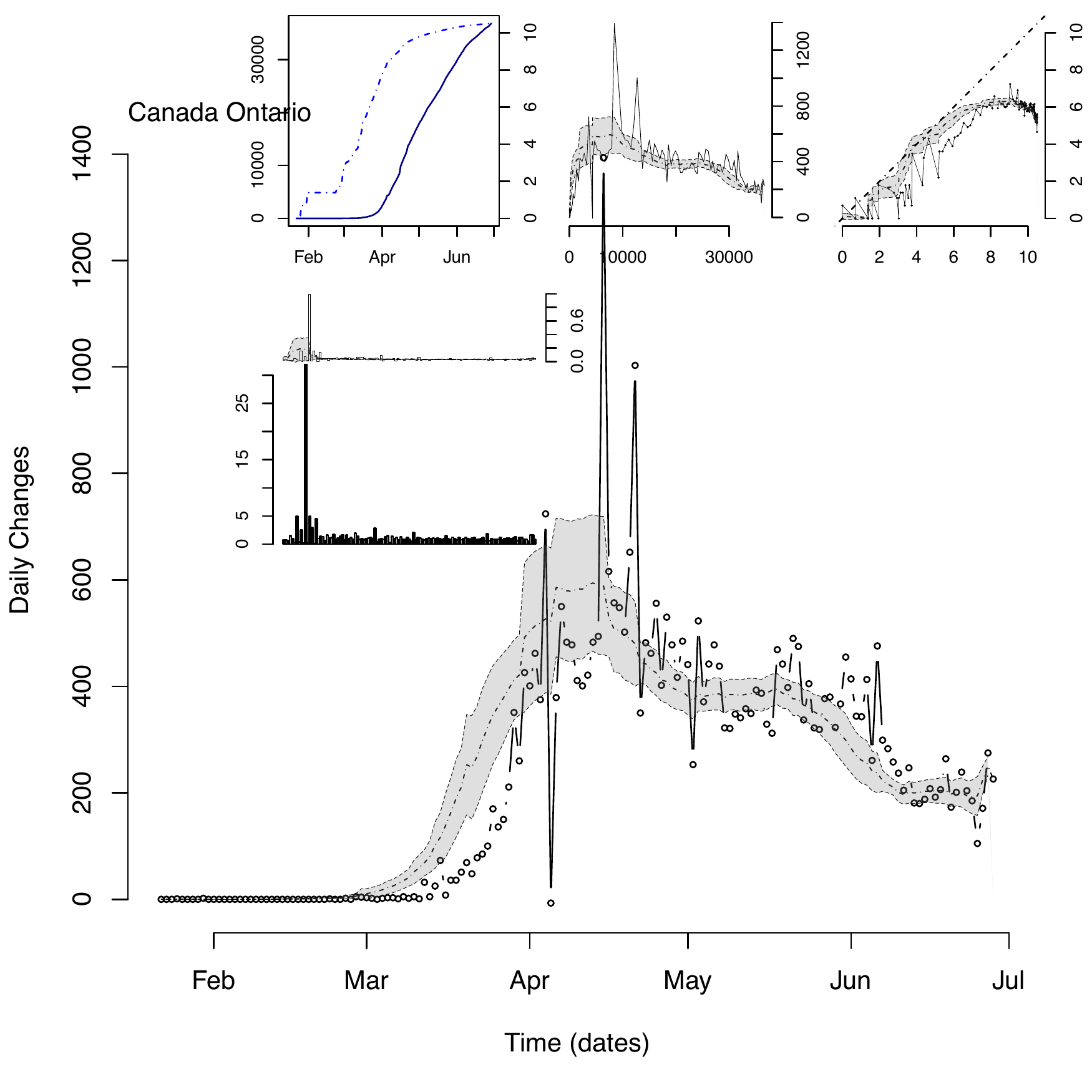}
	\includegraphics[width=0.6\textwidth]{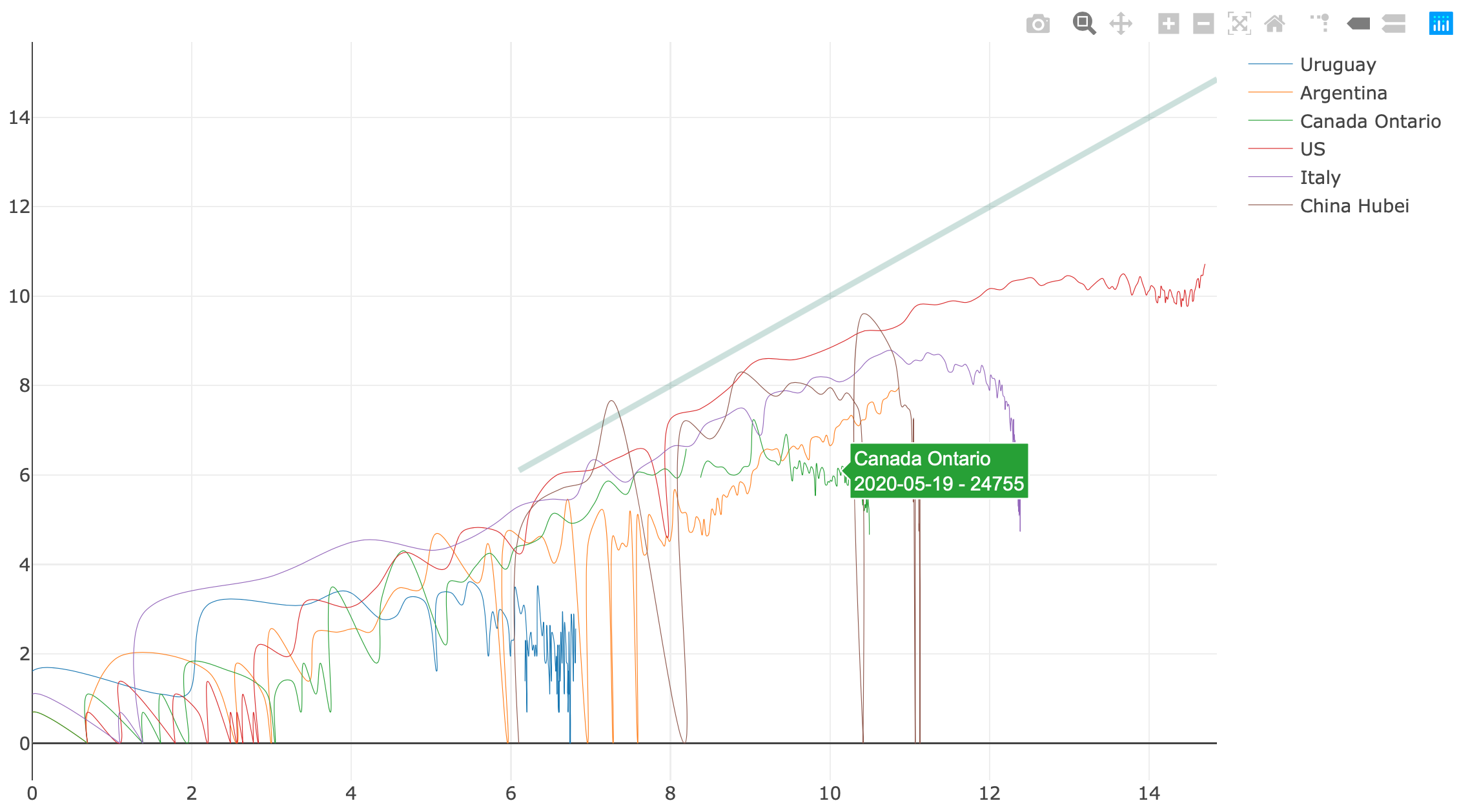}

	\backcaption{Static (left) and interactive (right) figures generated by the
		\code{single.trend}/\code{mtrends} and \code{itrends} functions respectively.
		The static figure includes several representations of the daily changes as described in Sec.~\ref{sec:Fns}.
		The interactive figure, offers a quick overview of the trend in particular
		compared to the straight diagonal line included which represents "exponential growth".
	}
	{fig:ex_trends}
	\label{fig:ex_trends}
\end{figure}

In addition to the cumulative indicators described above, it is possible to
estimate the global trends per location employing the functions
\code{single.trend}, \code{mtrends} and \code{itrends}.
The first two functions generate static plots of different quantities that can
be used as indicators, while the third function generates an interactive
representation of a normalized a-dimensional trend.
The Lst.~\ref{lst:ex_trends} shows examples of the use of these functions.
Fig.~\autobackref{fig:ex_trends} displays the graphical output produced by these functions.

\Rlisting
\begin{lstlisting}[caption={Calculation of trends for different cases, utilizing
			the \code{single.trend}, \code{mtrends} and \code{itrends} functions.
			The typical representations can be seen in Fig.~\ref{fig:ex_trends}.
			},
                captionpos=b,
                label={lst:ex_trends}]
# single location trend, in this case using data from the City of Toronto
tor.data <- covid19.Toronto.data()
single.trend(tor.data[tor.data$status=="Active Cases",])

# or data from the province of Ontario
ts.data <- covid19.data("ts-confirmed")
ont.data <- ts.data[ ts.data$Province.State == "Ontario",]
single.trend(ont.data)

# or from Italy
single.trend(ts.data[ ts.data$Country.Region=="Italy",])


# multiple locations
ts.data <- covid19.data("ts-confirmed")
mtrends(ts.data, geo.loc=c("Canada","Ontario","Uruguay","Italy"))

# multiple cases
mtrends(tor.data)


# interactive plot of trends
# for all locations and all type of cases
itrends(covid19.data("ts-ALL"),geo.loc="ALL")

# or just for confirmed cases and some specific locations, saving the result in an HTML file named "itrends_ex.html"
itrends(covid19.data("ts-confirmed"), geo.loc=c("Uruguay","Argentina","Ontario","US","Italy","Hubei"), fileName="itrends_ex")

# interactive trend for Toronto cases
itrends(tor.data[,-ncol(tor.data)])
\end{lstlisting}

\subsubsection{Interactive Visualization Tools}
\label{sec:ex_viz}

Most of the analysis functions in the \covidPckg package have already plotting
and visualization capabilities.
In addition to the previously described ones, the package has also specialized
visualization functions as shown in Lst.\ref{lst:ex_viz}.
Many of them will generate static and interactive figures, see Table~\ref{table:covid19Fns}
for details of the type of output.
In particular the \code{live.map} function is an utility function which allows
to plot the location of the recorded cases around the world.
This function in particular allows for several customizable features, such as,
the type of projection used in the map or to select different types of
projection operators in a pull down menu, displaying or not the legend of the
regions, specify rescaling factors for the sizes representing the number of cases,
among others.
The function will generate a live representation of the cases, utilizing the
\CRANpkg{plotly} package and ultimately open the map in a browser, where the user
can explore the map, drag the representation, zoom in/out, turn on/off legends, etc.

\Rlisting
\begin{lstlisting}[caption={Examples of some of the interactive and visualization capabilities of plotting functions.
                        The typical representations can be seen in Fig.~\ref{fig:ex_viz}.
                        },
                captionpos=b,
                label={lst:ex_viz}]
# retrieve time series data
TS.data <- covid19.data("ts-ALL")

# static and interactive plot 
totals.plt(TS.data)
# totals for Ontario and Canada, without displaying totals and one plot per page
totals.plt(TS.data, c("Canada","Ontario"), with.totals=FALSE,one.plt.per.page=TRUE)

# totals for Ontario, Canada, Italy and Uruguay; including global totals with the linear and semi-log plots arranged one next to the other
totals.plt(TS.data, c("Canada","Ontario","Italy","Uruguay"), with.totals=TRUE,one.plt.per.page=FALSE)

# totals for all the locations reported on the dataset, interactive plot will be saved as "totals-all.html"
totals.plt(TS.data, "ALL", fileName="totals-all")
# retrieve aggregated data
data <- covid19.data("aggregated")

# interactive map of aggregated cases -- with more spatial resolution
live.map(data)

# or
live.map()

# interactive map of the time series data of the confirmed cases with less spatial resolution, ie. aggregated by country
live.map(covid19.data("ts-confirmed"))
\end{lstlisting}

\begin{figure}
	\centering
        \includegraphics[height=0.45\textwidth]{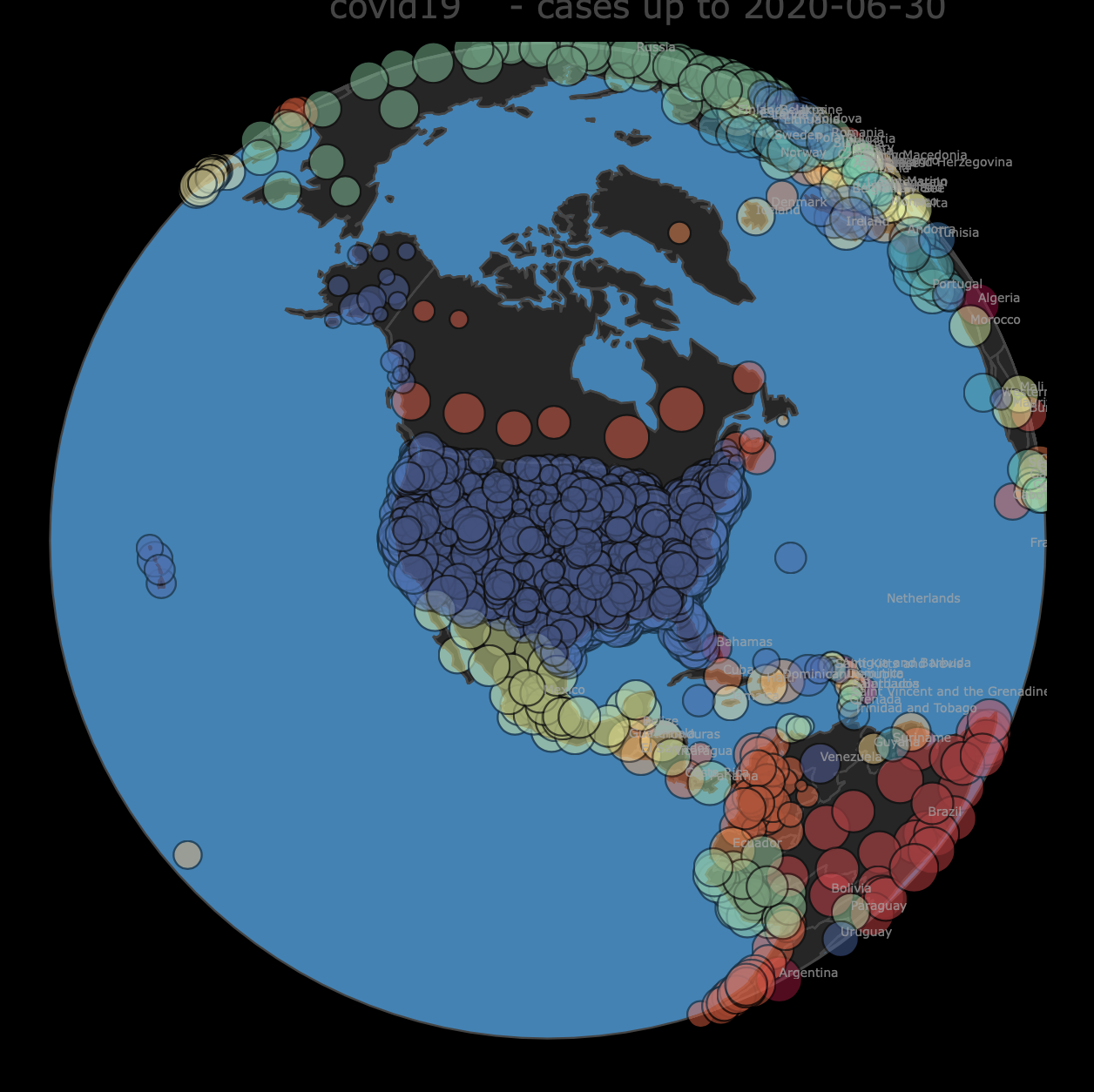}
        \includegraphics[height=0.45\textwidth]{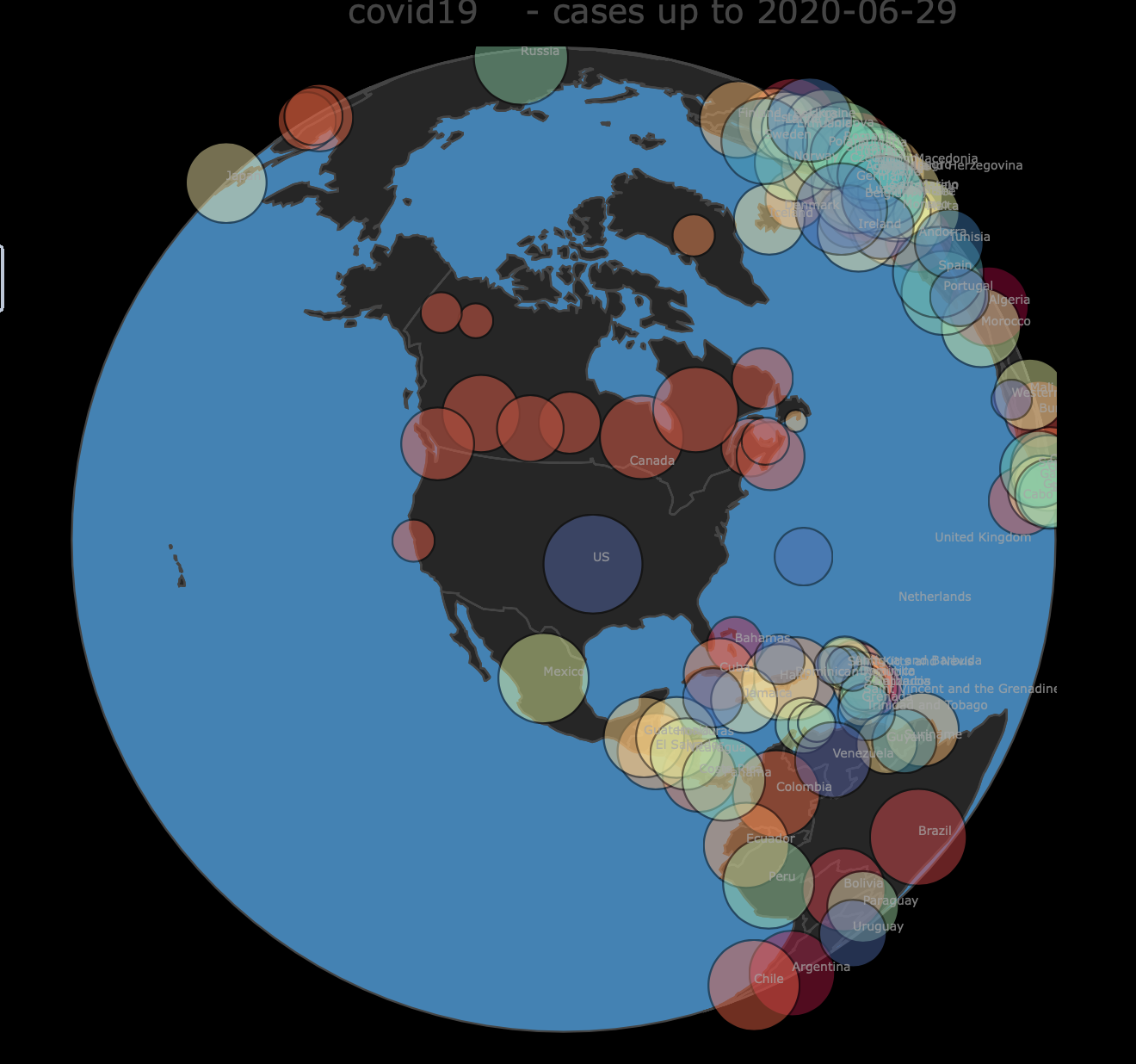}

	\includegraphics[width=0.95\textwidth]{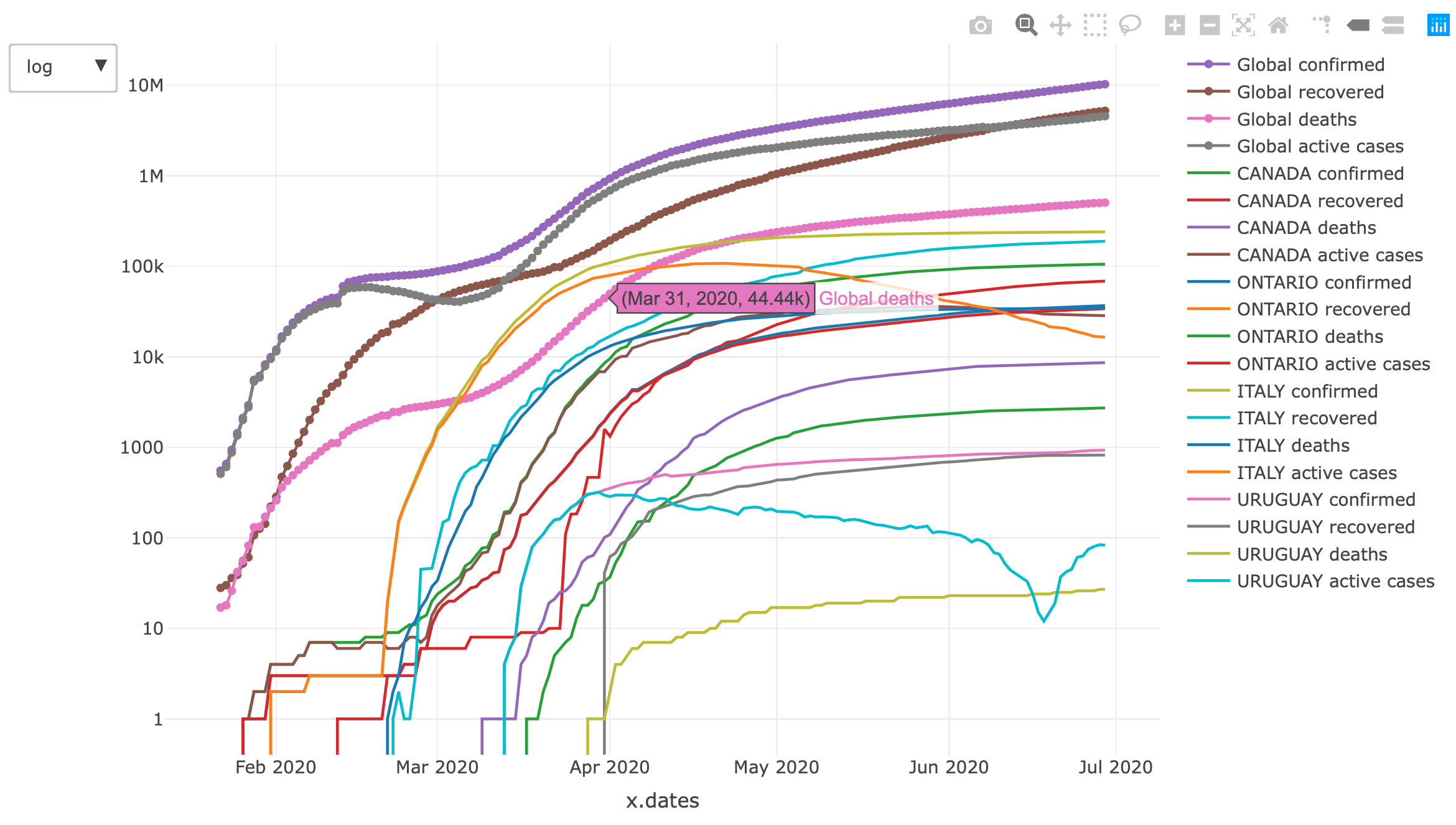}

	\backcaption{Examples of some of the interactive figures generated using the
		\code{live.map} (upper row, for time series and aggregated data respectively)
		and \code{totals.plt} functions (lower row) representing the total number of
		cases vs the reported dates for the selected regions.
		The latter visualization also allows the user to switch between
		a linear and a log-scale representation via a pull-down menu.
	}{fig:ex_viz}
        \label{fig:ex_viz}
\end{figure}

\subsection{Modeling the Virus Spread}
\label{sec:ex_models}

\begin{figure}
        \includegraphics[height=0.5\textwidth]{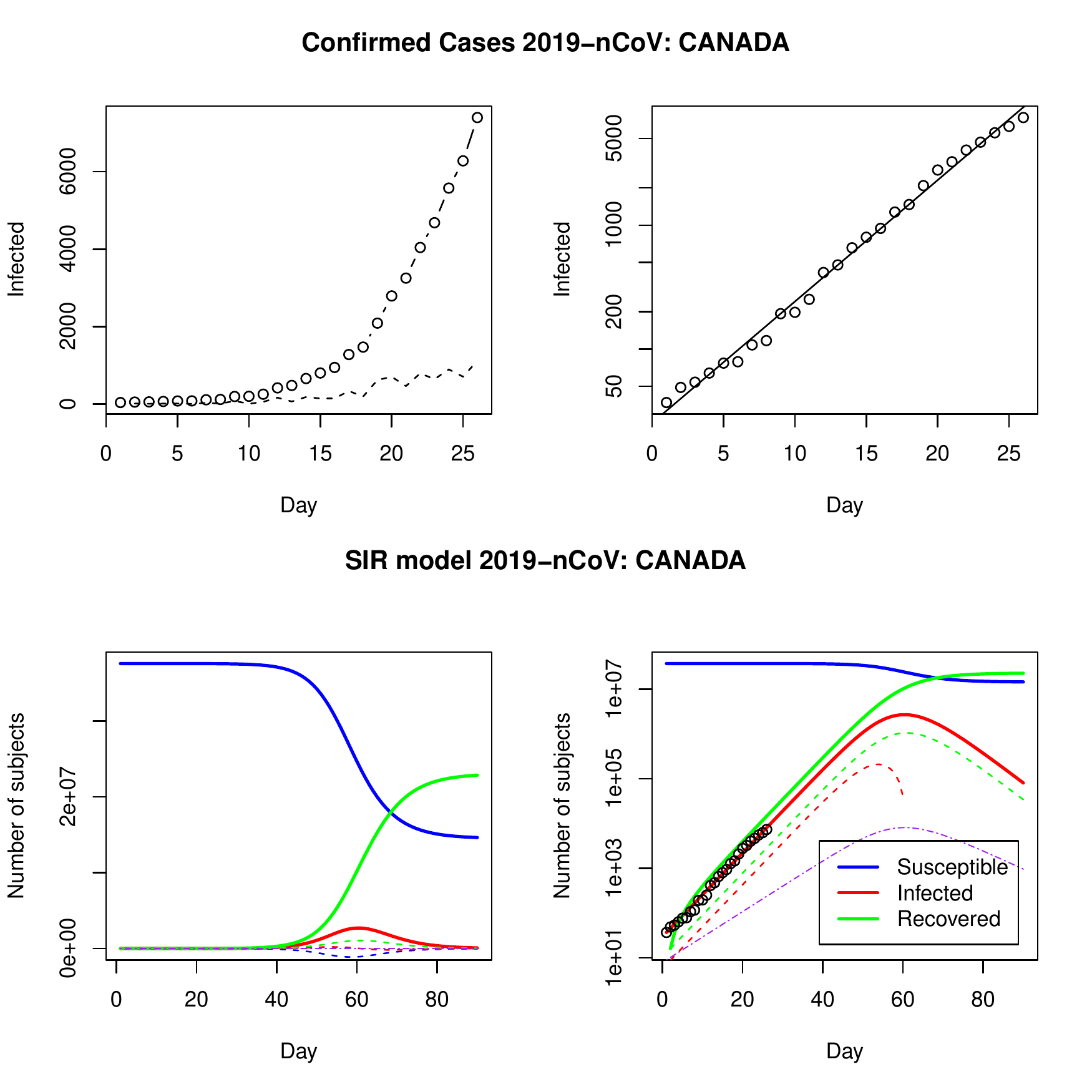}
	\includegraphics[height=0.5\textwidth]{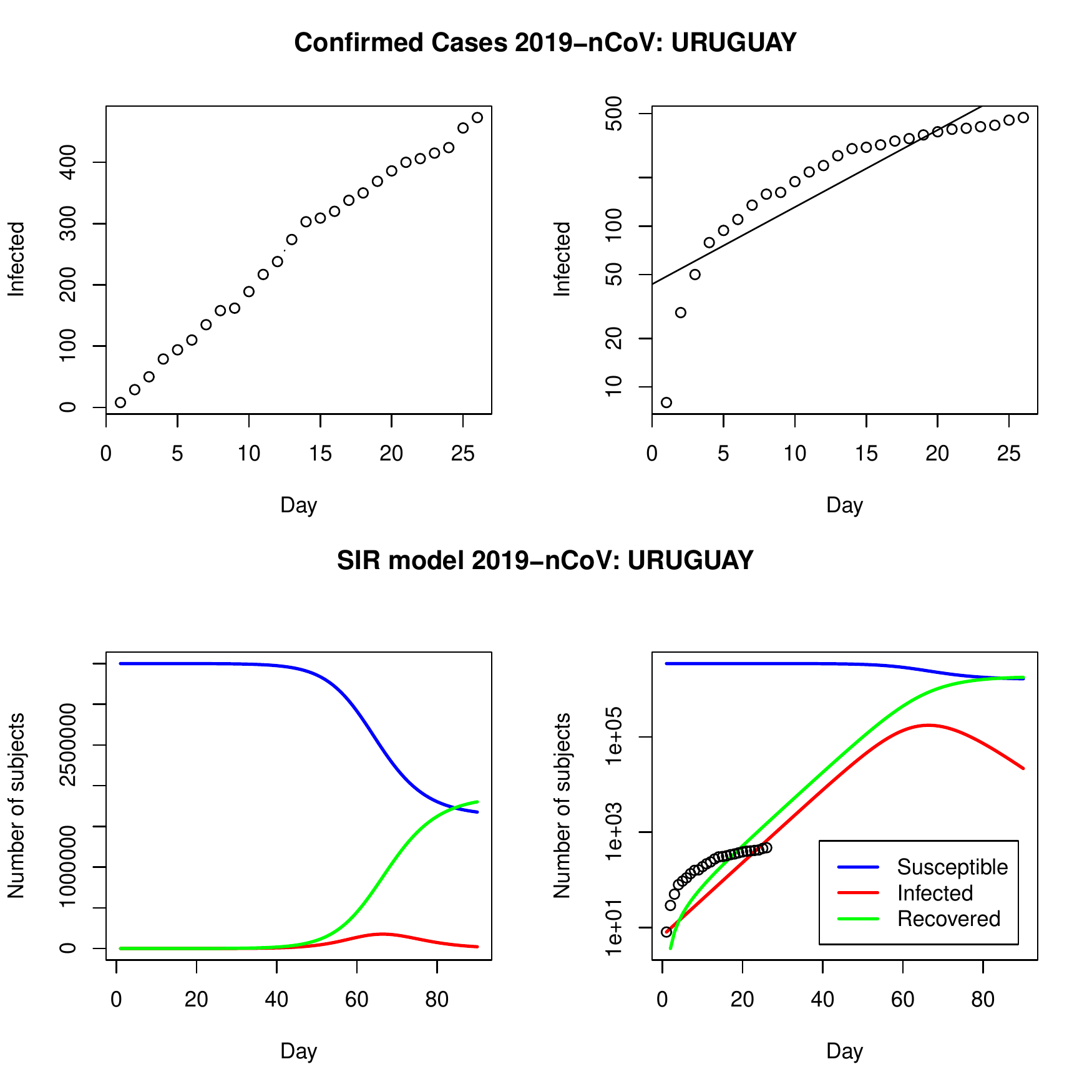}

        \caption{Graphical output of the SIR model produced by the examples from
		Lst.~\ref{lst:ex_SIRmodel}, applied to the cases of Canada (left)
		and Uruguay (right).
		Each figure contains 4 plots, the upper ones display the actual
		data for the number of infected individuals for the region
		in linear (left) and log-scale (right) as a function of time.
		The bottom panels, show the solution to the SIR model defined in
		Eq.(\ref{eqn:SIR_model}) in linear (left panel) and log-scale (right panel),
		in conjunction with the data points employed to determine
		the \textit{transition rates} parameters, $\beta$ and $\gamma$, for the model.
		On the left figure (Canada): dashed lines represent time derivatives,
		while the purple dash-dotted line represents the force of infection;
		these additional indicators can be added using the \code{add.extras=TRUE} flag
		when invoking the \code{generate.SIR.model}/\code{plt.SIR.model} functions.
		An interestng observation, is that in some cases the model can clearly trace a quite remarkable
		trend in accordance to the data, in particularly when an exponential growth is present,
		while in others not -- i.e. when the particular region has somehow
		manage to --usually the so-called--
		``\textit{flatten} (the growth of the) \textit{curve}''.}
        \label{fig:ex_SIRmodel}
\end{figure}

Last but not least, one the novel features added by the \covidPckg package,
is the ability of model the spread of the virus by incorporating real data.
As described in Sec.~\ref{sec:Fns}, the \code{generate.SIR.model} function,
implements a simple SIR model employing the data reported from an specified
dataset and a particular location.
Examples of this are shown in Lst.\ref{lst:ex_SIRmodel}.
The \code{generate.SIR.model} function is complemented with the 
\code{plt.SIR.model} function which can be used to generate static or
interactive figures as shown in Fig.~\ref{fig:ex_SIRmodel}.

The \code{generate.SIR.model} function as described in Sec.\ref{sec:covid19Pckg}
will attempt to obtain proper values for the parameters $\beta$ and $\gamma$,
by inferring the onset of the epidemic using the actual data.
This is also listed in the output of the function (see Lst.\ref{lst:ex_SIRmodel_output}),
and it can be controlled by setting the parameters \code{t0} and \code{t1} or
\code{deltaT}, which are used to specify the range of dates to be considered
for using when determining the values of $\beta$ and $\gamma$.
The fatality rate (constant) can also be indicated via the \code{fatality.rate}
argument, as well, as the total population of the region with
\code{tot.population}.

\Rlisting
\begin{lstlisting}[caption={Examples of SIR model generation using different
			datasets and locations via the \code{generate.SIR.model} function.},
                captionpos=b,
                label={lst:ex_SIRmodel}]
# read time series data for confirmed cases
data <- covid19.data("ts-confirmed")

# run a SIR model for a given geographical location
generate.SIR.model(data,"Hubei", t0=1,t1=15)
generate.SIR.model(data,"Germany",tot.population=83149300)
generate.SIR.model(data,"Uruguay", tot.population=3500000)
generate.SIR.model(data,"Ontario",tot.population=14570000, add.extras=TRUE)

# the function will aggregate data for a geographical location, like a country with multiple entries
generate.SIR.model(data,"Canada",tot.population=37590000, add.extras=TRUE)

#####

# modelling the spread for the whole world, storing the model and generating an interactive visualization
world.SIR.model <- generate.SIR.model(data,"ALL", t0=1,t1=15, tot.population=7.8e9, staticPlt=FALSE)
# plotting and visualizing the model
plt.SIR.model(world.SIR.model,"World",interactiveFig=TRUE,fileName="world.SIR.model", add.extras=TRUE)

\end{lstlisting}

\outputlisting        
\begin{lstlisting}[caption={Output from the \code{generate.SIR.model} function applied to the time series data of Canada.
			The output shows the original data points, and the selected ones to be used to determine the
			transition rate parameters of the model, $\beta$ and $\gamma$ from Eq.(\ref{eqn:SIR_model}).},
		captionpos=b,label={lst:ex_SIRmodel_output}]
################################################################################ 
################################################################################ 
Processing...  CANADA 
  [1]      0      0      0      0      1      1      2      2      2      4
 [11]      4      4      4      4      5      5      7      7      7      7
 [21]      7      7      7      7      7      7      8      8      8      8
 [31]      9      9      9     10     11     11     13     14     20     24
 [41]     27     30     33     37     49     54     64     77     79    108
 [51]    117    193    198    252    415    478    657    800    943   1277
 [61]   1469   2088   2790   3251   4042   4682   5576   6280   7398   8527
 [71]   9560  11284  12437  12978  15756  16563  17872  19141  20654  22059
 [81]  23316  24299  25680  27035  28209  30809  32814  34356  35633  37658
 [91]  39402  41663  43299  44919  46371  48033  49616  51150  52865  54457
[101]  56343  57926  60504  61957  63215  64694  66201  67674  68918  70091
[111]  71264  72419  73568  74781  75959  77206  78332  79411  80493  81575
[121]  82742  83947  85151  86106  87119  88090  88989  89976  90909  91681
[131]  92479  93288  93960  94641  95269  95947  96475  97178  97779  98241
[141]  98720  99159  99595 100043 100404 100763 101087 101491 101877 102314
[151] 102762 103078 103418 103767 104087 104463 104629 104878 105193 105830
[161] 106097 106288 106643 106962 107185
[1] 44
 [1]   37   49   54   64   77   79  108  117  193  198  252  415  478  657  800
[16]  943 1277 1469 2088 2790 3251 4042 4682 5576 6280 7398
------------------------  Parameters used to create model ------------------------ 
  	Region: CANADA 
  	Time interval to consider: t0=44 - t1= ; tfinal=90 
  		t0: 2020-03-06 -- t1:  
  	Number of days considered for initial guess: 26 
  	Fatality rate: 0.02 
  	Population of the region: 37590000 
-------------------------------------------------------------------------------- 
[1] "ERROR: ABNORMAL_TERMINATION_IN_LNSRCH"
     beta     gamma 
0.6078298 0.3921703 
  R0 = 1.54991315147533 
  Max of infecteded: 2706321.58  ( 7.2 %) 
  Max nbr of casualties, assuming  2% fatality rate: 54126.43 
  Max reached at day : 60 ==>  2020-05-05 
================================================================================ 
\end{lstlisting}

Fig.(\ref{fig:ex_SIRmodel}), also raises an interesting point regarding the
accuracy of the SIR model.
We should recall that this is the simplest approach one could take in order to
model the spread of diseases and usually more refined and complex models are used
to incorporate several factors, such as, vaccination, quarantines, effects of social
clusters, etc.
However, in some cases, specially when the spread of the disease appears to have
enter the so-called exponential growth rate, this simple SIR model can capture the
main trend of the dispersion (e.g. left plot from Fig.\ref{fig:ex_SIRmodel}).
While in other cases, when the rate of spread is slower than the freely exponential
dispersion, the model clearly fails in tracking the actual evolution of cases
(e.g. right plot from Fig.\ref{fig:ex_SIRmodel}).

Finally, Lst.~\ref{lst:ex_sweepSIRmodel} shows an example of the generation of
a sequence of values for $R_0$, and actually any of the parameteres
$(\beta,\gamma)$ describing the SIR model.
In this case, the function takes a range of values for the initial date \code{t0}
and generates different date intervals, this allows the function to generate
multiple SIR models and return the corresponding parameters for each model.
The results are then bundle in a "matrix"/"array" object which can be accessed
by column for each model or by row for each paramter sets.

\Rlisting
\begin{lstlisting}[caption={Example of SIR model generation using different
                        ranges of dates for the initial date, \code{t0},
			using the \code{sweep.SIR.models} function.},
                captionpos=b,
                label={lst:ex_sweepSIRmodel}]
# read TimeSeries data
TS.data <- covid19.data("TS-confirmed")

# select a location of interest, eg. France
# France has many entries, just pick "France"
FR.data <- TS.data[ (TS.data$Country.Region == "France") & (TS.data$Province.State == ""),]

# sweep values of R0 based on range of dates to consider for the model
ranges <- 15:25
deltaT <- 35
params_sweep <- sweep.SIR.models(data=FR.data,geo.loc="France", t0_range=ranges, deltaT=deltaT)

# the parameters --beta,gamma,R0-- are returned in a "matrix" "array" object
print(params_sweep)
#       [,1]      [,2]      [,3]      [,4]      [,5]      [,6]      [,7]    
# beta  0.5231031 0.5250777 0.5323438 0.5217565 0.5355503 0.5473388 0.559132
# gamma 0.4768969 0.4749223 0.4676562 0.4782435 0.4644497 0.4526611 0.440868
# R0    1.096889  1.105608  1.138323  1.090985  1.153086  1.209158  1.268253
#       [,8]      [,9]      [,10]     [,11]   
# beta  0.5668948 0.5753911 0.5835743 0.592407
# gamma 0.4331052 0.4246089 0.4164257 0.407593
# R0    1.308908  1.355108  1.401389  1.453428

# obtain the R0 values from the parameters
R0s <- unlist(params_sweep["R0",])
# nbr of infected cases
FR.infs<- preProcessingData(FR.data,"France")

# average per range
# define ranges
lst.ranges <- lapply(ranges, function(x) x:(x+deltaT))

# compute averages
avg.FR.infs <- lapply(lst.ranges, function(x) mean(FR.infs[x]))

# plots
plot(R0s, type='b')
# plot vs average number of infected cases
plot(avg.FR.infs, R0s, type='b')

\end{lstlisting}

\subsection{Working with your own data}
\label{sec:syntheticData}
As mentioned before, the functions from the \covidPckg package also allow
users to work with their own data, when the data is formated in the
\textit{Time Series} strucutre as discussed in Sec.\ref{sec:dataStructure}.
This opens a large range of possibilities for users to import their own
data into R and use the functions already defined in the \covidPckg package.
A concrete example of how the data has to be formatted is shown in
Lst.~\ref{lst:ex_syntheticData}.
The example shows how to structure the data in a TS format from ``synthetic''
data generated from randomly sampling different distributions.
However this could be actual data from other places or locations
not accesible from the datasets provided by the package, or some
researchers may have access to their own private sets of data too.
The example also shows two cases, where the data can include the "status"
column or not, and whether it could be more than one location.
As a matter of fact, we left the "Long" and "Lat" fields empty but if one
includes the actual coordinates, the maping function \code{live.map}
can also be used with these structured data.

\Rlisting
\begin{lstlisting}[caption={Example of structuring data in a TS format, so that it
		can be used with any of the TS functions from the \covidPckg package.}, 
                captionpos=b,
                label={lst:ex_syntheticData}]
# TS data structure:
#       "Province.State" "Country.Region" "Lat" "Long"  dates . . .

# First let's create a 'fake' location
fake.locn <- c(NA,NA,NA,NA)

# names for these columns
names(fake.locn) <- c("Province.State","Country.Region","Lat","Long")

# let's set the dates
dates.vec <- seq(as.Date("2020/1/1"), as.Date("2020/4/09"), "days")

# data.vecX would be the actual values/cases
data.vec1 <- rpois(length(dates.vec),lambda=25)
# can also add more cases
data.vec2 <- abs(rnorm(length(dates.vec),mean=135,sd=15))
data.vec3 <- abs(rnorm(length(dates.vec),mean=35,sd=5))

# this will names the columns as your dates
names(data.vec1) <- dates.vec
names(data.vec2) <- dates.vec
names(data.vec3) <- dates.vec

# merge them into a data frame with multiple entries
synthetic.data <- as.data.frame(rbind(
			rbind(c(fake.locn,data.vec1)),
			rbind(c(fake.locn,data.vec2)),
			rbind(c(fake.locn,data.vec3))
			))

# finally set you locn to somethign unqiue, so you can use it in the generate.SIR.model fn
synthetic.data$Country.Region <- "myLocn"

# one could even add "status"
synthetic.data$status <- c("confirmed","death","recovered")

# OR just one case per locn
synthetic.data2 <- synthetic.data[,-ncol(synthetic.data)]
synthetic.data2$Country.Region <- c("myLocn","myLocn2","myLocn3")

# now we can use this 'synthetic' dataset with any of the TS functions
# data checks
integrity.check(synthetic.data)
consistency.check(synthetic.data)
data.checks(synthetic.data)

# quantitative indicators
tots.per.location(synthetic.data)
growth.rate(synthetic.data)
single.trend(synthetic.data2[3,])
mtrends(synthetic.data)

# SIR models
synthSIR <- generate.SIR.model(synthetic.data2,geo.loc="myLocn")
plt.SIR.model(synthSIR, interactiveFig=TRUE)
sweep.SIR.models(synthetic.data2,geo.loc="MyLocn")
\end{lstlisting}


\subsection{Studying Pandemics Trends}
\label{sec:ex-pandemics}

\begin{figure}
	\centering
	\includegraphics[width=.7\textwidth]{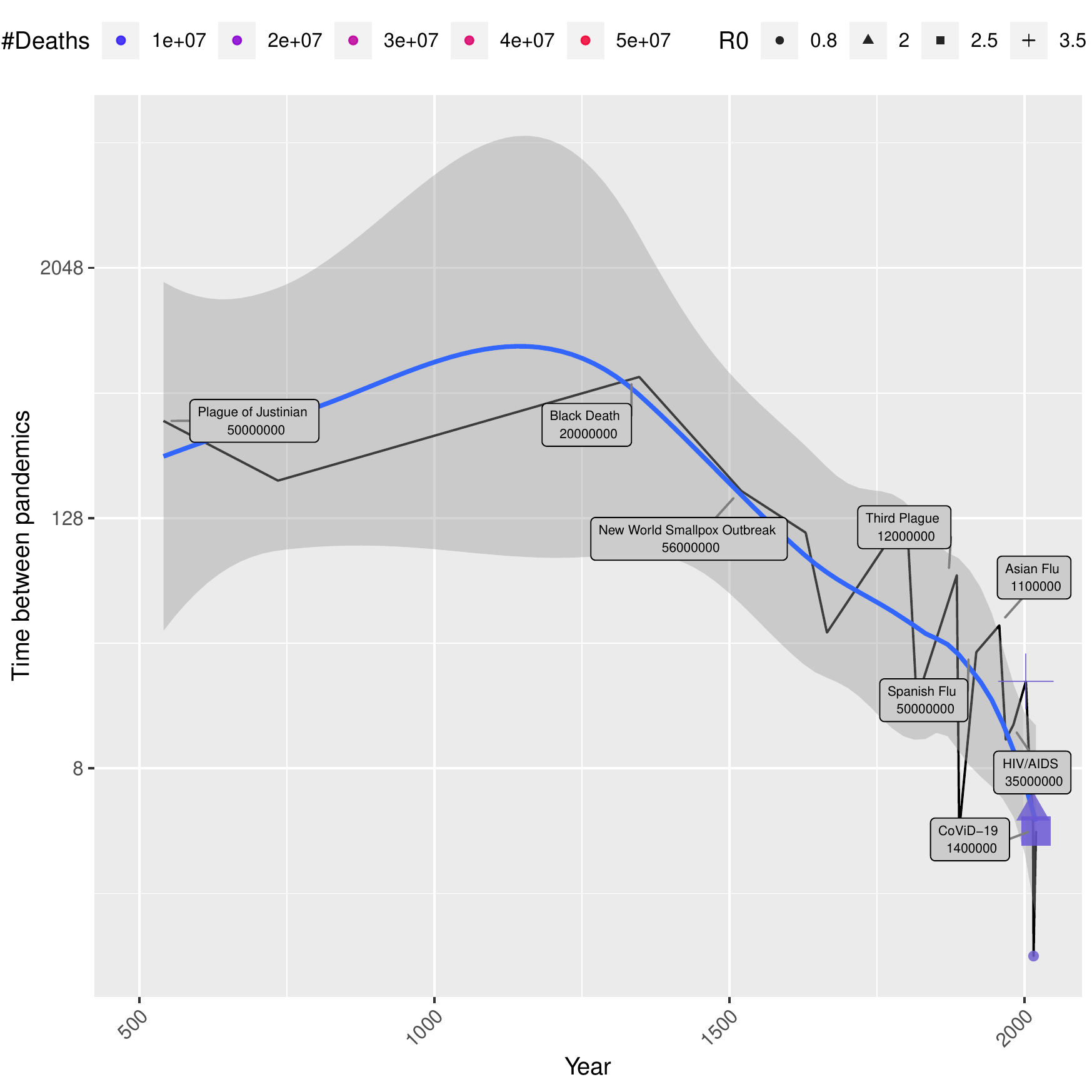}
	\includegraphics[width=.7\textwidth]{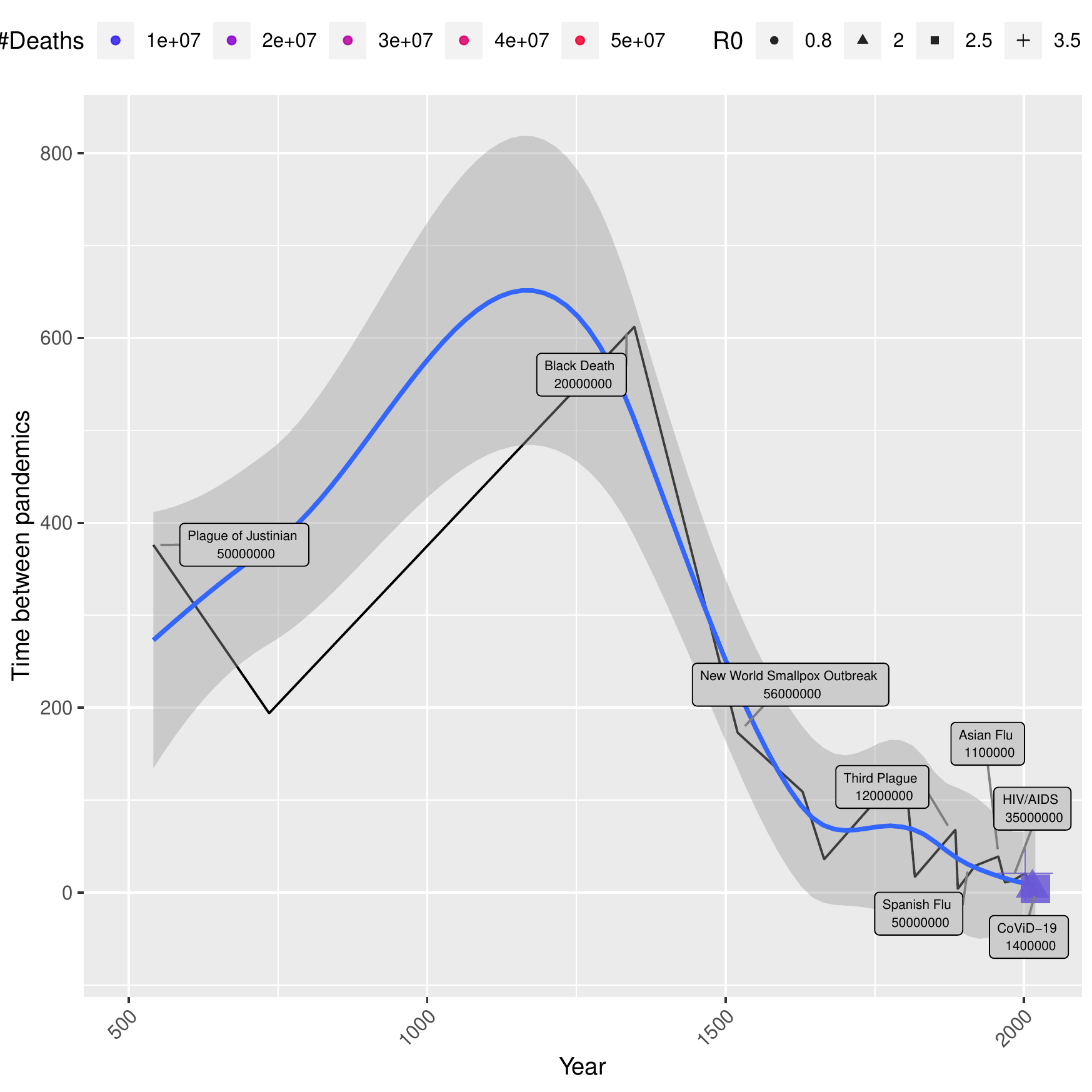}

	\caption{Historical Pandemics trend, as generate by the example in Lst.\ref{lst:ex_pandemics}.
		The vertical axis shows the time interval between consecutive pandemics in human history as a function of the reported year of ocurrence.
		Interestingly, this appears to indicate a clear trend to having more frequent pandemics in contemporary times.}
	\label{fig:pandemics_trend}
\end{figure}

The \covidPckg provides access to previous historical pandemic records.
Lst.~\ref{lst:ex_pandemics} and Fig.\ref{fig:pandemics_trend}, show an example
of how this data can be used to gain insights into past pandemics ocurrences.

\Rlisting
\begin{lstlisting}[caption={Analysis of historical pandemic records using the \covidPckg package.},
		captionpos=b,
		label={lst:ex_pandemics}]
# obtain Pandemic historical records
pnds <- pandemics.data(tgt="pandemics")

# obtain Pandemics vaccine development times
pnds.vacs <- pandemics.data(tgt="pandemics_vaccines")

###################

# pandemic's dates
date.pnd <- as.numeric(substr(pnds$Time.period,1,4))

# time between pandemics
pnd.interval <- y <- (diff(date.pnd))


z <- date.pnd[2:length(date.pnd)]

# remove possible NAs
filter <- !is.na(pnd.interval) & !is.na(z)

w <- pnds$Death.toll[filter]	#!is.na(pnd.interval) & !is.na(z)]
#w <- as.integer((w/abs(max(w)-min(w)))*1)

# generate dataframe with date, delta-time, death toll, pandemic's name and R0
pnd.df <- cbind(as.data.frame(na.omit(cbind(Year=date.pnd,delta=pnd.interval,z=z,Death.toll=w))) ,
		Name=pnds$Name.of.Pandemic[filter], R0=pnds$R0[filter])

# R0s
Rnaught <- as.factor(pnd.df$R0)

#################

# sort by year
pnd.df <- pnds[order(pnds$Year),] 
pnd.df <- cbind(pnd.df,delta=c(NA,diff(pnd.df$Year)))

# get the latest number of deaths for covid19
x <- covid19.analytics::covid19.data()
pnd.df$Death.toll[pnd.df$Name=="COVID-19"] <- sum(x$Deaths)


##################

# Visualizations


## using ggplot
library(ggplot2)
library(ggrepel)

# limits for x-axis
min.yr <- 500
max.yr <- 2021

# basic plot
plt <- ggplot(pnd.df, aes(x=Year,y=delta)) + geom_line() + geom_smooth() + xlim(min.yr,max.yr)
ggsave("pandemics.pdf",plt)

# tilt text 45 degrees
plt0 <-	plt + scale_x_continuous(breaks = pnd.df$Year) +
		theme(axis.text.x = element_text(angle = 45, hjust = 1)) +
		xlim(min.yr,max.yr)
ggsave("pandemics0.pdf",plt0)

# Add details about R0 and deaths.toll to plot
plt1 <-	plt0 + geom_point(size=as.integer(2.5*pnd.df$R0), aes(colour=(pnd.df$Death.toll),shape=as.factor(pnd.df$R0)), alpha=0.85) +
	#geom_text(aes(label=Name),hjust=0, vjust=0) +
	labs(y="Time between pandemics") +
	labs(shape="R0", colour="R0s") +
	geom_label_repel(aes(label = paste(Name.of.Pandemic,'\n',Death.toll)),
		data = subset(pnd.df, Death.toll > 1.0e+6),
		size = 2,
		box.padding   = 0.35, fill="grey80", #alpha=.85, 
		point.padding = 0.5,
		segment.color = 'grey50') +
	scale_color_gradient2(name="#Deaths", low = "green", mid = "blue",high = "red",
		midpoint = (mean(pnd.df$Death.toll,na.rm=T)) , guide = guide_legend(direction = "horizontal") ) +
	theme(legend.position = "top") +
	#+ scale_x_continuous(trans='log2')
	xlim(min.yr,max.yr)

# log scales
plt.log <- plt1 + scale_y_continuous(trans='log2') #+ xlim(min.yr,max.yr)
plt.loglog <- plt.log + scale_x_continuous(trans='log2') #+ xlim(min.yr,max.yr)

# save plots
ggsave('pandemics_semilog.pdf',plt.log)
ggsave('pandemics_loglog.pdf',plt.loglog)
ggsave("pandemics1.pdf",plt1)

# display on screen
print(plt0)
dev.new()
print(plt1)
dev.new()
print(plt.log)
#dev.new()
#print(plt.loglog)


#####

## using R basics plottings capabilities
pnds.df <- data.frame(na.omit(cbind(y,z)))
print(pnds.df)
p5 <- poly(pnds.df$z,degree=6)
lm.p5 <- lm(y ~ p5, data=pnds.df)
yp5 <- predict(lm.p5, as.data.frame(seq(min(pnds.df$z),max(pnds.df$z),100)))

dev.new()
plot(z,y)
lines(pnds.df$z,yp5, lwd=3)
\end{lstlisting}


\subsection{Testing and Vaccination}
\label{sec:ex-vaccination-testing}

\begin{figure}
	\includegraphics[width=0.495\textwidth]{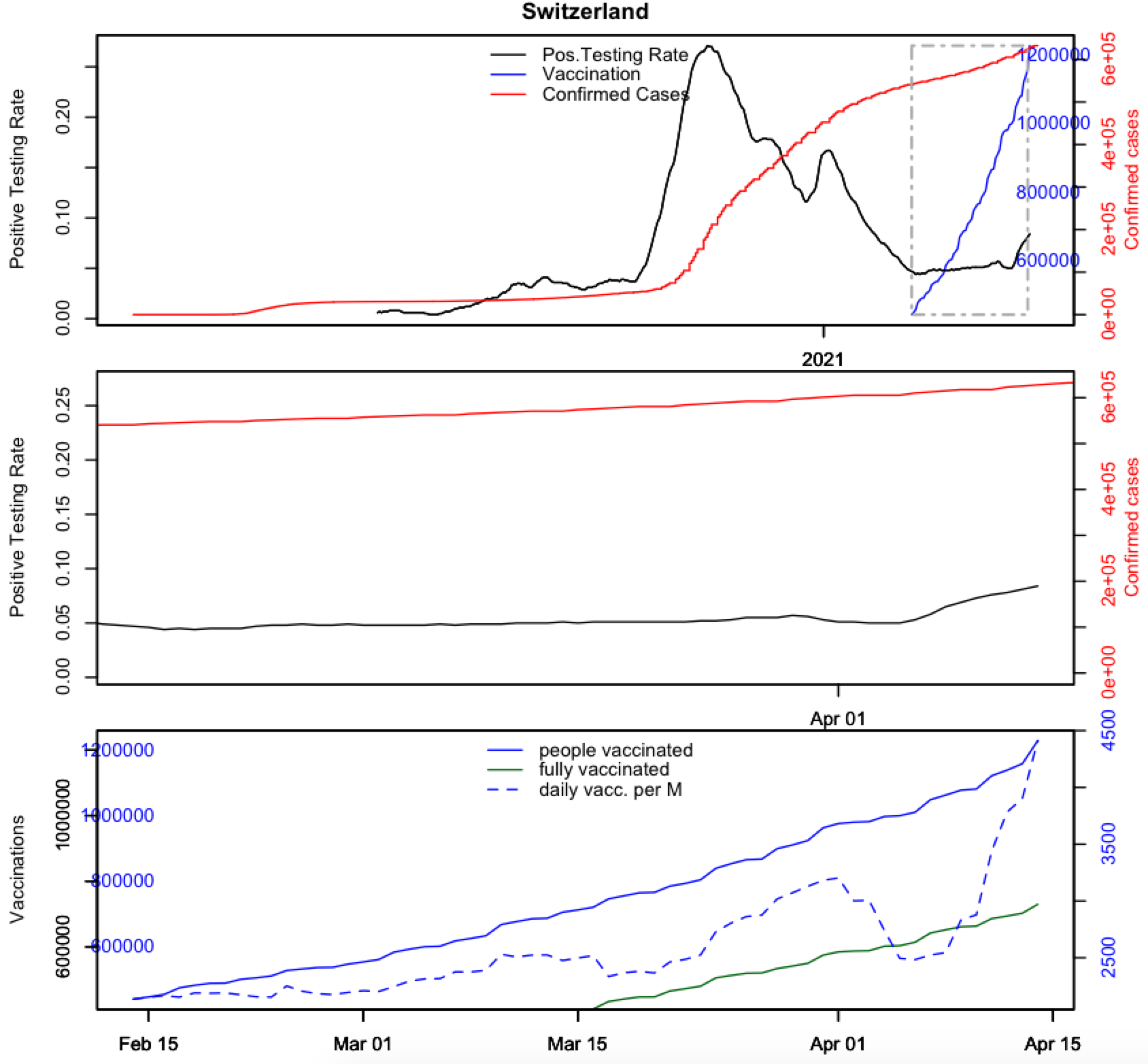}
	\includegraphics[width=0.495\textwidth]{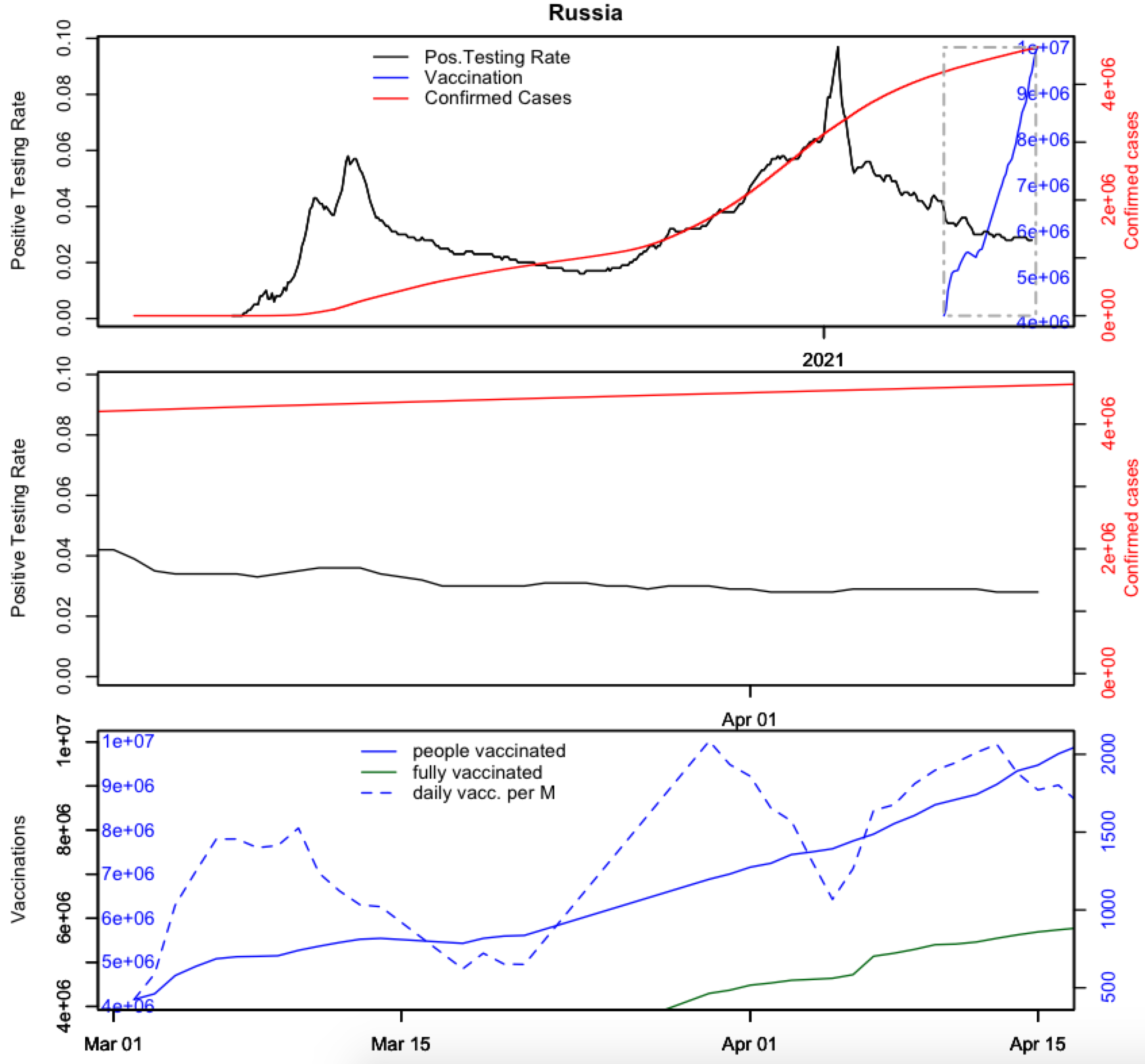}

	\caption{Plots generated using the function shown in Lst.\ref{lst:ex_testing-vaccn}
		for the cases of Switzerland (left) and Russia (right).
		The consecutive panels in each column display:
			i) the ``positive testing rate'', vaccinations and confirmed number of cases;
			ii) a zoom-in region demarked in the previous plot, into the period since the vaccination started;
			iii) vaccination records for total number the people vaccinated, fully vaccinated individuals and daily vaccinations per millon.}
	\label{fig:testing-vaccn}
\end{figure}

Using the testing and vaccination data can get insights into the way in which
the pandemic is spreading.  For instance, it could be used as indicator of how
efficient the vaccination campaing are being against the virus spread.
Lst.\ref{lst:ex_testing-vaccn} shows the example of a function that visualize the
positive testing rate and vaccinations for a given country.
Examples of the application of this function are shown in Fig.~\ref{fig:testing-vaccn}.
Some interesting observations can be done by considering the information presented in these plots,
for instance, how effective are being the vaccinations, are vacciantions changing the rate of 
contagion, are vaccinations curbing the pandemics, what is the effect of
single vaccination vs fully vaccinated individuals, etc.

\Rlisting
\begin{lstlisting}[caption={Analysis of vaccination and testing data using the \covidPckg package.},
                captionpos=b,
                label={lst:ex_testing-vaccn}]
testVaccCtry <- function(Ctry="Canada"){
   # fn that process testing/vaccination/and/confirmed cases for a given Ctry
   # and generates graphical representations to aid their comparison

   # read DATA
   ## reading COVID19 testing data
   c19.testing.data <- covid19.testing.data()

   ## reading COVID19 vaccination data
   c19.vacc.data <- covid19.vaccination()

   ## reading COVID19 cases
   c19.cases.data <- covid19.data("ts-confirmed")


   # data processing
   ## testing data
   ### select cases for a particular country, given by Ctry
   filter.ctry <- grepl(Ctry,c19.testing.data[,"Entity"])

   ### select columns: date and (short term) positive rate
   cols <- c("Date","Short.term.positive.rate")
   tstDta <- na.omit(c19.testing.data[filter.ctry, cols])
   names(tstDta) <- cols

   ### sort by date
   tstDta <- tstDta[order(tstDta[,1]),]

   ## vaccination data
   ### remove NAs
   vaccs <- na.omit(c19.vacc.data)
   ### select specific "Ctry"
   vacc.Ctry <- vaccs[vaccs$location==Ctry,]

   ## confirmed cases
   conf.Ctry <- c19.cases.data[c19.cases.data$Country.Region==Ctry,5:(length(c19.cases.data)-1)]


   ##### Graphics #####


   # plot daily vaccination per million, for every location
   #par(mfrow=c(5,5))
   #tapply(vaccs$daily_vaccinations_per_million,vaccs$location, plot)
   #par(mfrow=c(1,1))

   ###

   # mosaic plot combining testing/vaccination and confirmed cases data

   par(mfrow=c(3,1))
   par(mar=c(1,5,2,5))

   ### subplot #1
   minX <- as.Date(names(conf.Ctry)[1])
   maxX <- as.Date(names(conf.Ctry)[length(conf.Ctry)])
   # plot positive testing rate vs date
   plot(as.Date(tstDta$Date), tstDta[,2], 'l', ylab="Positive Testing Rate", xlim=c(minX,maxX))
   title(Ctry)
   par(new=TRUE)
   # add vaccination data 
   plot(as.Date(vacc.Ctry$date),(vacc.Ctry$people_vaccinated),
   	type='l', col='blue', xlab=NA, xaxt='n', ylab=NA, yaxt='n', xlim=c(minX,maxX))
   axis(4,col.axis='blue', line=-3.5, las=1, lwd=0)
   #axis.Date(3, as.Date(vacc.Ctry$date), col.axis='blue')
   par(new=TRUE)
   plot(as.Date(tstDta$Date), tstDta[,2], 'l', ylab=NA,yaxt='n', xlim=c(minX,maxX))
   #axis.Date(3, as.Date(vacc.Ctry$date), col.axis='red')
   par(new=TRUE)
   plot(as.Date(names(conf.Ctry)), as.numeric(conf.Ctry), type='s', col='red', ylab=NA,yaxt='n')
   #axis.Date(3, as.Date(vacc.Ctry$date), col.axis='red')
   #axis.Date(1,as.Date(names(conf.Ctry)))	#, at = seq(as.Date(names(conf.Ctry[1])),as.Date(names(conf.Ctry[length(conf.Ctry)]))) )
   axis(4,col.axis='red')
   mtext("Confirmed cases", 4, line=2, cex=.65, col='red')
   legend("top",c("Pos.Testing Rate","Vaccination","Confirmed Cases"),col=c('black','blue','red'),lty=c(1,1,1), bty='n')
   rect(as.Date(vacc.Ctry$date)[1],1, as.Date(vacc.Ctry$date)[length(vacc.Ctry$date)-1], as.numeric(conf.Ctry)[length(conf.Ctry)-1],
	border='darkgray', lty=4, lwd=1.5)

   par(mar=c(1,5,1,5))

   ### subplot #2
   # adjust limits to match testing/vaccination ranges...
   minX <- max(as.Date(tstDta$Date)[1],as.Date(vacc.Ctry$date)[1])
   maxX <- min(as.Date(vacc.Ctry$date)[length(vacc.Ctry$date)], as.Date(tstDta$Date)[length(tstDta$Date)])

   plot(as.Date(tstDta$Date), tstDta[,2], 'l', xlim=c(minX,maxX), xlab=NA,ylab="Positive Testing Rate")
   par(new=TRUE)
   plot(as.Date(names(conf.Ctry)),(as.numeric(conf.Ctry)), xlim=c(minX,maxX), type='l', col='red', xlab=NA,ylab=NA,yaxt='n')
   axis(4, col.axis="red")
   mtext("Confirmed cases", 4, line = 2, cex=.65, col='red')

   par(mar=c(2.5,5,1,5))

   ### subplot 3
   minY=min(vacc.Ctry$people_vaccinated)
   maxY=max(vacc.Ctry$people_vaccinated)
   plot(as.Date(vacc.Ctry$date),(vacc.Ctry$people_vaccinated),
   	xlim=c(minX,maxX), ylim=c(minY,maxY),
   	type='l', col='blue', ylab="Vaccinations")
   axis(2, col.axis='blue', line=-3.5, las=1, lwd=0)

   par(new=TRUE)
   plot(as.Date(vacc.Ctry$date),vacc.Ctry$people_fully_vaccinated,
   		xlim=c(minX,maxX), ylim=c(minY,maxY),
		type='l', col='darkgreen',
   		yaxt = "n", ylab = NA)
   axis(2,col.axis='black')

   par(new=TRUE)
   plot(as.Date(vacc.Ctry$date),vacc.Ctry$daily_vaccinations_per_million,
	xlim=c(minX,maxX),
	type='l', col='blue', lty=2,
   	yaxt = "n", ylab = NA)
   axis(4, col.axis='blue')

   legend("top",c("people vaccinated","fully vaccinated","daily vacc. per M"),col=c('blue','darkgreen','blue'),lty=c(1,1,2), bty='n')

   par(new=FALSE)
   par(mfrow=c(1,1))

}

##########

# Apply the fn to a set of countries...
lapply(c("Argentina","Uruguay","Italy","Spain","Switzerland"), testVaccCtry)
\end{lstlisting}

\subsection{Genomics}
\label{sec:ex-genomics}

The \covidPckg package provides access to genomics data available at the 
NCBI databases \cite{NCBI,NCBIdatabases}.
The \texttt{covid19.genomic.data} is the master function for accesing the
different variations of the genomics information available as shown in
Table~\ref{table:genomicTypes}.
In addition to that there are a few helper functions which can also be used for
retrieving specific types, such as:
	\texttt{c19.refGenome.data, c19.ptree.data, c19.fasta.data, c19.NPs.data, c19.NP\_fasta.data}.
Lst.~\ref{lst:ex_genomicsData} show examples of how to use these functions
for retrieving the different type of genomics data for the SARS-CoV-2 virus.

\Rlisting
\begin{lstlisting}[caption={Example of how to retrieve the different types of genomics
			datasets provided by the \covidPckg package.},
                captionpos=b,
                label={lst:ex_genomicsData}]
# test all sources...
srcs <- c('livedata','repo','local')

results <- list()

for (sr in srcs) {
	print(paste("##### ",sr," #####"))

	# obtain reference genome data
	results[['refGenome']][sr] <- c19.refGenome.data(src=sr)

	# obtain phylogenetic tree
	results[['tree']][sr] <- c19.ptree.data(src=sr)

	# obtain FASTA sequences
	results[['fasta']][sr] <- c19.fasta.data(src=sr)

	# combine several types
	if (sr != 'livedata') {
		results[['genomics']][sr] <- c19.genomic.data(src=sr)
	}

	# obtain nucleotides data
	results[['nucs']][sr] <- c19.NPs.data(src=sr, DB='nucleotide')
	if (sr != 'livedata') {
		results[['ptns']][sr] <- c19.NPs.data(src=sr, DB='protein')
	}

	# obtain FASTA sequences for nucleotides and proteins
	results[['FASTAs']][['nucs']][sr] <- c19.NP_fasta.data(target='nucleotide')
	results[['FASTAs']][['ptns']][sr] <- c19.NP_fasta.data(target='protein')
}



#####

gtypes <- c("genome","fasta","tree",
		"nucleotide","protein",
		"nucleotide-fasta","protein-fasta",
		"genomic")

results2 <- list()

for (gt in gtypes) {
	print(gt)

	results2[[gt]] <- covid19.genomic.data(type=gt)
}
\end{lstlisting}

Each of these functions return different objects, Lst.~\ref{lst:ex_genTypes}
shows an example of the different structures for some of the objects.
The most involved object is obtained from the \texttt{covid19.genomic.data}
when combining different types of datasets.

\outputlisting
\begin{lstlisting}[caption={Objects composition for the example presented in Lst.~\ref{lst:ex_genomicsData}
			},
		breaklines,
                captionpos=b,
                label={lst:ex_genTypes}]
# str(results)
List of 7
 $ refGenome:List of 3
  ..$ livedata: chr [1:29903] "a" "t" "t" "a" ...
  ..$ repo    : chr [1:29903] "a" "t" "t" "a" ...
  ..$ local   : chr [1:29903] "a" "t" "t" "a" ...
 $ tree     :List of 3
  ..$ livedata: int [1:13319, 1:2] 8113 8114 8115 8116 8117 8118 8119 8120 8121 8122 ...
  ..$ repo    : int [1:13319, 1:2] 8113 8114 8115 8116 8117 8118 8119 8120 8121 8122 ...
  ..$ local   : int [1:13319, 1:2] 8113 8114 8115 8116 8117 8118 8119 8120 8121 8122 ...
 $ fasta    :List of 3
  ..$ livedata: raw [1:29903] 88 18 18 88 ...
  ..$ repo    : raw [1:29903] 88 18 18 88 ...
  ..$ local   : raw [1:29903] 88 18 18 88 ...
 $ nucs     :List of 3
  ..$ livedata: chr "MT821795"
  ..$ repo    : chr [1:11142] "NC_045512" "MT772210" "MT772215" "MT772237" ...
  ..$ local   : chr [1:11142] "NC_045512" "MT772210" "MT772215" "MT772237" ...
 $ FASTAs   :List of 2
  ..$ nucs:List of 3
  .. ..$ livedata: raw [1:29903] 88 18 18 88 ...
  .. ..$ repo    : raw [1:29903] 88 18 18 88 ...
  .. ..$ local   : raw [1:29903] 88 18 18 88 ...
  ..$ ptns:List of 3
  .. ..$ livedata: raw [1:118] a0 60 e0 48 ...
  .. ..$ repo    : raw [1:118] a0 60 e0 48 ...
  .. ..$ local   : raw [1:118] a0 60 e0 48 ...
 $ genomics :List of 2
  ..$ repo :List of 1
  .. ..$ NC_045512.2: chr [1:29903] "a" "t" "t" "a" ...
  .. ..- attr(*, "species")= chr "Severe_acute_respiratory_syndrome_coronavirus_2"
  ..$ local:List of 1
  .. ..$ NC_045512.2: chr [1:29903] "a" "t" "t" "a" ...
  .. ..- attr(*, "species")= chr "Severe_acute_respiratory_syndrome_coronavirus_2"
 $ ptns     :List of 2
  ..$ repo : chr [1:117619] "YP_009742608" "YP_009742609" "YP_009742610" "YP_009742611" ...
  ..$ local: chr [1:117619] "YP_009742608" "YP_009742609" "YP_009742610" "YP_009742611" ...

# str(results['genomics'])
List of 7
 $ genome     :List of 1
  ..$ NC_045512.2: chr [1:29903] "a" "t" "t" "a" ...
  ..- attr(*, "species")= chr "Severe_acute_respiratory_syndrome_coronavirus_2"
 $ fasta      :List of 1
  ..$ gi|1798174254|ref|NC_045512.2| Severe acute respiratory syndrome coronavirus 2 isolate Wuhan-Hu-1, complete genome: raw [1:29903] 88 18 18 88 ...
  ..- attr(*, "class")= chr "DNAbin"
 $ annotation :'data.frame':	59 obs. of  9 variables:
  ..$ seqid     : Factor w/ 1 level "NC_045512.2": 1 1 1 1 1 1 1 1 1 1 ...
  ..$ source    : Factor w/ 1 level "RefSeq": 1 1 1 1 1 1 1 1 1 1 ...
  ..$ type      : Factor w/ 7 levels "CDS","five_prime_UTR",..: 5 2 3 1 1 4 4 4 4 4 ...
  ..$ start     : int [1:59] 1 1 266 266 13468 266 806 2720 8555 10055 ...
  ..$ end       : int [1:59] 29903 265 21555 13468 21555 805 2719 8554 10054 10972 ...
  ..$ score     : num [1:59] NA NA NA NA NA NA NA NA NA NA ...
  ..$ strand    : Factor w/ 1 level "+": 1 1 1 1 1 1 1 1 1 1 ...
  ..$ phase     : Factor w/ 1 level "0": NA NA NA 1 1 NA NA NA NA NA ...
  ..$ attributes: chr [1:59] ...
 $ nucleotides:'data.frame':	11142 obs. of  20 variables:
  ..$ Accession       : chr [1:11142] "NC_045512" "MT772210" "MT772215" "MT772237" ...
  ..$ SRA_Accession   : chr [1:11142] "" "SRR12181112" "SRR12181117" "SRR12181117" ...
  ..$ Release_Date    : chr [1:11142] "2020-01-13T00:00:00Z" "2020-07-17T00:00:00Z" "2020-07-17T00:00:00Z" "2020-07-17T00:00:00Z" ...
  ..$ Species         : chr [1:11142] "Severe acute respiratory syndrome-related coronavirus" "Severe acute respiratory syndrome-related coronavirus" "Severe acute respiratory syndrome-related coronavirus" "Severe acute respiratory syndrome-related coronavirus" ...
  ..$ Genus           : chr [1:11142] "Betacoronavirus" "Betacoronavirus" "Betacoronavirus" "Betacoronavirus" ...
  ..$ Family          : chr [1:11142] "Coronaviridae" "Coronaviridae" "Coronaviridae" "Coronaviridae" ...
  ..$ Length          : int [1:11142] 29903 29800 29800 29800 29800 29800 29791 29800 29800 29800 ...
  ..$ Sequence_Type   : chr [1:11142] "RefSeq" "GenBank" "GenBank" "GenBank" ...
  ..$ Nuc_Completeness: chr [1:11142] "complete" "complete" "complete" "complete" ...
  ..$ Genotype        : logi [1:11142] NA NA NA NA NA NA ...
  ..$ Segment         : logi [1:11142] NA NA NA NA NA NA ...
  ..$ Authors         : chr [1:11142] ...
  ..$ Publications    : chr [1:11142] "15680415, 15630477, 10482585" "" "" "" ...
  ..$ Geo_Location    : chr [1:11142] "China" "India: Himatnagar" "India: Choryasi" "India: Choryasi" ...
  ..$ U.S._State      : chr [1:11142] "" "" "" "" ...
  ..$ Host            : chr [1:11142] "Homo sapiens" "Homo sapiens" "Homo sapiens" "Homo sapiens" ...
  ..$ Isolation_Source: chr [1:11142] "" "" "" "" ...
  ..$ Collection_Date : chr [1:11142] "2019-12" "2020-06-14" "2020-06-13" "2020-06-13" ...
  ..$ BioSample       : chr [1:11142] "" "SAMN15488404" "SAMN15488426" "SAMN15488426" ...
  ..$ GenBank_Title   : chr [1:11142] "Severe acute respiratory syndrome coronavirus 2 isolate Wuhan-Hu-1, complete genome" "Severe acute respiratory syndrome coronavirus 2 isolate SARS-CoV-2/human/IND/GBRC242/2020, complete genome" "Severe acute respiratory syndrome coronavirus 2 isolate SARS-CoV-2/human/IND/GBRC264a/2020, complete genome" "Severe acute respiratory syndrome coronavirus 2 isolate SARS-CoV-2/human/IND/GBRC264b/2020, complete genome" ...
 $ proteins   :'data.frame':	117619 obs. of  20 variables:
  ..$ Accession       : chr [1:117619] "YP_009742608" "YP_009742609" "YP_009742610" "YP_009742611" ...
  ..$ SRA_Accession   : chr [1:117619] "" "" "" "" ...
  ..$ Release_Date    : chr [1:117619] "2020-03-30T00:00:00Z" "2020-03-30T00:00:00Z" "2020-03-30T00:00:00Z" "2020-03-30T00:00:00Z" ...
  ..$ Species         : chr [1:117619] "Severe acute respiratory syndrome-related coronavirus" "Severe acute respiratory syndrome-related coronavirus" "Severe acute respiratory syndrome-related coronavirus" "Severe acute respiratory syndrome-related coronavirus" ...
  ..$ Genus           : chr [1:117619] "Betacoronavirus" "Betacoronavirus" "Betacoronavirus" "Betacoronavirus" ...
  ..$ Family          : chr [1:117619] "Coronaviridae" "Coronaviridae" "Coronaviridae" "Coronaviridae" ...
  ..$ Length          : int [1:117619] 180 638 1945 500 306 290 83 198 113 139 ...
  ..$ Sequence_Type   : chr [1:117619] "RefSeq" "RefSeq" "RefSeq" "RefSeq" ...
  ..$ Genotype        : logi [1:117619] NA NA NA NA NA NA ...
  ..$ Segment         : logi [1:117619] NA NA NA NA NA NA ...
  ..$ Authors         : chr [1:117619] "Baranov,P.V., Henderson,C.M., Anderson,C.B., Gesteland,R.F., Atkins,J.F., Howard,M.T., Robertson,M.P., Igel,H.,"| __truncated__ "Baranov,P.V., Henderson,C.M., Anderson,C.B., Gesteland,R.F., Atkins,J.F., Howard,M.T., Robertson,M.P., Igel,H.,"| __truncated__ "Baranov,P.V., Henderson,C.M., Anderson,C.B., Gesteland,R.F., Atkins,J.F., Howard,M.T., Robertson,M.P., Igel,H.,"| __truncated__ "Baranov,P.V., Henderson,C.M., Anderson,C.B., Gesteland,R.F., Atkins,J.F., Howard,M.T., Robertson,M.P., Igel,H.,"| __truncated__ ...
  ..$ Publications    : chr [1:117619] "15680415, 15630477, 10482585" "15680415, 15630477, 10482585" "15680415, 15630477, 10482585" "15680415, 15630477, 10482585" ...
  ..$ Protein         : chr [1:117619] "leader protein" "nsp2" "nsp3" "nsp4" ...
  ..$ Geo_Location    : chr [1:117619] "China" "China" "China" "China" ...
  ..$ U.S._State      : chr [1:117619] "" "" "" "" ...
  ..$ Host            : chr [1:117619] "Homo sapiens" "Homo sapiens" "Homo sapiens" "Homo sapiens" ...
  ..$ Isolation_Source: chr [1:117619] "" "" "" "" ...
  ..$ Collection_Date : chr [1:117619] "2019-12" "2019-12" "2019-12" "2019-12" ...
  ..$ BioSample       : chr [1:117619] "" "" "" "" ...
  ..$ GenBank_Title   : chr [1:117619] "leader protein [Severe acute respiratory syndrome coronavirus 2]" "nsp2 [Severe acute respiratory syndrome coronavirus 2]" "nsp3 [Severe acute respiratory syndrome coronavirus 2]" "nsp4 [Severe acute respiratory syndrome coronavirus 2]" ...
 $ SRA        :List of 2
  ..$ sra_info: chr [1:6] "This download (via FTP) provides Coronaviridae family-containing SRA runs detected with NCBI's kmer analysis (STAT) tool. " "It provides corresponding SRA run (SRR), sample (SRS), and submission (SRA) accessions, as well as BioSample an"| __truncated__ "The STAT kmer analysis was performed via a two-step process with a 32-mer coarse database and a 64-mer fine database. " "The database is generated from RefSeq genomes and the viral genome set from nt using a minhash-based approach. " ...
  ..$ sra_runs:'data.frame':	26776 obs. of  5 variables:
  .. ..$ acc       : chr [1:26776] "ERR1857120" "ERR1857132" "ERR3955908" "ERR3955909" ...
  .. ..$ sample_acc: chr [1:26776] "ERS1572626" "ERS1572627" "ERS2783634" "ERS2783635" ...
  .. ..$ biosample : chr [1:26776] "SAMEA100883668" "SAMEA100884418" "SAMEA4965217" "SAMEA4965218" ...
  .. ..$ sra_study : chr [1:26776] "ERP021740" "ERP021740" "ERP111280" "ERP111280" ...
  .. ..$ bioproject: chr [1:26776] "" "" "" "" ...
 $ references :List of 2
  ..$ : chr "covid19.analytics -- local data"
  ..$ : chr "/Users/marcelo/Library/R/4.0/library/covid19.analytics/extdata/"

\end{lstlisting}

One aspect that should be mentioned with respect to the genomics data is that,
in general, these are large datasets which are continuously being updated
hence increasing theirs sizes even more.
These would ultimately present pragmatical challenges, such as, long processing
times or even starvation of memory resources.
We will not dive into major interesting examples, like DNA sequencing analysis
or building phylogenetics trees; but packages such as \CRANpkg{ape},
\CRANpkg{apegenet}, \CRANpkg{phylocanvas}, and others can be used for these and other analysis.

One simple example we can present is the creation of dynamical categorization
trees based on different elements of the sequencing data.
We will consider for instance the data for the nucleotides as reported from NCBI.

The example on Lst.~\ref{lst:ex_genomicsTrees} shows how to retrieve either
nucleotides (or proteins) data and generate categorization trees based on
different elements, such as, hosting organism, geographical location, sequences
length, etc.
In the examples we employed the \CRANpkg{collapsibleTree} package, that generates
interactive browsable trees through web browsers.

\Rlisting
\begin{lstlisting}[caption={Example of how to generate a dynamic browsable tree
			using some of information included in the nucleotides dataset.
			Some of these trees representations are shown in Fig.~\ref{fig:trees}.},
                captionpos=b,
                label={lst:ex_genomicsTrees}]
library(collapsibleTree)

# retrieve the nucleotides data
nucx <- covid19.genomic.data(type='nucleotide',src='repo')

# identify specific fields to look at
len.fld <- "Length"
acc.fld <- "Accession"
geoLoc.fld <- "Geo_Location"
seq.fld <- "Sequence_Type"
host.fld <- "Host"

seq.limit1 <- 25000

seq.limit1 <- 29600
seq.limit2 <- 31000

# selection criteria, nucleotides with seq.length between 29600 and 31000
selec.ctr.1 <- nucx$Length<seq.limit2 & nucx$Length>seq.limit1

# remove nucletoides without specifying a "host"
targets <- nucx[selec.ctr.1 & nucx$Host!='',]

# categorization tree based on Host-type, geographical location, ...
collapsibleTreeSummary(targets,
                hierarchy=c(host.fld,geoLoc.fld,acc.fld,c(len.fld,seq.fld)),
                attribute=len.fld,
                zoomable=FALSE, collapsed=TRUE)

#  categorization tree for nucleotides with seq. lengths smaller than 29600 or bigger than 31000
collapsibleTreeSummary(nucx[!selec.ctr.1,],
                hierarchy=c(host.fld,geoLoc.fld,acc.fld,c(len.fld,seq.fld)),
                attribute=len.fld,
                zoomable=FALSE, collapsed=TRUE)

#  categorization tree for "human hosted" nucleotides with seq. lengths smaller than 29600 or bigger than 31000
collapsibleTreeSummary(nucx[!selec.ctr.1 &  nucx$Host=='Homo sapiens',],
                hierarchy=c(host.fld,geoLoc.fld,acc.fld,c(len.fld,seq.fld)),
                attribute=len.fld,
                zoomable=FALSE, collapsed=TRUE)
\end{lstlisting}

\begin{figure}
	\includegraphics[width=.5\textwidth]{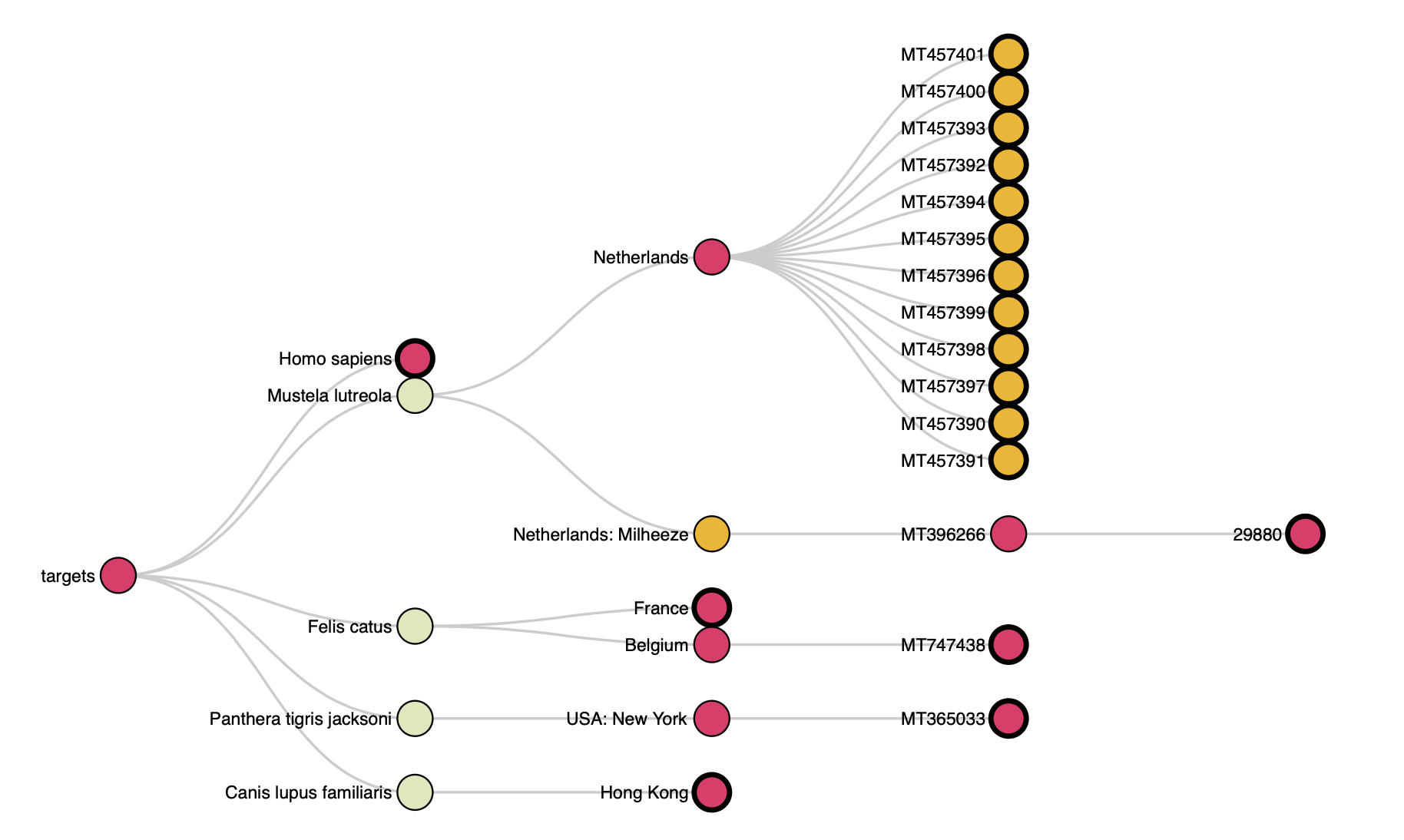}
	\includegraphics[width=.5\textwidth]{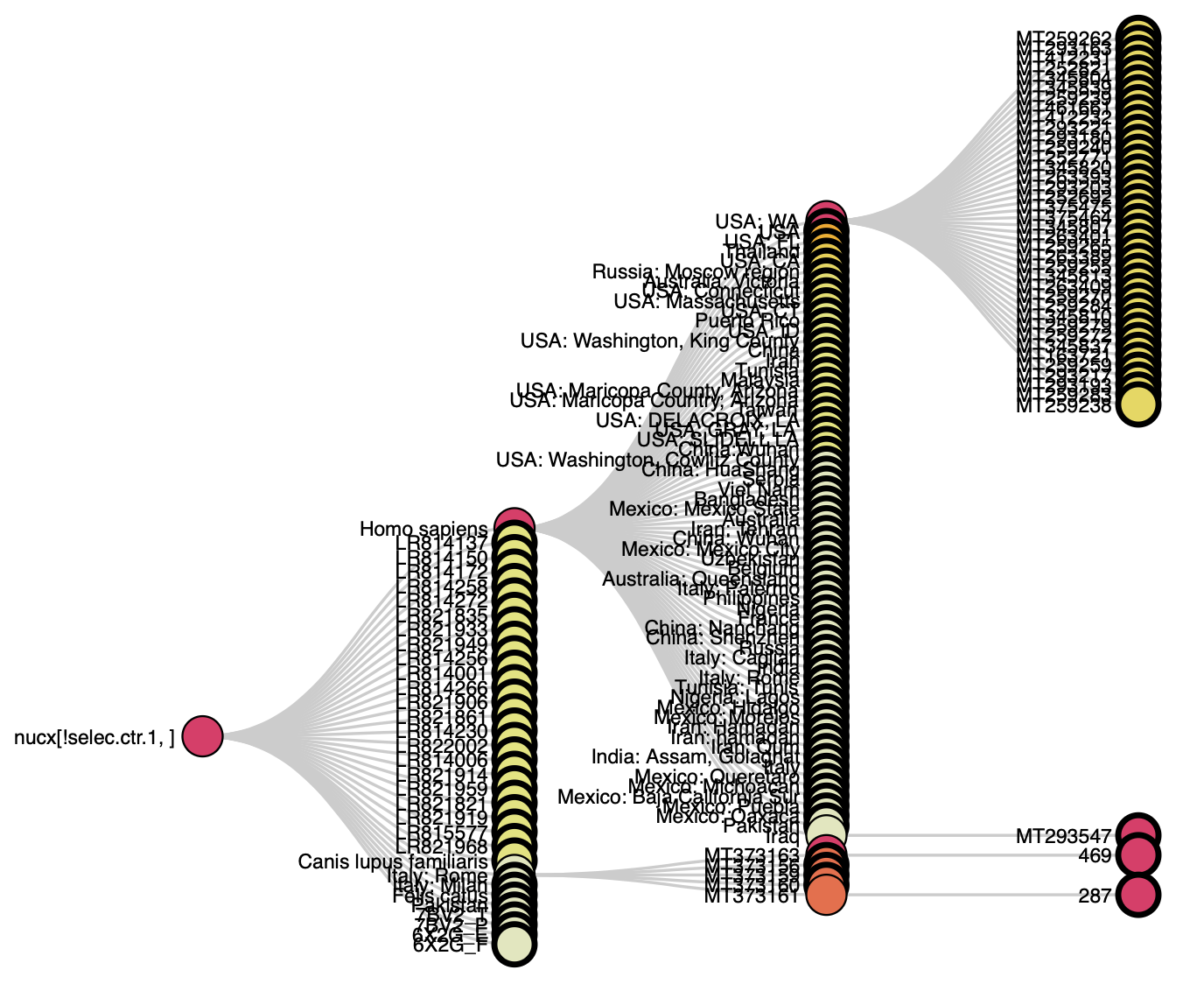}
	\caption{Browsable trees implemented using the nucleotides dataset in combination
		with the \CRANpkg{collapsibleTree} package, as presented in
		Lst.~\ref{lst:ex_genomicsTrees}.}
	\label{fig:trees}
\end{figure}


\section{Case Study: The \covidPckg Dashboard Explorer}
\label{sec:studyCase}


In this section we will present and discuss, how the \emph{\covidPckg Dashboard
Explorer} is implemented.
The main goal is to provide enough details about how the dashboard is implemented and works, 
so that users could modify it if/as they seem fit or even
develop their own.
For doing so, we will focus in three main points:

\begin{itemize}
	\item the \textit{front end} implementation, also know as the user
interface, mainly developed using the \CRANpkg{shiny} package

	\item the \textit{back end} implementation, mostly using the \covidPckg package

	\item the web \textit{server} installation and configuration where the dashboard is hosted
\end{itemize}

\subsection{Dashboard's Front End Implementation}
\label{sec:dashboard-frontend}
The \emph{covid19.analytics Dashboard Explorer} is built using the \CRANpkg{Shiny} package
\cite{shinyPckg} in combination with the \covidPckg package.
Shiny allows users to build interactive dashboards that can work through a web interface.
The dashboard mimics the \covidPckg package commands and features but enhances
the commands as it allows users to use dropdowns and other control widgets to
easily input the data rather than using a command terminal.
In addition the dashboard offers some unique features, such as
a \textit{Personal Protective Equipment} (PPE) model estimation, based on
realistic projections developed by the US \textit{Centers for Disease Control
and Prevention} (CDC).

The dashboard interface offers several features:
\begin{enumerate}
	\item The dashboard can be run on the cloud/web allowing for multiple
users to simultaneously analyze the data with no special software or hardware
requirements. The Shiny package makes the dashboard mobile and tablet
compatible as well.

	\item It aids researchers to share and discuss analytical findings.

	\item The dashboard can be run locally or through the web server.

	\item No programming or software expertise is required which reduces technical barriers to analyzing the data. Users can interact and analyze the data without any software expertise therefore users can focus on the modeling and analysis. In these times the dashboard can be a monumental tool as it removes barriers and allows a wider and diverse set of users to have quick access to the data.

	\item Interactivity. One feature of Shiny and other graphing packages, such as Plotly, is interactivity, i.e. the ability to interact with the data. This allows one to display and show complex data in a concise manner and focus on specific points of interest. Interactive options such as zoom, panning and mouse hover all help in making the user interaction enjoyable and informative.

	\item Fast and easy to compare. One advantage of a dashboard is that users can easily analyze and compare the data quickly and multiple times. For example users can change the slider or dropdown to select multiple countries to see the total daily count effortlessly. This allows the data to be displayed and changed as users analysis requirements change.
\end{enumerate}

\subsubsection{Accessing the \emph{covid19.analytics Dashboard Explorer}}
The dashboard can be laucnhed locally in a machine with R, either through an
interactive R session or in batch mode using \code{Rscript} or
\code{R CMD BATCH} or through the web server accessing the following URL
                \url{https://covid19analytics.scinet.utoronto.ca}.

For running the dashboard locally, the \covidPckg package has also to be installed.
For running the dashboard within an R session the package has to be loaded and
then it should be invoked using the following sequence of commands,

        \begin{verbatim}
        > library(covid19.analytics)

        > covid19Explorer()

        \end{verbatim}

The batch mode can be executed using an R script containing the commands listed
above.
When the dashboard is run locally the browser will open a \emph{port} in the
local machine --\textit{localhost:port}-- connection, i.e. \code{http://127.0.0.1}.
It should be noted, that if the dashboard is launched interactively within an R
session the port used is 5481 --\code{http://127.0.0.1:5481}--, while if this is
done through an R script in batch mode the port used will be different.

\subsubsection{Specific Libraries Needed in the Dasbhoard}

To implement the dashboard and enhance some of the basic functionalities
offered, the following libraries were specifically used in the implementation
of the dashboard:
\begin{itemize}
	\item \CRANpkg{shiny} \cite{shinyPckg}: The main package that builds the dashboard.
	\item \CRANpkg{shinydashboard} \cite{shinydashboard}: This is a package that assists us to build the dashboard with respect to themes, layouts and structure.
	\item \CRANpkg{shinycssloaders} \cite{shinycssloaders}: This package adds loader animations to shiny outputs such as plots and tables when they are loading or (re)-calculating. In general, this are wrappers around base CSS-style loaders.
	\item \CRANpkg{plotly} \cite{plotly}: Charting library to generate interactive charts and plots. Although extensively used in the core functions of the \covidPckg, we reiterate it here as it is great tool to develop interactive plots.
	\item \CRANpkg{DT} \cite{DT}: A DataTable library to generate interactive table output.
	\item \CRANpkg{dplyr} \cite{dplyr}: A library that helps to apply functions and operations to data frames. This is important for calculations specifically in the PPE calculations.  
\end{itemize}

The R Shiny package makes developing dashboards easy and seamless and 
removes challenges. For example setting the layout of a dashboard typically is
challenging as it requires knowledge of frontend technologies such as HTML,
CSS3 and Bootstrap4 to be able to position elements and change there asthetic
properties.
Shiny simplifies this problem by using a built in box controller widget which
allows developers to easily group elements, tables, charts and widgets together. Many of the
CSS properties, such as, widths or colors are input parameters to the 
functions of interest. The sidebar feature is simple to implement and the
Shiny package makes it easy to be compatible across multiple devices such as
tablets or cellphones.
The Shiny package also has built in layout properties such as FluidRow or
Columns making it easy to position elements on a page.
The library does have some challenges as well. One challenge faced is theme
design. ShinyDashboard does not make it easy to change the whole color theme of the dashboard outside of the white or blue
theme that is provided by default.
The issue is resolved by having the developer write custom CSS and change each of the various
properties manually.

\subsubsection{Dashboard Layout}
The dashboard contains two main components a sidebar and a main body.
The sidebar contains a list of all the menu options. Options which are similar in nature are grouped in a nested format. For example the dashboard menu section called "Datasets and Reports", when selected, displays a nested list of further options the user can choose such as the World Data or Toronto Data. Grouping similar menu options together is important for making the user understand the data.
The main body displays the content of a page. The content a main body displays depends on the sidebar and the selected menu option the user selects.

There are three main generic elements needed to develop a dashboard: layouts, control widgets and output widgets.
The layout options are components needed to layout the features or components on a page. In this dashboard the layout widgets used are the following:
\begin{itemize}
	\item Box: The boxes are the main building blocks of a dashboard. This allows us to group content together.
	\item TabPanels: TabPanels allow us to create tabs to divide one page into several sections. This allows for multiple charts or multiple types of data to be displayed in a single page. For example in the Indicators page there are four tabs which display four different charts with mosaic tab displaying the charts in different configurations. 
	\item Header and Title: these are used to display text, title pages in the appropriate sizes and fonts. 
\end{itemize}

An example describing these elements and its implementation is shown in Lst.~\ref{lst:box_example}.

\Rlisting
\begin{lstlisting}[caption={Snippet of a code that describes the various features used in generating a dashboard.
				The \code{ns(id)} is a namespaced id for inputs/outputs.
				\code{WithSpinner} is the shiny cssloaders which generates a
				loading gif while the chart is being loaded.},
		captionpos=b,
		label={lst:box_example}]
tableWorldDataUI <- function(id) {
	ns <- NS(id)
	box(width=12,
		h1('WORLD TABLE'), h4('World data of all covid cases across the globe'),
		column(4, selectInput(ns("category_list3"), label=h4("Category"), choices=category_list)),		
		column(4, downloadButton(ns('downloadData'), "Download")),
		withSpinner(DT::dataTableOutput(ns("table_contents")))
	)
}
\end{lstlisting}

Shiny modules are used when a shiny application gets larger and complicated, they can also be used to fix the namespacing problem.
However shiny modules also allow for code reusability and code modularity as the code can be broken into several pieces called modules.
Each module can then be called in different applications or even in the same application multiple times.
In this dashboard we break the code into two main groups \textit{user interface} (UI) modules and \textit{server} modules.
Each menu option has there own dedicated set of UI and associated server modules.
This makes the code easy to build and expand. For each new menu option a new set of UI and sever module functions will be built.
%
%
Lst.~\ref{lst:box_example} is also an example of an UI module, where it
specifies the desing and look of the element and connect with the active parts
of the application.

Lst.~\ref{lst:report_dashboard} shows an example of a server function called \code{reportServer}.
This type of module can update and display charts, tables and valueboxes based on the user selections.
This same scenario occurs for all menu options with the UI/server paradigm.

%
%

Another way to think about the UI/server separation, is that the UI modules are
in charge of laying down the look of a particular element in the dahboard,
while the sever is in charge of dynamically `filling' the dynamical elements
and data to populate this element.

\begin{figure}
	\includegraphics[width=\textwidth]{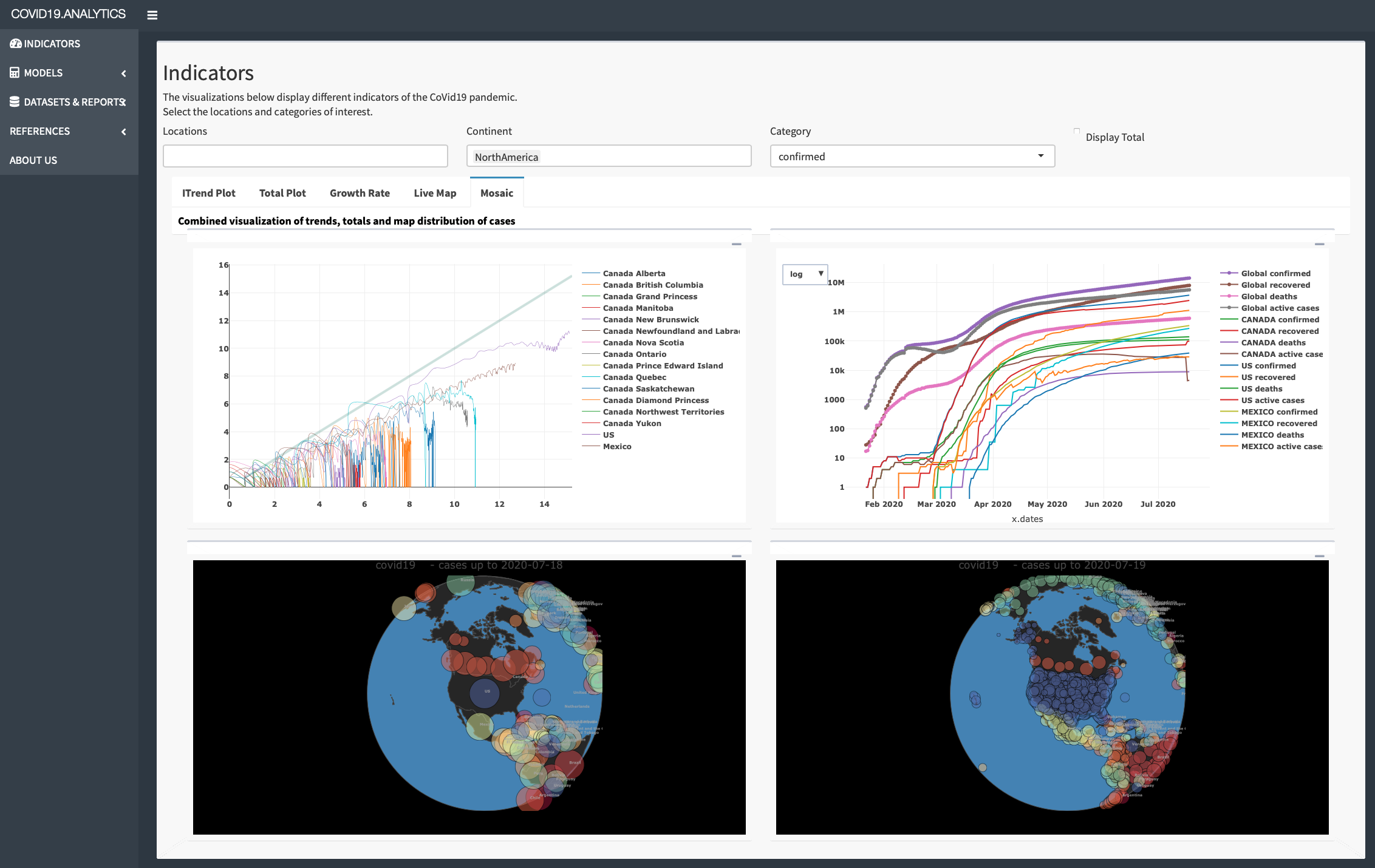}
	\caption{Screenshot from the ``\covidPckg Dashboard Explorer'', "Mosaic" tab from the 'Indicators' category.
		Four interactive figures are shown in this case:
			the trends (generated using the \code{itrends} function),
			totals (genrated using the \code{totals.plt} function)
			and two world global representations of CoViD19 reported cases (generated using the \code{live.map} function).
		The two upper plots are adjusted and re-rendered according to the selection of
		the country, category of the data from the input boxes.}
	\label{fig:dashboard_mosaic}
\end{figure}

Control widgets, also called input widgets, are widgets which users use to input data, information or settings to update charts, tables and other output widgets.
The following control widgets were used in this dashboard: 
\begin{itemize}
\item \code{NumericalInput}: A textbox that only allows for numerical input which is used to select a single numerical value. 
\item \code{SelectInput}: a dropdown that may be multi select for allowing users to select multiple options as in the case of the country dropdown. 
\item \code{Slider}: The slider in our dashboard is purely numerical but used to select a single numerical value from a given range of a min and max range. 
\item \code{Download Button}: This is a button which allows users to download and save data in various formats such as csv format. 
\item \code{Radiobuttons}: Used to select only one from a limited number of choices. 
\item \code{Checkbox}: Similar in purpose to RadioButtons that also allow users to select one option from a limited number of options. 
\end{itemize}

Output control widgets are widgets that are used to display content/information back to the user.
There are three main ouput widgets used in this dashboard:
\begin{itemize}
	\item \code{PlotlyOuput}: This widget output and creates Plotly charts. Plotly is a graphical package library used to generate interactive charts. 
	\item \code{RenderTable}: Is an output that generates the output as an interactive table with search, filter and sort capabilities provided out of the box. 
	\item \code{ValueBox}: This is a fancy textbox with border colors and descriptive font text to generate descriptive text to users such as the total number of deaths.
\end{itemize}

\begin{minipage}[c]{.85\textwidth}
\outputlisting
\begin{lstlisting}[caption={Map of the menues and functionalities offered through the \emph{\covidPckg Dashboard Explorer},
			available at \url{https://covid19analytics.scinet.utoronto.ca}},
                captionpos=b,label={lst:dashboard_map}]
* Indicators
     +----> iTrend Plot
     +----> Growth Rate
     +----> Total Plot
     +----> Live Map
     +----> Mosaic

* Models
     +----> SIR Model
     +----> Hospital PPE
		+----> PPE
		+----> PPE Advanced Settings
		+----> Burn Rate Analysis
				+----> Daily Usage
				+----> Remaining Supply
				+----> PPE per Patient
				+----> Advanced Settings

* Datasets & Repots
     +----> World Data
     +----> Toronto Data
     +----> Data Integrity
     +----> Report

* References
     +----> covid19.analytics repo
     +----> Documentation
     +----> Data Sources
     +----> External Dashboards
     +----> Resources

* About Us
\end{lstlisting}
\end{minipage}

The dashboard contains the menus and elements shown in \ref{lst:dashboard_map}
and described below:
\begin{itemize}
	\item \textit{Indicators}:
		This menu section displays different CoViD19 indicators to analyze the pandemic.
		There are four notable indicators Itrend, Total Plot, Growth Rate and Live map, which are displayed in each of the various tabs.
		Itrend displays the ``trend'' in a log-log plot,
		Total plot shows a line graph of total number,
		Growth rate displays the daily number of changes and growth rate (as defined in Sec.~\ref{sec:Fns}),
		Live map shows a world map of infections in an aggregated or timeseries format.
		These indicators are shown together in the "Mosaic" tab.

	\item \textit{Models}:
		This menu option contains a sub-menu presenting models related to the pandemic. 
		The first model is the SIR (Susceptible Infection Recovery) which is implemented in the \covidPckg package.
		SIR is a compartmental model to model how a disease will infect a population.
		The second two models are used to estimate the amount of PPE needed due to infectious diseases, such as Ebola and CoViD19.

	\item \textit{Datasets and Reports}:
		This section provides reporting capability which outputs reports as csv and text files for the data. The World Data subsection displays all the world data as a table which can be filtered, sorted and searched. The data can also be saved as a csv file. The Toronto data displays the Toronto data in tabular format while also displaying the current pandemic numbers. Data Integrity section checks the integrity and consistency of the data set such as when the raw data contains negative numbers or if the cumulative quantities decrease. The Report section is used to generate a report as a text file.

	\item \textit{References}: The reference section displays information on the github repo and documentation along with an External Dashboards section which contains hyperlinks to other dashboards of interest. Dashboards of interest are the Vaccine Tracker which tracks the progress of vaccines being tested for covid19, John Hopkins University and the Canada dashboard built by the Dall Lana School of Epidemiology at the University of Toronto. 

	\item \textit{About Us}:
		Contact information and information about the developers.
\end{itemize}

\subsubsection{Additional Elements of the Dashboard}
In addition to implementing some of the functionalities provided by the \covidPckg package,
the dashboard also includes a PPE calculator.
The hospital PPE is a qualitative model which is designed to analyze the
amount of PPE equipment needed for a single CoViD19 patient over a hospital
duration.
The PPE calculation implemented in the \textit{\covidPckg Dashboard Explorer}
is derived from the CDC's studies for infectious diseases, such as Ebola and CoViD19.
The rationality is that Ebola and CoViD19 are both contagious infections and
PPE is used to protect staff and patients and prevent transmission for both of these contagious diseases.

The Hospital PPE calculation estimates and models the amount of PPE a hospital will need during the CoViD19 pandemic.
There are two analysis methods a user can choose to determine Hospital PPE requirement.
The first method to analyze PPE is to determine the amount of PPE needed for a single hospitalized CoViD19 patient.
This first model requires two major component: the size of the healthcare team needed to take care of a single CoViD19 patient and the amount of PPE equipment used by hospital staff per shift over a hosptialization duration.
The model is based off the CDC Ebola crisis calculation \cite{CDCppeEbola}.
Alhough Ebola is a different disease compared to CoViD19, there is one major similarity.
Both CoViD19 and Ebola are diseases which both require PPE for protection of healthcare staff and the infected patient.
To compensate the user can change the amount of PPE a healthcare staff uses per shift. That information can be adjusted by changing the slider values in the Advanced Setting tab.
The calculation is pretty straightforward as it takes the PPE amount used per shift and multiplies it by the number of healthcare staff and then by the hospitalization duration.

The first model has two tabs.
The first tab displays a stacked bar chart displaying the amount of PPE equipment user by each hospital staff over the total hospital duration of a single patient.
It breaks each PPE equipment by stacks. The second tab panel called Advanced settings has a series of sliders for each hospital staff example nurses where users can use the slider to change the amount of PPE that the hospital staff will user per shift. 

The second model is a more recent calculation developed by the CDC \cite{CDCburnrate}.
The model calculates the \textit{burn rate} of PPE equipment for hospitals for a one week time period.
This model is designed specifically for Covid19.
The CDC has created an excel file for hospital staff to input their information and also an android app as well which can be utilized. 

This model also implemented in our dashboard, is simplified to calculate the PPE for a one week setting.
The one week limit was implemented for two reasons, first to limit the amount of input data a user has to enter into the system as too much data can overwhelm and confuse a user;
second because the CoViD19 pandemic is a highly fluidic situation and for hospital staff to forecast their PPE and resource equipments greater than a one week period may not be accurate.

Note that this model is not accurate if the facilitiy recieves a resupply of PPE.
For resupplied PPE start a new calculation. There are four tab panels to the burn rate calculation which displays charts and settings.
The first tab Daily Usage displays a multi-line chart displaying the amount of PPE used daily, $\Delta PPE_{daily}$.
The calculation for this is a simple subtraction between two consecutive days,
i.e. the second day ($j+1$) from the first day ($j$) as noted in Eq.(\ref{eqn:duration}). 

\begin{equation} 
	\Delta PPE_{daily} \equiv (PPE_{j+1})-PPE_j 
	\label{eqn:duration}
\end{equation}

The tab panel called Remaining Supply shows a multi line chart the number of days the remaining PPE equipment will last in the facility.
The duration of how long the PPE equipment can last in a given facility, inversely depends on the amount of CoViD19 patients admitted to the hospital.
To calculate the remaining PPE one calculates the average amount of PPE used over the one week duration and then divides the amount of PPE at the beginning of the day by the average PPE usage, as shown in Eq.(\ref{eqn:remPPE}),

\begin{equation}
	remainingPPE \equiv \frac{\Delta PPE_{daily}}{\langle PPE \rangle_{1week}}
	\label{eqn:remPPE}
\end{equation}

where $\langle \rangle_{T}$ denotes the time average over a $T$ period of time.

The third panel called PPE per patient displays a multi line chart of the burn
rate, i.e. the amount of PPE used per patient per day.
Eq.(\ref{eqn:PPEpatient}) represents the calculation as the remaining PPE supply
divided by the number of CoViD19 patients in the hospital staff during that exact day.

\begin{equation}
	\frac{PPE}{Patient} \equiv \frac{remainingPPE}{Npatient}
	\label{eqn:PPEpatient}
\end{equation}

The fourth tab called Advanced settings is a series of show and hide ``accordians'' 
where users can input the amount of PPE equpiment they have at the start of
each day.
There are six collapsed boxes for each PPE equipment type and for CoViD19 patient count.
Expanding a box displays seven numericalInput textboxes
which allows users to input the number of PPE or patient count for each day.

The equations describing the PPE needs,
Eqs.(\ref{eqn:duration},\ref{eqn:remPPE},\ref{eqn:PPEpatient})
are implemented in the shiny dashboard using the \CRANpkg{dplyr} library.
The \CRANpkg{dplyr} library allows users to work with dataframe like objects in a quick and efficient manner.
The three equations are implemented using a single dataframe.
The Advanced Setting inputs of the Burn Rate Analysis tab are saved into a dataframe.
The PPE equations --Eqs.(\ref{eqn:duration}-\ref{eqn:PPEpatient})-- are implemented
on the dataframe by creating new columns that are then appended to the existing dataframe.
This results in the final version of the dataframe as shown in Lst.~\ref{lst:PPE_dataframe}
which is then used to generate the corresponding charts.

\outputlisting
\begin{lstlisting}[caption={Dataframe object employed in the PPE calculation.},
                captionpos=b,
                label={lst:PPE_dataframe}]
d.Name glovediff n95diff maskdiff gownsdiff respdiff gloveremaining
1   day1         0       0        0         0        0            6.0
2   day2       200      25       50       100       20            5.2
3   day3       300      75       50        20       20            4.0
4   day4        NA      NA       NA        NA       NA             NA
5   day5        NA      NA       NA        NA       NA             NA
6   day6        NA      NA       NA        NA       NA             NA
7   day7        NA      NA       NA        NA       NA             NA
  n95remaining maskremaining gownremaining respremaining glove_per_patient
1         10.0            10      8.333333            15           0.00000
2          9.5             9      6.666667            14          10.00000
3          8.0             8      6.333333            13          10.71429
4           NA            NA            NA            NA                NA
5           NA            NA            NA            NA                NA
6           NA            NA            NA            NA                NA
7           NA            NA            NA            NA                NA
  n95_per_patient mask_per_patient gown_per_patient resp_per_patient
1        0.000000         0.000000        0.0000000        0.0000000
2        1.250000         2.500000        5.0000000        1.0000000
3        2.678571         1.785714        0.7142857        0.7142857
4              NA               NA               NA               NA
5              NA               NA               NA               NA
6              NA               NA               NA               NA
7              NA               NA               NA               NA
\end{lstlisting}

\subsection{Dashboard's Back End Implementation}
\label{sec:dashboard-backend}

The back-end implementation of the dashboard is achieved using the functions
presented in Sec.\ref{sec:covid19Pckg} on the \textit{server} module of the dashboard.
The main strategy is to use a particular function and \textit{connect} it with the
input controls to feed the needed arguments into the function and then
\textit{capture} the output of the function and \textit{render} it accordingly.

Let's consider the example of the globe map representation shown in the dashboard
which is done using the \code{live.map} function.
Lst.~\ref{lst:livemap_dashboard} shows how this function connects with the other
elements in the dashboard: the input elements are accessed using
\code{input\$...} which in this are used to control the particular options for
the displaying the legends or projections based on checkboxes.
The output returned from this function is captured through the
\code{renderPlotly(\{...\})} function, that is aimed to take plotly type of plots
and integrate them into the dashboard.

\Rlisting
\begin{lstlisting}[caption={Example of how the \code{live.map} function is used to render the ineractive figures display on the dashboard.},
                captionpos=b,
                label={lst:livemap_dashboard}]
#livemap plot charts on the three possible commbinations
output$ts_livemap  <- output$ts2_livemap <- output$ts3_livemap <- output$ts4_livemap <- renderPlotly({
	legend <- input$sel_legend
	projections <- input$sel_projection
	live.map(covid19.data("ts-confirmed"), interactive.display=FALSE, no.legend=legend, select.projctn=projections)
})
\end{lstlisting}

Another example is the report generation capability using the
\code{report.summary} function which is shown in Lst.~\ref{lst:report_dashboard}.
As mentioned before, the input arguments of the function are obtained from the input controls.
The output in this case is rendered usign the \code{renderText(\{...\})} function, as
the output of the original function is plain text.
Notice also that there are two implementations of the \code{report.summary},
one is for the rendering in the screen and the second one is for making the report
available to be downloaded which is handled by the \code{downloadHandler} function.

\Rlisting
\begin{lstlisting}[caption={Report capabilites implemented in the dashboard using the \code{report.summary} function.},
	captionpos=b,
	label={lst:report_dashboard}]
reportServer <- function(input, output, session, result) {
        output$report_output_default <- renderText({
                #extract the vairables of the inputs
                nentries <- input$txtbox_Nentries
                geo_loc <- input$geo_loc_select6
                ts <- input$ddl_TS
                capture.output(report.summary(graphical.output=FALSE, Nentries=nentries, geo.loc=geo_loc, cases.to.process=ts))}
                , sep='\n')

        report <- reactive({
                nentries <- input$txtbox_Nentries
                geo_loc <- input$geo_loc_select6
                ts <- input$ddl_TS
                report <- capture.output(report.summary(graphical.output=FALSE, Nentries=nentries, geo.loc=geo_loc, cases.to.process=ts))
                return (report)
        })

    output$downloadReport <- downloadHandler(
            filename = function() {
              paste("Report-", Sys.Date(),".txt", sep = "")
            },
            content = function(file) {
              writeLines(paste(report() ), file)
            }
        )
}

\end{lstlisting}

\subsection{Dashboard's Server Configuration}
\label{sec:dashboard-serverConfig}

The final element in the deployment of the dashboard is the actual set up and
configuration of the \textit{web server} where the application will run.
The actual implementation of our web dashboard, accessible through
\url{https://covid19analytics.scinet.utoronto.ca}, relies on a
\textit{virtual machine} (VM) in a physical server located at SciNet
headquarters.

We should also notice that there are other forms or ways to "publish" a
dashboard, in particular for shiny based-dashboards, the most common way and
perhaps straighforward one is to deploy the dashboard on \url{https://www.shinyapps.io}.
Alternatively one could also implement the dashboard in a cloud-based solution,
e.g. \url{https://aws.amazon.com/blogs/big-data/running-r-on-aws/}.
Each approach has its own advantages and disadvantages, for instance,
depending on a third party solution (like the previous mentioned) implies
some cost to be paid to or dependency on the provider but will certainly
eliminate some of the complexity and special attention one must take when
using its own server.
On the other hand, a self-deployed server will allow you for full control,
in principle cost-effective or cost-controled expense and full integration with
the end application.

In our case, we opted for a self-controlled and configured server as mentioned
above.
Moreover, it is quite a common practice to deploy (multiple) web services
via VMs or ``containers''.
The VM for our web server runs on CentOS 7 and has installed R version 4.0 from
sources and compiled on the VM.

After that we proceeded to install the shiny server from sources, i.e.
\url{https://github.com/rstudio/shiny-server/wiki/Building-Shiny-Server-from-Source}.

After the installation of the shiny-server is completed, we proceed by creating
a new user in the VM from where the server is going to be run.
For security reasons, we recommend to avoid running the server as root.
In general, the shiny server can use a user named "shiny".
Hence a local account is created for this user, and then logged as this user,
one can proceed with the installation of the required R packages in a local library
for this user.
All the neeeded packages for running the dashboard and the \covidPckg package
need to be installed.

Lst.~\ref{lst:shiny-config} shows the commands used for creating the shiny user and 
finalizing the configuration and details of the log files.

\configlisting
\begin{lstlisting}[caption={List of commands used on the VM to finalize the setup of the shiny user and server.
			Source: \url{https://github.com/rstudio/shiny-server}.},
                captionpos=b, language=bash,
                label={lst:shiny-config}]
# Place a shortcut to the shiny-server executable in /usr/bin
sudo ln -s /usr/local/shiny-server/bin/shiny-server /usr/bin/shiny-server

# Create shiny user
sudo useradd -r -m shiny

# Create log, config, and application directories
sudo mkdir -p /var/log/shiny-server
sudo mkdir -p /srv/shiny-server
sudo mkdir -p /var/lib/shiny-server
sudo chown shiny /var/log/shiny-server
sudo mkdir -p /etc/shiny-server
\end{lstlisting}

The R script containing the shiny app to be run should be placed in
\code{/etc/shiny-server} and confiurations details about the shiny 
interface are adjusted in the \code{/etc/shiny-server/shiny-server.conf} file.

Permissions for the application file has to match the identity of the user
launching the server, in this case the shiny user.

At this point if the installation was sucessful and all the pieces were
placed properly, when the \texttt{shiny-server} command is executed, a shiny
hosted app will be accessible from \texttt{localhost:3838}.


Since the shiny server listens on port 3838 in plain \texttt{http}, it is
necessary to setup an Apache web server to act as a \textit{reverse proxy} to
receive the connection requests from the Internet on ports 80 and 443,
the regular \texttt{http} and \texttt{https} ports,
and redirect them to port 3838 on the same host (\texttt{localhost}).


For dealing with the \textit{Apache configuration on port 80},
we added the file \texttt{/etc/httpd/conf.d/rewrite.conf} as shown in Lst.~\ref{lst:apache_port_80}.

\configlisting
\begin{lstlisting}[caption={Modifications to the Apache configurations, specified in the file
			\texttt{rewrite.conf}.
			These three lines rewrite any incoming request from \texttt{http} to \texttt{https}.},
                captionpos=b,
		frame=single, language=HTML,
                label={lst:apache_port_80}]
RewriteEngine On
RewriteCond %{REQUEST_SCHEME} =http
RewriteRule ^ https://%{SERVER_NAME}%{REQUEST_URI} [QSA,R=permanent]
\end{lstlisting}


For handling the \textit{Apache configuration on port 443},
we added this file \texttt{/etc/httpd/conf.d/shiny.conf}, as shown in Lst.\ref{lst:apache_port_443}.

\configlisting
\begin{lstlisting}[caption={Content of the \texttt{shiny.conf} file,
			for the creation of a \textit{VirtualHost} listening on port 443.},
                captionpos=b,
                frame=single, language=HTML,
                label={lst:apache_port_443}]
<VirtualHost _default_:443>
SSLEngine on

ProxyPreserveHost On
ProxyPass / http://0.0.0.0:3838/app1/ 
ProxyPassReverse / http://0.0.0.0:3838/app1/ 
</VirtualHost>
\end{lstlisting}

This VirtualHost receives the \texttt{https} requests from the Internet on port 443,
establishes the secure connection, and redirects all input to port 3838 using
plain \texttt{http}.
All requests to "/" are redirected to "\texttt{http://0.0.0.0:3838/app1/}", where
\texttt{app1} in this case is a subdirectory where a particular shiny app is located.

There is an additional configuration file, \texttt{/etc/httpd/conf.d/ssl.conf},
which contains the configuration for establishing secure connections such as
protocols, certificate paths, ciphers, etc.

\subsubsection{Decentralized Systems and Services}
\begin{figure}
	\includegraphics[width=\textwidth]{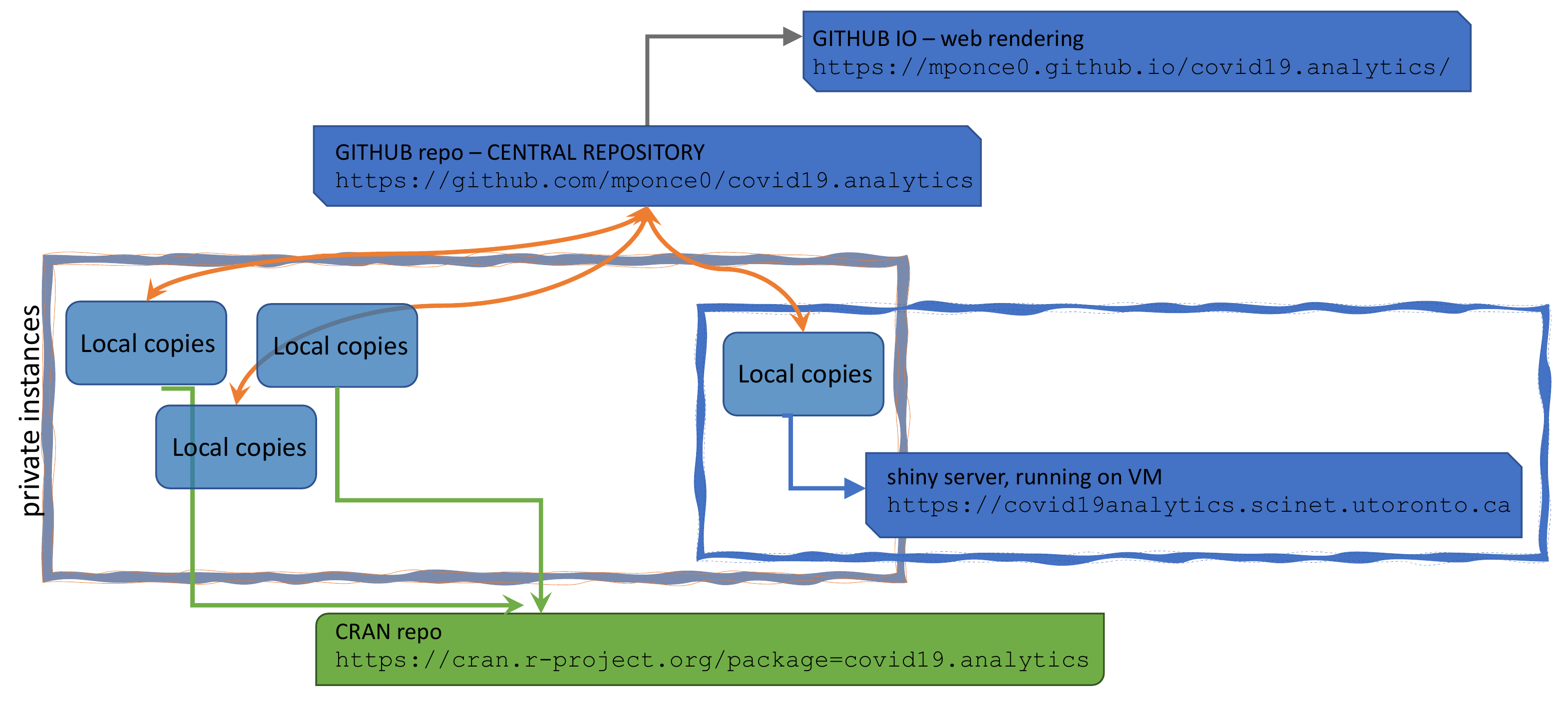}
	\caption{Schematic of the different repositories and systems employed by the \covidPckg package and dashboard interface.}
	\label{fig:repos}
\end{figure}

The main tool we use in order to communicate updates between the different
elements we use in the development and mantainance of the \covidPckg package
and dashboard web interface is orchestrated via GIT repositories.
In this way, we have in place version control systems but also offer decentralized
with multiple replicas.
Fig.~\ref{fig:repos} shows an schematic of how our network of repositories and
service is connected.  The central hub for our package, is located at the
GIThub repo \url{htttps://github.com/mponce0/covid19.analytics};
we then have (and users can too) our own clones of local copies of this repo
--we usually use this for development and testing--.
When a stable and substantial contribution to the package is reached, we submit
this to the CRAN repository.
Similarly, when an update is done on the dashboard we can synchronize the VM
via git pulls and deploy the updates on the server side.

\section{Conclusions}
\label{sec:conc}

In this paper we have presented and discussed the R \covidPckg package, which
is an open source tool to obtain, analyze and visualize data of the CoViD19
pandemic.
The package also incorporates a dashboard to facilitate the access to its
functionalities to less experienced users.

As today, there are a few dozen other packages also in the CRAN repository that
allow users to gain access to different datasets of the CoViD19 pandemic.
In some cases, some packages just provide access to data from specific
geographical locations or the approach to the data structure in which the
data is presented is different from the one presented here.
Nevertheless, having a variety of packages from which users can try and
probably combine, is an important and crucial element in data analysis.
Moreover it has been reported different cases of data misuse/misinterpretation
due to different issues, such as, erroneous metadata or data formats
\cite{Ledford_2020} and in some cases ending in articles' retractions \cite{Schriml:2020aa}.
Therefore providing additional functionalities to check the integrity and
consistency of the data, as our the \covidPckg package does is paramount.
This is specially true in a situation where the unfolding of events and data
availability is flowing so fast that sometimes is even hard to keep track of
all the changes.

Moreover, the \covidPckg package offers a modular and versatile approach to
the data, by allowing users to input their own data for which most of the
package functions can be applied when the data is structured using a
\textit{time series} format as described in this manuscript.

The \covidPckg is also capable of retrieving genomics data, and it does that
by incorporating a novel, more reliable and robust way of accessing and
designing different pathways to the data sources.

Another unique feature of this package is the ability of incorporating models
to estimate the disease spread by using the actual data.
Although a simple model, it has shown some interesting results in agreement
for certain cases.
Of course there are more sophisticated approaches to shred light in analyzing
this pandemic; in particular novel ``community'' approaches have been catalyzed
by this too \cite{Luengo-Oroz:2020aa}.
However all of these approaches face new challenges as well\cite{Hu:2020aa},
and on that regards counting with a variety, in particular of open source tools
and direct access to the data might help on this front.


\section*{Acknowledgments}
MP wants to thank all his colleagues at SciNet, especially Daniel Gruner for
his continous and unconditional support, and Marco Saldarriaga who helped us
setting up the VM for installing the shiny server.



\bibliography{references}

\begin{thebibliography}{10}
\expandafter\ifx\csname url\endcsname\relax
  \def\url#1{\texttt{#1}}\fi
\expandafter\ifx\csname urlprefix\endcsname\relax\def\urlprefix{URL }\fi
\expandafter\ifx\csname href\endcsname\relax
  \def\href#1#2{#2} \def\path#1{#1}\fi

\bibitem{citeR}
{R Core Team}, \href{https://www.R-project.org/}{{R: A Language and Environment
  for Statistical Computing}}, R Foundation for Statistical Computing, Vienna,
  Austria (2016).
\newline\urlprefix\url{https://www.R-project.org/}

\bibitem{ihaka:1996}
R.~Ihaka, R.~Gentleman,
  \href{https://doi.org/10.1080/10618600.1996.10474713}{R: A language for data
  analysis and graphics}, Journal of Computational and Graphical Statistics
  5~(3) (1996) 299--314.
\newline\urlprefix\url{https://doi.org/10.1080/10618600.1996.10474713}

\bibitem{covid19analytics}
M.~Ponce,
  \href{https://CRAN.R-project.org/package=covid19.analytics}{covid19.analytics:
  Load and Analyze Live Data from the CoViD-19 Pandemic}, r package version 2.1
  (2020).
\newline\urlprefix\url{https://CRAN.R-project.org/package=covid19.analytics}

\bibitem{Cyranoski_2020}
D.~Cyranoski, \href{http://dx.doi.org/10.1038/d41586-020-01541-z}{The biggest
  mystery: what it will take to trace the coronavirus source}, Nature\href
  {http://dx.doi.org/10.1038/d41586-020-01541-z}
  {\path{doi:10.1038/d41586-020-01541-z}}.
\newline\urlprefix\url{http://dx.doi.org/10.1038/d41586-020-01541-z}

\bibitem{Mallapaty_2020}
S.~Mallapaty, \href{http://dx.doi.org/10.1038/d41586-020-01449-8}{Animal source
  of the coronavirus continues to elude scientists}, Nature\href
  {http://dx.doi.org/10.1038/d41586-020-01449-8}
  {\path{doi:10.1038/d41586-020-01449-8}}.
\newline\urlprefix\url{http://dx.doi.org/10.1038/d41586-020-01449-8}

\bibitem{Zhou_2020}
P.~Zhou, X.-L. Yang, X.-G. Wang, B.~Hu, L.~Zhang, W.~Zhang, H.-R. Si, Y.~Zhu,
  B.~Li, C.-L. Huang, et~al.,
  \href{http://dx.doi.org/10.1038/s41586-020-2012-7}{A pneumonia outbreak
  associated with a new coronavirus of probable bat origin}, Nature 579~(7798)
  (2020) 270--273.
\newblock \href {http://dx.doi.org/10.1038/s41586-020-2012-7}
  {\path{doi:10.1038/s41586-020-2012-7}}.
\newline\urlprefix\url{http://dx.doi.org/10.1038/s41586-020-2012-7}

\bibitem{origin-SARS-CoV-2}
K.~G. Andersen, A.~Rambaut, W.~I. Lipkin, E.~C. Holmes, R.~F. Garry,
  \href{https://doi.org/10.1038/s41591-020-0820-9}{The proximal origin of
  sars-cov-2}, Nature Medicine 26~(4) (2020) 450--452.
\newblock \href {http://dx.doi.org/10.1038/s41591-020-0820-9}
  {\path{doi:10.1038/s41591-020-0820-9}}.
\newline\urlprefix\url{https://doi.org/10.1038/s41591-020-0820-9}

\bibitem{Letko:2020aa}
M.~Letko, S.~N. Seifert, K.~J. Olival, R.~K. Plowright, V.~J. Munster,
  \href{https://doi.org/10.1038/s41579-020-0394-z}{Bat-borne virus diversity,
  spillover and emergence}, Nature Reviews Microbiology\href
  {http://dx.doi.org/10.1038/s41579-020-0394-z}
  {\path{doi:10.1038/s41579-020-0394-z}}.
\newline\urlprefix\url{https://doi.org/10.1038/s41579-020-0394-z}

\bibitem{Gupta_2020}
A.~Gupta, M.~V. Madhavan, K.~Sehgal, N.~Nair, S.~Mahajan, T.~S. Sehrawat,
  B.~Bikdeli, N.~Ahluwalia, J.~C. Ausiello, E.~Y. Wan, et~al.,
  \href{http://dx.doi.org/10.1038/s41591-020-0968-3}{Extrapulmonary
  manifestations of covid-19}, Nature Medicine 26~(7) (2020) 1017--1032.
\newblock \href {http://dx.doi.org/10.1038/s41591-020-0968-3}
  {\path{doi:10.1038/s41591-020-0968-3}}.
\newline\urlprefix\url{http://dx.doi.org/10.1038/s41591-020-0968-3}

\bibitem{Williamson_2020}
E.~J. Williamson, A.~J. Walker, K.~Bhaskaran, S.~Bacon, C.~Bates, C.~E. Morton,
  H.~J. Curtis, A.~Mehrkar, D.~Evans, P.~Inglesby, et~al.,
  \href{http://dx.doi.org/10.1038/s41586-020-2521-4}{Opensafely: factors
  associated with covid-19 death in 17 million patients}, Nature\href
  {http://dx.doi.org/10.1038/s41586-020-2521-4}
  {\path{doi:10.1038/s41586-020-2521-4}}.
\newline\urlprefix\url{http://dx.doi.org/10.1038/s41586-020-2521-4}

\bibitem{Scully:2020aa}
E.~P. Scully, J.~Haverfield, R.~L. Ursin, C.~Tannenbaum, S.~L. Klein,
  \href{https://doi.org/10.1038/s41577-020-0348-8}{Considering how biological
  sex impacts immune responses and covid-19 outcomes}, Nature Reviews
  Immunology\href {http://dx.doi.org/10.1038/s41577-020-0348-8}
  {\path{doi:10.1038/s41577-020-0348-8}}.
\newline\urlprefix\url{https://doi.org/10.1038/s41577-020-0348-8}

\bibitem{Willyard_2020}
C.~Willyard, \href{http://dx.doi.org/10.1038/d41586-020-01403-8}{Coronavirus
  blood-clot mystery intensifies}, Nature 581~(7808) (2020) 250--250.
\newblock \href {http://dx.doi.org/10.1038/d41586-020-01403-8}
  {\path{doi:10.1038/d41586-020-01403-8}}.
\newline\urlprefix\url{http://dx.doi.org/10.1038/d41586-020-01403-8}

\bibitem{Silvermaneabc1126}
J.~D. Silverman, N.~Hupert, A.~D. Washburne,
  \href{https://stm.sciencemag.org/content/early/2020/06/22/scitranslmed.abc1126}{Using
  influenza surveillance networks to estimate state-specific prevalence of
  sars-cov-2 in the united states}, Science Translational Medicine\href
  {http://arxiv.org/abs/https://stm.sciencemag.org/content/early/2020/06/22/scitranslmed.abc1126.full.pdf}
  {\path{arXiv:https://stm.sciencemag.org/content/early/2020/06/22/scitranslmed.abc1126.full.pdf}},
  \href {http://dx.doi.org/10.1126/scitranslmed.abc1126}
  {\path{doi:10.1126/scitranslmed.abc1126}}.
\newline\urlprefix\url{https://stm.sciencemag.org/content/early/2020/06/22/scitranslmed.abc1126}

\bibitem{airborneLancet}
T.~Greenhalgh, J.~L. Jimenez, K.~A. Prather, Z.~Tufekci, D.~Fisman,
  R.~Schooley, \href{https://doi.org/10.1016/S0140-6736(21)00869-2}{{Ten
  scientific reasons in support of airborne transmission of SARS-CoV-2}}, {The
  Lancet}\href {http://dx.doi.org/10.1016/S0140-6736(21)00869-2}
  {\path{doi:10.1016/S0140-6736(21)00869-2}}.
\newline\urlprefix\url{https://doi.org/10.1016/S0140-6736(21)00869-2}

\bibitem{Fry_2020}
C.~V. Fry, X.~Cai, Y.~Zhang, C.~S. Wagner,
  \href{http://dx.doi.org/10.1371/journal.pone.0236307}{Consolidation in a
  crisis: Patterns of international collaboration in early covid-19 research},
  PLOS ONE 15~(7) (2020) e0236307.
\newblock \href {http://dx.doi.org/10.1371/journal.pone.0236307}
  {\path{doi:10.1371/journal.pone.0236307}}.
\newline\urlprefix\url{http://dx.doi.org/10.1371/journal.pone.0236307}

\bibitem{Singh_Chawla_2020}
D.~Singh~Chawla, \href{http://dx.doi.org/10.1038/d41586-020-01685-y}{Critiqued
  coronavirus simulation gets thumbs up from code-checking efforts}, Nature
  582~(7812) (2020) 323--324.
\newblock \href {http://dx.doi.org/10.1038/d41586-020-01685-y}
  {\path{doi:10.1038/d41586-020-01685-y}}.
\newline\urlprefix\url{http://dx.doi.org/10.1038/d41586-020-01685-y}

\bibitem{Duque:2020aa}
D.~Duque, D.~Morton, B.~Singh, Z.~Du, R.~Pasco, L.~Meyers, Timing social
  distancing to avert unmanageable covid-19 hospital surges, PNAS; Proceedings
  of the National Academy of Sciences\href
  {http://dx.doi.org/10.1073/pnas.2009033117}
  {\path{doi:10.1073/pnas.2009033117}}.

\bibitem{Adam_2020}
D.~Adam, \href{http://dx.doi.org/10.1038/d41586-020-01003-6}{Special report:
  The simulations driving the world's response to covid-19}, Nature 580~(7803)
  (2020) 316--318.
\newblock \href {http://dx.doi.org/10.1038/d41586-020-01003-6}
  {\path{doi:10.1038/d41586-020-01003-6}}.
\newline\urlprefix\url{http://dx.doi.org/10.1038/d41586-020-01003-6}

\bibitem{Hotez:2020ab}
P.~J. Hotez, D.~B. Corry, M.~E. Bottazzi,
  \href{https://doi.org/10.1038/s41577-020-0323-4}{Covid-19 vaccine design: the
  janus face of immune enhancement}, Nature Reviews Immunology 20~(6) (2020)
  347--348.
\newblock \href {http://dx.doi.org/10.1038/s41577-020-0323-4}
  {\path{doi:10.1038/s41577-020-0323-4}}.
\newline\urlprefix\url{https://doi.org/10.1038/s41577-020-0323-4}

\bibitem{Ahmed:2020aa}
S.~F. Ahmed, A.~A. Quadeer, M.~R. McKay,
  \href{https://doi.org/10.1038/s41596-020-0358-9}{Covidep: a web-based
  platform for real-time reporting of vaccine target recommendations for
  sars-cov-2}, Nature Protocols\href
  {http://dx.doi.org/10.1038/s41596-020-0358-9}
  {\path{doi:10.1038/s41596-020-0358-9}}.
\newline\urlprefix\url{https://doi.org/10.1038/s41596-020-0358-9}

\bibitem{Block:2020aa}
P.~Block, M.~Hoffman, I.~J. Raabe, J.~B. Dowd, C.~Rahal, R.~Kashyap, M.~C.
  Mills, \href{https://doi.org/10.1038/s41562-020-0898-6}{Social network-based
  distancing strategies to flatten the covid-19 curve in a post-lockdown
  world}, Nature Human Behaviour 4~(6) (2020) 588--596.
\newblock \href {http://dx.doi.org/10.1038/s41562-020-0898-6}
  {\path{doi:10.1038/s41562-020-0898-6}}.
\newline\urlprefix\url{https://doi.org/10.1038/s41562-020-0898-6}

\bibitem{doi:10.1063/5.0008834}
D.~Faranda, I.~P. Castillo, O.~Hulme, A.~Jezequel, J.~S.~W. Lamb, Y.~Sato,
  E.~L. Thompson, \href{https://doi.org/10.1063/5.0008834}{Asymptotic estimates
  of sars-cov-2 infection counts and their sensitivity to stochastic
  perturbation}, Chaos: An Interdisciplinary Journal of Nonlinear Science
  30~(5) (2020) 051107.
\newblock \href {http://arxiv.org/abs/https://doi.org/10.1063/5.0008834}
  {\path{arXiv:https://doi.org/10.1063/5.0008834}}, \href
  {http://dx.doi.org/10.1063/5.0008834} {\path{doi:10.1063/5.0008834}}.
\newline\urlprefix\url{https://doi.org/10.1063/5.0008834}

\bibitem{KORBER2020}
B.~Korber, W.~M. Fischer, S.~Gnanakaran, H.~Yoon, J.~Theiler, W.~Abfalterer,
  N.~Hengartner, E.~E. Giorgi, T.~Bhattacharya, B.~Foley, K.~M. Hastie, M.~D.
  Parker, D.~G. Partridge, C.~M. Evans, T.~M. Freeman, T.~I. {de Silva},
  A.~Angyal, R.~L. Brown, L.~Carrilero, L.~R. Green, D.~C. Groves, K.~J.
  Johnson, A.~J. Keeley, B.~B. Lindsey, P.~J. Parsons, M.~Raza,
  S.~Rowland-Jones, N.~Smith, R.~M. Tucker, D.~Wang, M.~D. Wyles, C.~McDanal,
  L.~G. Perez, H.~Tang, A.~Moon-Walker, S.~P. Whelan, C.~C. LaBranche, E.~O.
  Saphire, D.~C. Montefiori,
  \href{http://www.sciencedirect.com/science/article/pii/S0092867420308205}{Tracking
  changes in sars-cov-2 spike: Evidence that d614g increases infectivity of the
  covid-19 virus}, Cell\href {http://dx.doi.org/10.1016/j.cell.2020.06.043}
  {\path{doi:10.1016/j.cell.2020.06.043}}.
\newline\urlprefix\url{http://www.sciencedirect.com/science/article/pii/S0092867420308205}

\bibitem{Boni_2020}
M.~F. Boni, P.~Lemey, X.~Jiang, T.~T.-Y. Lam, B.~W. Perry, T.~A. Castoe,
  A.~Rambaut, D.~L. Robertson,
  \href{http://dx.doi.org/10.1038/s41564-020-0771-4}{Evolutionary origins of
  the sars-cov-2 sarbecovirus lineage responsible for the covid-19 pandemic},
  Nature Microbiology\href {http://dx.doi.org/10.1038/s41564-020-0771-4}
  {\path{doi:10.1038/s41564-020-0771-4}}.
\newline\urlprefix\url{http://dx.doi.org/10.1038/s41564-020-0771-4}

\bibitem{Dong:2020aa}
E.~Dong, H.~Du, L.~Gardner,
  \href{https://doi.org/10.1016/S1473-3099(20)30120-1}{An interactive web-based
  dashboard to track covid-19 in real time}, The Lancet Infectious Diseases
  20~(5) (2020) 533--534.
\newblock \href {http://dx.doi.org/10.1016/S1473-3099(20)30120-1}
  {\path{doi:10.1016/S1473-3099(20)30120-1}}.
\newline\urlprefix\url{https://doi.org/10.1016/S1473-3099(20)30120-1}

\bibitem{Palayew:2020aa}
A.~Palayew, O.~Norgaard, K.~Safreed-Harmon, T.~H. Andersen, L.~N. Rasmussen,
  J.~V. Lazarus, \href{https://doi.org/10.1038/s41562-020-0911-0}{Pandemic
  publishing poses a new covid-19 challenge}, Nature Human Behaviour\href
  {http://dx.doi.org/10.1038/s41562-020-0911-0}
  {\path{doi:10.1038/s41562-020-0911-0}}.
\newline\urlprefix\url{https://doi.org/10.1038/s41562-020-0911-0}

\bibitem{Callaway_2020}
E.~Callaway, \href{http://dx.doi.org/10.1038/d41586-020-01520-4}{Will the
  pandemic permanently alter scientific publishing?}, Nature 582~(7811) (2020)
  167--168.
\newblock \href {http://dx.doi.org/10.1038/d41586-020-01520-4}
  {\path{doi:10.1038/d41586-020-01520-4}}.
\newline\urlprefix\url{http://dx.doi.org/10.1038/d41586-020-01520-4}

\bibitem{Kwon_2020}
D.~Kwon, \href{http://dx.doi.org/10.1038/d41586-020-01394-6}{How swamped
  preprint servers are blocking bad coronavirus research}, Nature 581~(7807)
  (2020) 130--131.
\newblock \href {http://dx.doi.org/10.1038/d41586-020-01394-6}
  {\path{doi:10.1038/d41586-020-01394-6}}.
\newline\urlprefix\url{http://dx.doi.org/10.1038/d41586-020-01394-6}

\bibitem{Vabret2020}
N.~Vabret, R.~Samstein, N.~Fernandez, M.~Merad, A.~S. Schanoski, A.~Rodriguez,
  A.~Moreira, A.~Chan, A.~Charap, A.~Leader, A.~Magen, B.~Salom{\'e}, C.~C.
  Bozkus, C.~Moon, C.~Gruber, D.~F. Ruan, D.~Lozano-Ojalvo, D.~Jha, E.~Meritt,
  E.~Risson, E.~Dalla, E.~Humblin, E.~Cody, F.~Cossarini, G.~Lubitz,
  G.~Britton, G.~Martinez-Delgado, I.~R. Torres, J.~Mateus-Tique, J.~Redes,
  J.~Tan, J.~Shang, J.~Kim, J.~Grout, J.~Chung, J.~Catalan, J.~Kodysh, J.~Noel,
  K.~Lindblad, L.~Malle, L.~Pia, M.~Agrawal, M.~Casanova-Acebes, M.~Kuksin,
  M.~Suprun, M.~Aleynick, M.~Roberto, M.~Brown, M.~Lin, M.~Park, M.~Spindler,
  M.~Wilk, M.~Belabed, M.~Saffern, M.~Ota, M.~Centa, N.~Vaninov, P.~Hamon,
  R.~Levantovsky, S.~Hegde, S.~Chen, T.~Plitt, T.~O'Donnell, V.~Pothula, V.~Van
  Der~Heide, Z.~Mahmood, A.~Khamporst, A.~Horowitz, B.~Brown, D.~Homann,
  D.~Bogunovic, E.~Grasset, E.~Kenigsberg, J.~Faith, K.~Alexandropoulos,
  M.~Lafaille, N.~Bhardwaj, P.~Heeger, S.~Mehandru, U.~Laserson, T.~S. I.~R.
  Project, Trainees, Faculty,
  \href{https://doi.org/10.1038/s41577-020-0319-0}{Advancing scientific
  knowledge in times of pandemics}, Nature Reviews Immunology 20~(6) (2020)
  338--338.
\newblock \href {http://dx.doi.org/10.1038/s41577-020-0319-0}
  {\path{doi:10.1038/s41577-020-0319-0}}.
\newline\urlprefix\url{https://doi.org/10.1038/s41577-020-0319-0}

\bibitem{epidyHealth}
{Epidy Health Research Ltd.}, \href{https://corona.epidy.com}{{COVID-19 Risk
  Factors: literature database \& meta-analysis}}, accessed: 2020-07-21.
\newline\urlprefix\url{https://corona.epidy.com}

\bibitem{coronaWhy}
\href{http://coronawhy.org/}{{CoronaWhy}}, accessed: 2020-06-01 (2020).
\newline\urlprefix\url{http://coronawhy.org/}

\bibitem{coronaAbs}
V.~Tykhonov, A.~Polishko, A.~Kiulian, M.~Komar, {CoronaWhy: Building a
  Distributed, Credible and Scalable Research and Data Infrastructure for Open
  Science}, {SciNLP: Natural Language Processing and Data Mining for Scientific
  Text}, 2020.

\bibitem{cran}
K.~Hornik, \href{https://onlinelibrary.wiley.com/doi/abs/10.1002/wics.1212}{The
  comprehensive r archive network}, WIREs Computational Statistics 4~(4) (2012)
  394--398.
\newblock \href
  {http://arxiv.org/abs/https://onlinelibrary.wiley.com/doi/pdf/10.1002/wics.1212}
  {\path{arXiv:https://onlinelibrary.wiley.com/doi/pdf/10.1002/wics.1212}},
  \href {http://dx.doi.org/https://doi.org/10.1002/wics.1212}
  {\path{doi:https://doi.org/10.1002/wics.1212}}.
\newline\urlprefix\url{https://onlinelibrary.wiley.com/doi/abs/10.1002/wics.1212}

\bibitem{JHUCSSErepo}
\href{https://github.com/CSSEGISandData/COVID-19}{{COVID-19 Data Repository by
  the Center for Systems Science and Engineering (CSSE) at Johns Hopkins
  University}}, accessed: 2020-06-01 (2020).
\newline\urlprefix\url{https://github.com/CSSEGISandData/COVID-19}

\bibitem{HealthCanada}
\href{https://health-infobase.canada.ca/src/data/covidLive}{{Health Canada --
  covid19 cases in Canada}}, accessed: 2020-10-14 (2020).
\newline\urlprefix\url{https://health-infobase.canada.ca/src/data/covidLive}

\bibitem{TorontoData}
\href{https://www.toronto.ca/home/covid-19/covid-19-latest-city-of-toronto-news/covid-19-status-of-cases-in-toronto/}{{COVID-19:
  Status of Cases in Toronto}}, accessed: 2020-06-01 (2020).
\newline\urlprefix\url{https://www.toronto.ca/home/covid-19/covid-19-latest-city-of-toronto-news/covid-19-status-of-cases-in-toronto/}

\bibitem{OpenDataToronto}
\href{https://open.toronto.ca/dataset/covid-19-cases-in-toronto/}{{Open Data
  Toronto -- covid19 cases in the City of Toronto}}, accessed: 2020-10-14
  (2020).
\newline\urlprefix\url{https://open.toronto.ca/dataset/covid-19-cases-in-toronto/}

\bibitem{VCpandemics}
{Visual Capitalist},
  \href{https://www.visualcapitalist.com/history-of-pandemics-deadliest/}{{Visualizing
  the History of Pandemics \& The Race to Save Lives: Comparing Vaccine
  Development Timelines}}, accessed: 2021-01-20 (2020).
\newline\urlprefix\url{https://www.visualcapitalist.com/history-of-pandemics-deadliest/}

\bibitem{OWIDvaccination}
{Our World In Data}, \href{https://github.com/owid/covid-19-data/}{{Our World
  In Data -- CoVid19 Data Repository}}, accessed: 2021-01-20 (2020).
\newline\urlprefix\url{https://github.com/owid/covid-19-data/}

\bibitem{NCBI}
\href{https://www.ncbi.nlm.nih.gov/search/all/?term=sars-cov-2}{{National
  Center for Biotechnology Information}}, accessed: 2020-07-25.
\newline\urlprefix\url{https://www.ncbi.nlm.nih.gov/search/all/?term=sars-cov-2}

\bibitem{NCBIdatabases}
{NCBI Resource Coordinators},
  \href{https://doi.org/10.1093/nar/gkx1095}{{Database resources of the
  National Center for Biotechnology Information}}, Nucleic Acids Research
  46~(D1) (2017) D8--D13.
\newblock \href
  {http://arxiv.org/abs/https://academic.oup.com/nar/article-pdf/46/D1/D8/23162308/gkx1095.pdf}
  {\path{arXiv:https://academic.oup.com/nar/article-pdf/46/D1/D8/23162308/gkx1095.pdf}},
  \href {http://dx.doi.org/10.1093/nar/gkx1095}
  {\path{doi:10.1093/nar/gkx1095}}.
\newline\urlprefix\url{https://doi.org/10.1093/nar/gkx1095}

\bibitem{apePckg}
E.~Paradis, K.~Schliep, ape 5.0: an environment for modern phylogenetics and
  evolutionary analyses in {R}, Bioinformatics 35 (2019) 526--528.
\newblock \href {http://dx.doi.org/10.1093/bioinformatics/bty633}
  {\path{doi:10.1093/bioinformatics/bty633}}.

\bibitem{rentrezPckg}
D.~J. Winter, {rentrez}: an r package for the ncbi eutils api, The R Journal 9
  (2017) 520--526.
\newblock \href {http://dx.doi.org/10.7287/peerj.preprints.3179v2}
  {\path{doi:10.7287/peerj.preprints.3179v2}}.

\bibitem{kermack1927contribution}
W.~O. Kermack, A.~G. McKendrick, A contribution to the mathematical theory of
  epidemics, Proceedings of the royal society of london. Series A, Containing
  papers of a mathematical and physical character 115~(772) (1927) 700--721.

\bibitem{smith2004sir}
D.~Smith, L.~Moore, et~al., {The SIR model for spread of disease: the
  differential equation model}, Loci.(originally Convergence.)
  https://www.maa.org/press/periodicals/loci/joma/the-sir-model-for-spread-of-disease-the-differential-equation-model.

\bibitem{HARKO2014184}
T.~Harko, F.~S. Lobo, M.~Mak,
  \href{http://www.sciencedirect.com/science/article/pii/S009630031400383X}{{Exact
  analytical solutions of the Susceptible-Infected-Recovered (SIR) epidemic
  model and of the SIR model with equal death and birth rates}}, Applied
  Mathematics and Computation 236 (2014) 184 -- 194.
\newblock \href {http://dx.doi.org/10.1016/j.amc.2014.03.030}
  {\path{doi:10.1016/j.amc.2014.03.030}}.
\newline\urlprefix\url{http://www.sciencedirect.com/science/article/pii/S009630031400383X}

\bibitem{devtools}
H.~Wickham, J.~Hester, W.~Chang,
  \href{https://CRAN.R-project.org/package=devtools}{devtools: Tools to Make
  Developing R Packages Easier}, r package version 2.3.0 (2020).
\newline\urlprefix\url{https://CRAN.R-project.org/package=devtools}

\bibitem{shinyPckg}
W.~Chang, J.~Cheng, J.~Allaire, Y.~Xie, J.~McPherson,
  \href{https://CRAN.R-project.org/package=shiny}{shiny: Web Application
  Framework for R}, r package version 1.4.0.2 (2020).
\newline\urlprefix\url{https://CRAN.R-project.org/package=shiny}

\bibitem{shinydashboard}
W.~Chang, B.~{Borges Ribeiro},
  \href{https://CRAN.R-project.org/package=shinydashboard}{shinydashboard:
  Create Dashboards with 'Shiny'}, r package version 0.7.1 (2018).
\newline\urlprefix\url{https://CRAN.R-project.org/package=shinydashboard}

\bibitem{shinycssloaders}
A.~Sali, D.~Attali,
  \href{https://CRAN.R-project.org/package=shinycssloaders}{shinycssloaders:
  Add CSS Loading Animations to 'shiny' Outputs}, r package version 0.3 (2020).
\newline\urlprefix\url{https://CRAN.R-project.org/package=shinycssloaders}

\bibitem{plotly}
C.~Sievert, \href{https://plotly-r.com}{Interactive Web-Based Data
  Visualization with R, plotly, and shiny}, Chapman and Hall/CRC, 2020.
\newblock \href {http://dx.doi.org/10.1201/9780429447273}
  {\path{doi:10.1201/9780429447273}}.
\newline\urlprefix\url{https://plotly-r.com}

\bibitem{DT}
Y.~Xie, J.~Cheng, X.~Tan, \href{https://CRAN.R-project.org/package=DT}{DT: A
  Wrapper of the JavaScript Library 'DataTables'}, r package version 0.13
  (2020).
\newline\urlprefix\url{https://CRAN.R-project.org/package=DT}

\bibitem{dplyr}
H.~Wickham, R.~François, L.~Henry, K.~Müller,
  \href{https://CRAN.R-project.org/package=dplyr}{dplyr: A Grammar of Data
  Manipulation}, r package version 1.0.0 (2020).
\newline\urlprefix\url{https://CRAN.R-project.org/package=dplyr}

\bibitem{CDCppeEbola}
{Centers for Disease Control and Prevention},
  \href{https://www.cdc.gov/vhf/ebola/healthcare-us/ppe/calculator.html}{{Estimated
  Personal Protective Equipment (PPE) Needed for Healthcare Facilities}},
  accessed: 2020-07-14 (2020).
\newline\urlprefix\url{https://www.cdc.gov/vhf/ebola/healthcare-us/ppe/calculator.html}

\bibitem{CDCburnrate}
{Centers for Disease Control and Prevention},
  \href{https://www.cdc.gov/coronavirus/2019-ncov/hcp/ppe-strategy/burn-calculator.html}{{Personal
  Protective Equipment (PPE) Burn Rate Calculator}}, accessed: 2020-07-14
  (2020).
\newline\urlprefix\url{https://www.cdc.gov/coronavirus/2019-ncov/hcp/ppe-strategy/burn-calculator.html}

\bibitem{Ledford_2020}
H.~Ledford, R.~Van~Noorden,
  \href{http://dx.doi.org/10.1038/d41586-020-01695-w}{High-profile coronavirus
  retractions raise concerns about data oversight}, Nature 582~(7811) (2020)
  160--160.
\newblock \href {http://dx.doi.org/10.1038/d41586-020-01695-w}
  {\path{doi:10.1038/d41586-020-01695-w}}.
\newline\urlprefix\url{http://dx.doi.org/10.1038/d41586-020-01695-w}

\bibitem{Schriml:2020aa}
L.~M. Schriml, M.~Chuvochina, N.~Davies, E.~A. Eloe-Fadrosh, R.~D. Finn,
  P.~Hugenholtz, C.~I. Hunter, B.~L. Hurwitz, N.~C. Kyrpides, F.~Meyer, I.~K.
  Mizrachi, S.-A. Sansone, G.~Sutton, S.~Tighe, R.~Walls,
  \href{https://doi.org/10.1038/s41597-020-0524-5}{Covid-19 pandemic reveals
  the peril of ignoring metadata standards}, Scientific Data 7~(1) (2020) 188.
\newblock \href {http://dx.doi.org/10.1038/s41597-020-0524-5}
  {\path{doi:10.1038/s41597-020-0524-5}}.
\newline\urlprefix\url{https://doi.org/10.1038/s41597-020-0524-5}

\bibitem{Luengo-Oroz:2020aa}
M.~Luengo-Oroz, K.~Hoffmann~Pham, J.~Bullock, R.~Kirkpatrick, A.~Luccioni,
  S.~Rubel, C.~Wachholz, M.~Chakchouk, P.~Biggs, T.~Nguyen, T.~Purnat,
  B.~Mariano, \href{https://doi.org/10.1038/s42256-020-0184-3}{Artificial
  intelligence cooperation to support the global response to covid-19}, Nature
  Machine Intelligence 2~(6) (2020) 295--297.
\newblock \href {http://dx.doi.org/10.1038/s42256-020-0184-3}
  {\path{doi:10.1038/s42256-020-0184-3}}.
\newline\urlprefix\url{https://doi.org/10.1038/s42256-020-0184-3}

\bibitem{Hu:2020aa}
Y.~Hu, J.~Jacob, G.~J.~M. Parker, D.~J. Hawkes, J.~R. Hurst, D.~Stoyanov,
  \href{https://doi.org/10.1038/s42256-020-0185-2}{The challenges of deploying
  artificial intelligence models in a rapidly evolving pandemic}, Nature
  Machine Intelligence 2~(6) (2020) 298--300.
\newblock \href {http://dx.doi.org/10.1038/s42256-020-0185-2}
  {\path{doi:10.1038/s42256-020-0185-2}}.
\newline\urlprefix\url{https://doi.org/10.1038/s42256-020-0185-2}

\bibitem{ponce2020covid19analytics}
M.~Ponce, A.~Sandhel, covid19.analytics: An r package to obtain, analyze and
  visualize data from the corona virus disease pandemic (2020).
\newblock \href {http://arxiv.org/abs/2009.01091} {\path{arXiv:2009.01091}}.

\end{thebibliography}


\newpage

\appendix


\section{List of Figures, Tables and Listings}


\listoftables

\listoffigures

\lstlistoflistings


\section{Package Software Engineering Features}

\begin{enumerate}
	\item Version Control:
		\url{https://github.com/mponce0/covid19.analytics}

	\item Continuous Integration (CI), via TravisCI:
		\url{https://travis-ci.org/mponce0/covid19.analytics}

	\item Code Coverage, with \texttt{codecov}:
		\url{https://codecov.io/gh/mponce0/covid19.analytics}

	\item Unit Testing

	\item Publications:
		\begin{itemize}
			\item CRAN, \cite{covid19analytics}:
				\url{https://CRAN.R-project.org/package=covid19.analytics}

			\item Zenodo,
				\url{https://doi.org/10.5281/zenodo.4640306}

			\item JOSS, \cite{ponce2020covid19analytics}:
				\url{https://doi.org/10.21105/joss.02995}

			\item arXiv,
				\url{http://arxiv.org/abs/2009.01091}
		\end{itemize}
\end{enumerate}




\end{document}